\documentclass[final,3p,times,11pt,a4paper,authoryear]{elsarticle}

\usepackage{amssymb}
\usepackage{amsmath}
\usepackage{amsthm}

\usepackage{tabularx}
\usepackage{threeparttable}
\usepackage[utf8]{inputenc}   
\usepackage[T1]{fontenc}      
\usepackage{enumitem}
\setlist[itemize,enumerate]{
    leftmargin=*,        
    itemindent=\parindent, 
    listparindent=0pt,     
    labelsep=0.5em          
}
\usepackage{hyperref}
\newtheorem{lemma}{Lemma}
\newtheorem{definition}{Definition}
\newtheorem{problem}{Problem}

\newtheorem{assumption}{Assumption}

\allowdisplaybreaks
\usepackage{tikz}
\usetikzlibrary {arrows.meta}
\usepackage{float}
\usepackage{xcolor}
\usepackage{multirow}
\usepackage{comment}
\usepackage{adjustbox}
\usepackage{subcaption}
\usepackage{caption}
\usepackage{booktabs}
\newcommand{\STMMP}{{\texttt{2-STMMP}}}
\newcommand{\TMMP}{{\texttt{TMMP}}}
\newcommand{\MILP}{{\texttt{MILP}}}

\newcommand{\SAA}{{\texttt{SAA}}}
\newcommand{\DAA}{{\texttt{DAA}}}
\newcommand{\MIA}{{\texttt{MIA}}}
\newcommand{\MI}{{\texttt{MI}}}
\newcommand{\IOA}{{\texttt{IOA}}}
\newcommand{\UM}{{\texttt{UM}}}
\newcommand{\CEUM}{{\texttt{CEUM}}}
\newcommand{\IEUM}{{\texttt{IEUM}}}
\newcommand{\ASG}{{\texttt{ASG}}}
\newcommand{\LR}{{\texttt{LR}}}
\newcommand{\LS}{{\texttt{LS}}}
\newcommand{\SA}{{\texttt{SA}}}

\newcommand{\EV}{{\texttt{EV}}}
\newcommand{\EEV}{{\texttt{EEV}}}
\newcommand{\VSS}{{\texttt{VSS}}}
\newcommand{\VSSENTER}{{\texttt{VSS-E}}}
\newcommand{\VSSIMPROV}{{\texttt{VSS-I}}}
\newcommand{\DE}{\texttt{DE}}
\usepackage[ruled]{algorithm2e}
\SetKw{Return}{Return:}
\SetKw{Returns}{returns}
\SetKw{Output}{Output:}
\SetKw{Input}{Input:}
\SetKw{Stepfive}{Step 5:}
\SetKw{Stepfour}{Step 4:}
\SetKw{Stepthree}{Step 3:}
\SetKw{Steptwo}{Step 2:}
\SetKw{Stepone}{Step 1:}
\SetKw{StepA}{a)}
\SetKw{StepB}{b)}
\SetKwFor{While}{While}{do}{end}
\SetKwFor{ForEach}{For each}{do}{end}
\SetKwFor{For}{For}{do}{end}
\SetKwIF{If}{ElseIf}{Else}{If}{then}{Else if}{Else}{end}
\newcommand{\update}[1]{\textcolor{black}{#1}}

\newcommand{\changed}[1]{\textcolor{black}{#1}}
\journal{}

\begin{document}

\begin{frontmatter}



\title{Two-Stage Stochastic Capacity Expansion in Stable Matching under Truthful or Strategic Preference Uncertainty} 


\author[af1,af3]{Maria Bazotte\corref{corresp}} 
\ead{maria-carolina.bazotte-corgozinho@polymtl.ca}
\cortext[corresp]{Corresponding author.}
\affiliation[af1]{organization={MAGI, Polytechnique Montr\'{e}al},
            addressline={2500 Chem. de Polytechnique}, 
            city={Montr\'{e}al},
            postcode={H3T 1J4}, 
            state={QC},
            country={Canada}}

\author[af2,af3]{Margarida Carvalho} 
\ead{carvalho@iro.umontreal.ca}
\affiliation[af2]{organization={DIRO, Universit\'{e} de Montr\'{e}al},
            addressline={2920, Chem. de la Tour}, 
            city={Montr\'{e}al},
            postcode={H3T 1N8}, 
            state={QC},
            country={Canada}}
\affiliation[af3]{organization={CIRRELT},
            addressline={2920, Chem. de la Tour}, 
            city={Montr\'{e}al},
            postcode={H3T 1N8}, 
            state={QC},
            country={Canada}}

\author[af1,af3]{Thibaut Vidal} 
\ead{thibaut.vidal@polymtl.ca}

\begin{abstract}
Recent studies on many-to-one matching markets have explored agents with flexible capacity and truthful preferences, focusing on mechanisms that jointly design capacities and select a matching. However, in real-world applications such as school choice and residency matching, preferences are typically revealed \emph{after} capacity decisions are made, with matching occurring afterward; uncertainty about agents' preferences must be considered during capacity planning. Moreover, even under \emph{strategy-proof} mechanisms, agents may strategically misreport true preferences based on beliefs about admission chances. We introduce a two-stage stochastic matching problem with uncertain preferences, using school choice as a case study. In the first stage, the clearinghouse expands schools’ capacities before observing students’ \emph{reported} preferences. Students either act truthfully, reporting their \emph{true} preferences and producing exogenous uncertainty, or act strategically, submitting \emph{reported} preferences based on their \emph{true} preferences and perceived admission chances (which depend on capacities), introducing endogenous uncertainty. In the second stage, the clearinghouse computes the student-optimal stable matching based on schools' priorities and students' \emph{reported} preferences. \update{We investigate strategic behaviors based on widely-studied utility-maximization behavioral models. Thus,} endogenous \emph{reported} preferences are defined by these models as functions of capacity decisions and exogenous \emph{true} preferences. We handle uncertainty using sample average approximation ({\SAA}), develop behavior-based mathematical formulations and, due to problem complexity, propose Lagrangian- and local-search-based heuristics for near-optimal solutions. Our {\SAA}-based approaches outperform the average scenario approach on students' matching preferences and admission outcomes, emphasizing the impact of stochastic preferences on capacity decisions. Student behavior significantly influences capacity design, highlighting the need to consider misreporting.
\end{abstract}



\begin{keyword}
stochastic programming \sep endogenous (decision-dependent) uncertainty \sep many-to-one markets \sep stable matching \sep utility-preference behavior



\end{keyword}

\end{frontmatter}




\section{Introduction}

Two-sided matching markets often rely on centralized mechanisms to make assignment decisions among participating agents based on their preferences. Examples include school and college admission, where a clearinghouse matches students to schools or universities. Most research focuses on designing mechanisms that ensure \emph{stability}, discouraging agents from circumventing the matching~\citep{roth1990twosided}.

In the context of stable matching, numerous works investigate \emph{strategy-proof} mechanisms, where agents do not benefit by misrepresenting their preferences, regardless of the actions of other agents. For example, the student-proposing version of the Deferred Acceptance algorithm ({\DAA})~\citep{gale1962college} is known to be strategy-proof for students in school choice applications. However, empirical evidence shows that agents often misreport their preferences or act strategically even under such mechanisms~\citep{hassidim2017mechanism}. This occurs for several reasons. First, agents may have limited information about the mechanism, or mistrust its functionality and the truthfulness of other participants, prompting them to strategize~\citep{rees2018suboptimal}.~\citet{rees2018experimental} showed that many medical students misreported their preferences to the U.S. National Resident Matching, despite the process being designed to discourage such behavior. In school choice, students might not list selective schools despite preferring them due to uncertainty about admission chances~\citep{fack2019beyond}. Thus, students' subjective beliefs about admission chances can lead to divergences between reported and true preferences even in a strategy-proof environment~\citep{arteaga2022smart}. In higher-education admissions, students declined the opportunity to receive a tuition waiver in Hungary~\citep{shorrer2023dominated} and ranked unfunded programs above identical funded ones, despite the funding having no strings attached, in Israel \citep{hassidim2021limits}. Second, in many applications, the length of agents' preference lists is limited, compelling them to strategically select their reported preferences~\citep{haeringer2009constrained}. Third, non-strategy-proof mechanisms are often employed, leading agents to choose their preferences strategically~\citep{abdulkadirouglu2003school}.

Agents' strategic behavior is shaped by both their true preferences and their beliefs about admission chances~\citep[see, e.g.,][]{agarwal2020revealed}, with distinct models capturing this behavior, the most common being the portfolio choice problem introduced by~\citet{chade2006simultaneous}. As shown by~\citet{kapor2020heterogeneous}, these beliefs often differ from agents' actual admission chances, undermining the effectiveness of the mechanism. In many-to-one matching markets, the capacity of the many-sided agents affects both the real and perceived admission chances of the one-sided agents. These chances are defined by the minimum acceptance requirement of the many-sided agents~\citep{kapor2020heterogeneous}, which depends on their capacities. Misinformation regarding changes in the many-sided agent capacity can thus distort the one-sided agents' perceptions of their chances, leading to choices that differ from reality~\citep{agarwal2018demand}. 

Previous research on many-to-one matching markets has mainly addressed settings in which many-sided agents have fixed capacity. More recently, studies have begun to examine the flexible capacity case in which the clearinghouse can adjust capacities~\citep[see, e.g.,][]{chen2023optimal,bobbio2025capacity}. These works, often in the context of school choice, assume that students’ preferences are known in advance and also truthful, and thus that the clearinghouse jointly determines both school capacity allocations and student-stable assignments in response to deterministic preferences. However, capacity decisions must often be made before preferences are revealed, regardless of whether they are reported truthfully or strategically. In vaccine distribution, governments must establish inoculation venues and allocate vaccine doses before knowing citizens’ venue preferences~\citep{kumano2022quota}. In school systems, long-term investments in infrastructure must be planned under uncertainty about future student enrollment and preferences. For example, school districts in the U.S. often enroll more students than their physical capacity allows, leading to overcrowding. The survey by~\citet{survey} indicates that 22\% of schools experience overcrowding, with 8\% exceeding their capacity by 25\%. While temporary measures, such as portable classrooms and staggered schedules, offer short-term relief, persistent overcrowding requires permanent capacity expansion. Similar timing constraints arise in other public sector decisions: education ministries may allocate funding before knowing applicants’ preferences, healthcare authorities may increase residency positions without knowing candidates' choices, and refugee resettlement agencies often distribute regional support resources before assessing family preferences. In all such cases, agents' preferences are uncertain at the time of resource or capacity allocation and must be modeled accordingly.

For consistency, we adopt the terminology of school choice introduced by~\citet{abdulkadirouglu2003school}, referring to the two groups of agents as students and schools. The previous discussion illustrates three key points:
(i) Schools' capacities are often expanded to enhance future matchings when students' preferences are still unknown and, hence, uncertain;
(ii) Students often behave strategically, even in strategy-proof environments, with their reported preferences influenced by their true preferences and perceived admission chances, as captured by distinct models previously proposed to represent this behavior; and 
(iii) Students' perceived admission chances depend on the schools' capacity.
However, no study has investigated matching markets with the first point, nor examined them with all three points simultaneously. 

Hence, prior approaches to capacity design in stable matching face two key limitations. First, they often rely on deterministic models in which capacity decisions are based on fully known or deterministic aggregated (i.e., averaged) student preferences. Second, these approaches typically assume that students report their preferences truthfully. Ignoring preference uncertainty or potential strategic behavior may lead to capacity allocations that fail to achieve the intended outcomes in practice (e.g., to maximize expected student-optimal matching).

These modeling choices have both economic and societal consequences. Economically, ignoring uncertainty or misrepresenting behavior can lead to inefficient capacity investments, such as over- or under-expanding certain schools, resulting in wasted resources or missed opportunities. Societally, misaligned capacity can leave students unmatched or assigned to less-preferred schools, whereas well-informed capacity expansions can admit more students and assign them to more-preferred schools. Accounting for uncertainty and correct behavior modeling is therefore crucial to both efficient resource allocation and maximizing student welfare.

\paragraph{\textbf{Contributions}}
This work introduces a canonical model of a \emph{two-stage stochastic program} for a centralized many-to-one matching market, covering applications such as school choice, and job market, and proposes a mechanism to optimize, in expectation, student outcomes under preference uncertainty. In the first stage, the clearinghouse decides on school capacity expansions before students' preferences are revealed. Students' uncertain preferences are either exogenous (decision-independent), i.e., they report their true preferences, or endogenous (decision-dependent), i.e., they strategically adjust their preferences based on their perceived admission chances influenced by capacity decisions.
In the second stage, the clearinghouse implements a student-optimal stable matching based on the decided school capacities and students' realized reported preferences. We refer to this problem as the \emph{two-stage stochastic capacity expansion in two-sided many-to-one matching problem with uncertain preferences} ({\STMMP}). Our main contributions are as follows: 
\begin{enumerate}
\item We introduce the {\STMMP} and formally define it. In strategic cases, which induce endogenous uncertainty, we follow the random-variable transformation method of~\citet{bazotte2025solving} by defining the endogenous reported preferences as a transformation of the first-stage capacity decisions and the exogenous true preferences. Our approach thus captures any strategically reported preferences expressible in this form.
\item We explore three student preference-reporting behaviors, motivated by prior empirical and modeling work on agent behavior in school choice and matching markets: (i) Utility Maximization ({\UM}), where students report their true preferences, resulting in exogenous uncertainty (as commonly assumed in the literature on many-to-one matching markets); (ii) Conjoint Expected Utility Maximization ({\CEUM}); and Individual Expected Utility Maximization ({\IEUM}). In both {\CEUM} and {\IEUM}, students strategically report preferences by solving a utility maximization problem that accounts for their true preferences and perceived admission chances, which depend on capacity decisions. This utility maximization acts as the transformation from true preferences and capacity decisions to reported preferences. Both {\CEUM} and {\IEUM} build on the foundational portfolio choice problem of~\citet{chade2006simultaneous}, most common in prior works: {\CEUM} exactly represents it, while {\IEUM} is a simplified approximation. We provide mathematical formulations for all behaviors. 
%
\item Building on the sample average approximation ({\SAA}) method, as scenario enumeration is intractable, we design solution approaches for the {\STMMP} under each behavior. We develop mathematical formulations for the corresponding {\SAA} programs of the {\STMMP}. We model the second-stage program of the {\STMMP} by extending well-established formulations of many-to-one matching problems to incorporate uncertainty and to model students' reported preferences via the proposed utility-maximizing transformations. In the strategic cases, this leads to a bilevel program, where the inner optimization problem defines students' reported preferences. For both strategic behaviors, we derive an exact single-level formulation based on behavior-specific algorithms that yield an optimal solution to the inner problem. Solving the {\SAA} program under {\CEUM} and {\IEUM} with a mixed-integer linear solver is significantly more computationally demanding than under {\UM}, due to the added complexity of endogenous uncertainty. Even under {\UM}, solving large instances of the {\SAA} program is computationally challenging. Hence, we introduce behavior-specific heuristics and show in our experiments that they efficiently compute high-quality, near-optimal solutions within short runtimes:
\begin{enumerate}
\item \changed{For the {\SAA} program under {\UM} behavior, we develop a Lagrangian relaxation scheme that relaxes the stability constraints and provides stronger lower bounds than the linear relaxation. We combine this with the {\DAA} to propose a Lagrangian ({\LR}) heuristic that yields tight lower and upper bounds for the problem.}
\item For the {\SAA} program under all three behaviors, \changed{we propose local search heuristics: Greedy Local Search ({\LS}) and Simulated Annealing ({\SA}).} They explore the feasible space of capacity decisions, apply behavior-specific algorithms to define students' reported preferences, and use the {\DAA} to evaluate resulting matchings.
\end{enumerate}
Hence, on one hand, we propose an exact solution approach for the {\SAA} program under all behaviors, which is well-suited when capacity decisions are made well in advance of the matching process and computational time is not restrictive. On the other hand, we provide efficient heuristics, which are particularly relevant for large-scale instances and time-sensitive contexts where capacity decisions must be implemented rapidly. Thus, these approaches form a versatile toolkit for solving the {\STMMP} across different real-world situations.
\item We conduct extensive experiments to assess the impact of uncertainty and behavioral modeling on capacity decisions. Our results show that, overall, the clearinghouse selects significantly different allocations depending on student behavior. Capacity decisions made under the nonstrategic {\UM} behavior result in lower-quality assignments when students behave strategically, compared to decisions that account for their actual strategic behavior. This underscores the importance of accurately modeling student strategic reporting in capacity planning, in contrast to prior work, which assumes truthful reporting. This quality decline is more pronounced when decisions are based on an incorrect strategic behavior, i.e., assuming {\IEUM} when students follow {\CEUM}. Nonetheless, when the maximum number of schools that students can list in their preference report is large, decisions under {\UM} behavior better approximate those from actual {\CEUM} behavior; when it is small, they better approximate those from actual {\IEUM} behavior. In such cases, relying on {\UM} behavior offers a simpler problem with exogenous uncertainty as an effective approximation. Across all behaviors, {\SAA}-based approaches outperform the deterministic or average scenario approach ({\EV} problem) used in prior work, consistently producing capacity expansion decisions that yield higher-quality matchings and greater student welfare. This highlights the importance of stochastic preference modeling in capacity planning.
\end{enumerate}

 
\section{Related Works} \label{sec:literature}
 
\subsection{Matching Markets}

We focus on \emph{two-sided many-to-one matching problems} ({\TMMP}'s), a topic with extensive prior work. We underline studies most closely related to our setting, and refer to~\citet{manlove2013algorithmics} for a detailed background.

\paragraph{Mechanisms and Mathematical Formulations} 
The student-proposing version of the {\DAA} yields a student-optimal stable matching in polynomial time that is strategy-proof for the students~\citep{gale1962college}. This matching can also be obtained efficiently with linear programming~\citep{baiou2000stable}. However, many special features make the problem of finding and optimizing over the set of stable-matchings NP-hard~\citep{ronn1990np,biro2010college}. In these cases, integer programming techniques have proven effective. Formulations to optimize over stable matchings in problems with ties, lower and common quotas, couples, paired applications, and other extensions, have been proposed~\citep{Kwanashie2014,delorme2019mathematical,biro2014hospitals,agoston2016integer,rios2024stable}. Unlike our case, these studies assume agents' capacities and preferences are fixed and deterministic.

\paragraph{Capacity expansion} A small but growing body of work has examined stable matching problems with capacity expansion, primarily in the context of school choice. In these models, the clearinghouse simultaneously allocates additional seats to schools and identifies a student-optimal stable matching. These studies assume that student preferences are known before capacity planning, leading to capacity allocations that enhance the existing matching. Most papers consider that schools start with initial capacities, which can be increased within the limits of the maximum available extra seats. Unlike the classical student-optimal stable matching problem, the capacity expansion variant is NP-hard~\citep{bobbio2022capacity}. To address it,~\citet{bobbio2025capacity} offers a toolkit composed of mixed-integer programming-based methods.~\citet{afacan2024capacity} introduced a myopic algorithm in which each iteration obtains a student-optimal stable matching under a fixed capacity using the {\DAA}. Both studies showed that their mechanisms are not strategy-proof in the general case, though~\citet{bobbio2025capacity} also showed that their approach is \emph{strategy-proof in the large}.

\citet{abe2022anytime} introduced a heuristic based on upper-confidence Monte Carlo search trees, which explores the space of capacity expansion decisions and uses the {\DAA} to compute a student-optimal stable matching for each fixed capacity.~\citet{kumano2022quota} studied a related problem in which schools (or programs) are grouped into departments, each with a limit on student intake; capacity decisions determine the allocation of seats among schools within each department. Similar to~\citet{afacan2024capacity}, their method iteratively applies the {\DAA} for each fixed capacity expansion solution, but differs in its initialization and capacity update rules. Hence, while prior work has examined how capacity choices affect matching outcomes~\citep[e.g.,][]{che2016decentralized,romero2013games}, they typically assume that preferences are known at the time capacities are chosen. In contrast, to our knowledge, no study has considered capacity decisions made \emph{before} student preferences (whether truthful or strategic) are revealed.

\paragraph{Stochastic setting} Most research on {\TMMP} has focused on deterministic settings. A few studies consider that a stable matching is defined first, and uncertainty arises from agents leaving, declining assignments, or new agents entering the system~\citep[e.g.,][]{bampis2023online}. Others examine robust stable matchings against pair breakups or changes in preferences~\citep[e.g.,][]{bredereck2020adapting}. In contrast, our setting has uncertainty from the timing of students’ preference submissions, which occur after capacity decisions. 

\paragraph{Agents' behavior}
Agents' behavior in {\TMMP} has been extensively studied, with behavior models developed to estimate agents' distributional preferences from empirical data; see~\citet{rees2023behavioral} for a review. That work defines three main reasons for misreporting true preferences: (i)~incorrect incentive processing; (ii)~nonstandard preferences; and (iii)~biased beliefs about admission chances, which has received the most attention~\citep{benjamin2019errors}. A foundational model of strategic behavior is the portfolio choice problem proposed by~\citet{chade2006simultaneous}, where agents select a subset of ranked options within a limit to maximize expected utility based on their admission chances and true preferences~\citep[see, e.g.,][]{agarwal2018demand}. Agents' preferences have also been estimated under truthful reporting or mechanism stability~\citep[e.g.,][]{fack2019beyond}. For a review of behavior models estimating students' preferences in school choice, see~\citet{agarwal2020revealed}. Typically, agents' behavior and the {\TMMP} are treated separately: agents' preferences are defined first (e.g., via the portfolio choice problem) and then used as input for the {\TMMP}. In our case, school capacities affect perceived admission chances, which in turn influence reported preferences, motivating a joint optimization framework integrating behavior modeling with capacity design.

\subsection{Two-Stage Stochastic Programs with Exogenous or Endogenous Uncertainty}\label{subsec:literature-stochastic}

Most research on two-stage programs focuses on \emph{exogenous uncertainty}, where the random vector is independent of the decisions~\citep{li2021review}. \changed{\emph{Endogenous uncertainty} arises when either the random vector’s distribution (Type~1) or its realization time (Type~2) is decision-dependent~\citep{goel2006class}; in {\STMMP}, it is Type~1.} \changed{Applications with Type~1 endogenous uncertainty include infrastructure planning, such as network design and facility location with protection~\citep{peeta2010pre,bhuiyan2020stochastic}, risk management~\citep{krasko2017two,bhuiyan2019stochastic}, scheduling~\citep{karaesmen2004overbooking}, and newsvendor~\citep{ernst1995optimal,ekin2017augmented} problems.}

\changed{Prior work focuses on discrete endogenous random vectors, solving the deterministic equivalent ({\DE}) formulation of the stochastic program~\citep[e.g.,][]{hellemo2018decision,bhuiyan2020stochastic}. This is challenging due to the nonlinearity and nonconvexity of the {\DE}, arising from the endogenous probability distribution and its product with the second-stage objective, making stochastic programs with endogenous uncertainty substantially harder than those with exogenous uncertainty~\citep{dupacova2006optimization}. Solution approaches are typically problem-specific~\citep{bazotte2025solving}, including solving the mixed-integer nonlinear {\DE} with standard solvers~\citep{krasko2017two}, designing greedy algorithms when the objective is submodular~\citep{karaesmen2004overbooking}, or using an approximate problem based on first-order terms of the objective~\citep{peeta2010pre}. When first-stage decisions select a probability distribution from a predefined set (discrete selection of distributions), the {\DE} is often linearized using binary variables to indicate the chosen distribution, introducing many variables and big-M constraints~\citep[e.g.,][]{bhuiyan2020stochastic}. ~\citet{pantuso2025shaped} extended the (integer) L-shaped method to this case, with distribution-specific cuts activated via big-M terms from the indicator functions mapping first-stage decisions to selected distributions. These cuts can be weak, especially when few first-stage decisions induce the same distribution, substantially slowing convergence. For discrete endogenous vectors with numerous realizations or continuous random vectors, solving the {\DE} is intractable due to explicit scenario enumeration, while sampling is restricted by decision-dependent distributions. Research in this setting remains limited. Initial approaches rely on discretization~\citep{hellemo2018decision}, which is challenging when the distribution lacks a closed form, or on augmented probability simulation, applicable when the second-stage admits analytical expressions~\citep{ekin2017augmented}. A search procedure for a single decision was also proposed by~\citet{ernst1995optimal}.}

\changed{\citet{bazotte2025solving} then introduced a general-purpose method that models endogenous uncertainty by a transformation function of first-stage decisions and exogenous uncertainty. The transformed program is obtained by substituting the endogenous random vector with this function and thus has only exogenous uncertainty, allowing it to be solved with well-studied methods~\citep{birge2011introduction}, including sampling approaches such as the {\SAA}. The authors show that the transformed program often exhibits tractable properties, including a convex linear relaxation in many cases. They demonstrate that suitable transformations exist even when the endogenous distribution is complex or lacks a closed form (e.g., normal distribution).}

\changed{We adopt this approach to solve the {\STMMP}; however, transformations depend on the particular properties of the endogenous random vector. This is the first study of this type of uncertainty in {\TMMP}, so methodological advances in defining transformations and solving the transformed program are needed. Endogenous customer choices~\citep[e.g.,][]{paneque2021integrating} are the closest analogue, although customers select a single option, whereas students select and rank a subset of schools. Discrete selection of distributions is also relevant: each capacity expansion design induces a distinct distribution. However, the number of designs is large, and students’ reported preference distributions involve many scenarios and often lack a closed form.}

\section{Problem Statement} \label{sec:prob-stat}

Figure~\ref{fig:timeline-problem} illustrates the {\STMMP} timeline in three main steps. At Time Step~1 (first stage of the stochastic program), the clearinghouse determines a capacity expansion plan: it allocates additional seats to schools within a predefined budget. \changed{We focus on applications with excess student demand, where existing school seat infrastructure is fixed and cannot be repurposed, so that only capacity expansions are feasible to accommodate students.} At this point, schools also rank students based on their priorities, which are treated as deterministic. \changed{We note that schools’ priority orderings over students are determined by institutional rules, including, among others, priority given to students with siblings already enrolled or whose parents are employed by the school.} At Time Step~2, a scenario unfolds: students' random \emph{true} preferences over schools are realized (revealed). Students then submit their \emph{reported} preferences to the clearinghouse. We assume that students are aware of the capacity and their individual priority at each school, but lack complete information about the priorities assigned to other students. This reflects a realistic case where students have partial information about their standing but remain uncertain about their position in the broader applicant pool. They are also unaware of other students' true and reported preferences, as submissions occur simultaneously. This setting captures a common form of information asymmetry in large-scale market design~\citep{azevedo2019strategy}. We assume that students follow distinct behaviors: (i) truthful behavior, in which they report their true preferences directly, leading to exogenous uncertainty \changed{as students’ random true preferences are independent of school capacities, capturing variability in student-school affinities due to factors
such as educational quality, commute distance, or sibling enrollment;} and (ii) strategic behavior, in which their reported preferences depend on their true preferences and their perceived admission chances, defined by their priority at each school and the schools’ capacities, resulting in endogenous uncertainty. We consider that students can submit a maximum number of schools in their reported preferences, reflecting cases where agents’ preference lists are limited in length, incentivizing them to be strategic~\citep{haeringer2009constrained}. At Time Step~3 (second stage of the stochastic program), the clearinghouse computes a student-optimal stable matching based on reported preferences, priorities, and expanded capacities. Overall, the {\STMMP} seeks capacity decisions that respect the budget while minimizing the value of the second stage, the expected sum of the ranks that students assign to their matched schools in the stable matching, accounting for all possible realizations of students' preferences and their probabilities. \changed{We note that the {\STMMP} is a strategic or tactical-level decision problem: capacity expansion decisions are made prior to the actual admission process based on data from previous years. Therefore, once the admission process is executed, it may involve multiple phases of student acceptance or refusal, while capacities are already fixed.}

\changed{Beyond overcrowding in U.S. school districts,} a related application arises in the Chilean school choice system, which has been under severe strain due to persistent excess demand. \changed{In particular, the total number of seats available across all schools is smaller than the total number of students applying (e.g., roughly 0.9 seats per applying student in certain cases), highlighting the need to expand capacities to accommodate all students and increase student welfare.} To address this, Chile’s Ministry of Education introduced the ``Plan de Fortalecimiento de Matrícula''~\citep{mineduc2022plan}.~\citet{bobbio2025capacity} analyses this policy using a deterministic version of the {\STMMP} with truthful behavior, where student preferences are assumed to be known. In their setting, \changed{capacity expansion is} determined based on current-year applications, i.e., after students have submitted their choices, \update{and capacity and matching decisions are jointly designed.} However, as noted by the authors, policymakers may need to decide capacities before preferences are revealed or submitted, or may prefer permanent expansions rather than temporary solutions to accommodate multi-year enrollment fluctuations. Thus, uncertainty in student preferences must be explicitly considered, which motivates the framework developed in this work. \changed{Unlike previous works, the {\STMMP} allows us to model capacity planning decisions when reported preferences (truthful or strategic) are unknown or uncertain, an aspect that has not been explored before.}

\begin{figure}[t]
    \small
    \caption{Timeline of the {\STMMP}}\label{fig:timeline-problem}
    \begin{adjustbox}{width=.7\textwidth,center}
    \begin{tikzpicture}
        \draw[->, thick,>=Latex] (0,0) -- (14,0);
        \foreach \x in {2,7,12}
        \draw (\x cm,3pt) -- (\x cm,-3pt);
        \draw (2,0) node[below=3pt] {Time 1} node[above=16pt] {Clearinghouse decides extra capacity} node[above=3pt] {allocation and schools rank students} node[below=16pt] {First stage};
        \draw (7,0) node[below=3pt] {Time 2} node[above=16pt] {Students report} node[above=3pt] {their preferences} node[below=16pt] {Random variable realization};
        \draw (12,0) node[below=3pt] {Time 3} node[above=16pt] {Clearinghouse implements a} node[above=3pt] {student-optimal stable matching} node[below=16pt] {Second stage};
    \end{tikzpicture}
    \end{adjustbox}
\end{figure}

\section{Problem Formulation} \label{sec:formulation}

Let $\mathcal{C}=\{c_1, \ldots, c_m\}$ be the set of $m$ schools, and $\mathcal{S}=\{s_1, \ldots, s_n\}$ be the set of $n$ students, all assumed, without loss of generality, to belong to the same education level.\footnote{Assuming a fixed number of students $n$ is justified by the reliable availability of annual statistics tracking student progression across education levels.} We \changed{let} $\varnothing$ represent an unassigned student. Each school $c \in \mathcal{C}$ has a capacity of $q_c \in \mathbb{Z}_+$ seats, while $\varnothing$ has a sufficiently large capacity such that all students may remain unassigned (i.e., $q_{\varnothing} = n$). Let $B$ be the maximum number of extra seats allocated to schools, with $\mathcal{B}=\{1, \ldots, B\}$. Each school $c \in \mathcal{C}$ ranks all the students in $\mathcal{S}$ according to a deterministic strict order $\succ_c$, with $\succ_{\mathcal{C}} \, = \left\{ \succ_c \right\}_{c \in \mathcal{C}}$. \changed{This order is defined} by scores $e_{s,c}$, where school $c$ prefers student $s$ over student $s'$ (i.e., $s \succ_c s'$) iff $e_{s,c} > e_{s',c}$. We assume $e_{s,c}$ to be the reverse rank of student $s$ in school $c$ and thus $e_{s,c} \leq n$. Students can list at most $ K $ schools in their reported preferences, \changed{with $K < m$, and $\mathcal{K} = \{1, \dots, K\}$.}

The continuous exogenous random \changed{vector ${\boldsymbol{u}_s = \left\{\boldsymbol{u}_{s,c}\right\}_{c \in \mathcal{C} \cup \{\varnothing\}}}$} denotes the utility of a student $s \in \mathcal{S}$ upon admission to schools $c \in \mathcal{C}$. Let $\mathcal{W}$ be the set of their realizations (scenarios). \changed{This random utility vector specifies the random \emph{true} strict preference order of students $\boldsymbol{\succ}_{\mathcal{S}} \, =  \left\{\boldsymbol{\succ}_{s}\right\}_{s \in \mathcal{S}}$ over $\mathcal{C} \cup \{\varnothing\}$. For each scenario $w \in \mathcal{W}$, we define the realization ${u_s^w = \left\{u_{s,c}^w\right\}_{c \in \mathcal{C} \cup \{\varnothing\}}}$, leading to the \emph{true} preference realization $\succ_{\mathcal{S},w} = \left\{\succ_{s, w}\right\}_{s \in \mathcal{S}}$,} where student $s$ \emph{prefers} school $c$ over school $c'$ in scenario $w$ (i.e., $c \succ_{s,w} c'$) iff $u_{s,c}^w > u_{s,c'}^w$. We use $c' \succeq_{s,w} c$ to indicate that either $c' \succ_{s,w} c$ or $c' = c$. We set $u_{s,\varnothing}^w = 0$ for every student $s \in \mathcal{S}$ and scenario $w \in \mathcal{W}$, and allow for $u_{s,c}^w < 0$ for some $c \in \mathcal{C}$ such that $\varnothing \succ_{s, w} c$, i.e., students may truly prefer being unassigned over being matched with certain schools. Thus, in certain scenarios, fewer than $n$ students may apply to the matching, as those who prefer $\varnothing$ over all schools opt out of centralized assignments (e.g., choosing private options). We let $\mathcal{C}_{s,w} \subseteq \mathcal{C}$ be the set of schools in the true preference $\succ_{s,w}$, i.e., the set of schools that student $s$ \emph{truly} prefers over being unassigned in scenario $w$ ($c \in \mathcal{C}_{s,w}$ iff $c \succ_{s,w} \varnothing$), and $r_{s,c,w}$ be the rank of school $c \in \mathcal{C}_{s,w}$ in the true preference $\succ_{s,w}$, \changed{i.e., $r_{s,c,w}=1$ is student $s$'s true top preference in scenario $w$.} 

Nonetheless, students may behave differently, and their random \emph{reported} preference order \changed{based on behavior $a \in \mathcal{A}=\{{\CEUM},{\IEUM},{\UM}\}$ (see Section~\ref{subsec:students-behavior}),} may differ from their \emph{true} preference order $\boldsymbol{\succ}_{\mathcal{S}}$, while respecting the size limit of $K$ schools. Reported preferences are always shaped by students’ true preferences and, in strategic cases, also by their perceived admission chances, thus by capacity decisions. They are then exogenous in the nonstrategic case ($a = {\UM}$) and endogenous in the strategic cases ($a \in \{{\CEUM},{\IEUM}\}$). 

\changed{Let $x = \left\{x^j_c\right\}_{c \in \mathcal{C}, j \in \mathcal{B}}$ be the first-stage capacity expansion decisions of the {\STMMP}, where ${x_c^j \in \{0,1\}}$ equals one if at least $j \in \mathcal{B}$ extra seats are allocated to school $c \in \mathcal{C}$, and zero otherwise. Given $x$, the expanded capacities are ${\boldsymbol{\bar{q}}_x = \left\{q_c + \sum_{j \in \mathcal{B}} x_c^j\right\}_{c \in \mathcal{C}}}$. We thus denote the random \emph{reported} preference order under behaviors $a \in \mathcal{A}$ as ${\boldsymbol{\succ}^{a}_{\mathcal{S}} \left( \boldsymbol{\bar{q}}_x, \boldsymbol{u}, K \right) = \left\{\boldsymbol{\succ}_{s}^{a} \left( \boldsymbol{\bar{q}}_x, \boldsymbol{u}_s, K \right) \right\}_{s \in \mathcal{S}} }$, with ${\boldsymbol{u} = \left\{ \boldsymbol{u}_{s} \right\}_{s \in \mathcal{S}}}$, evidencing their dependence on students’ \emph{true} preferences $\boldsymbol{u}$, expanded capacities $\boldsymbol{\bar{q}}_x$ (for strategic behaviors), and list size limit $K$. This explicitly expresses the endogenous preference order as a function of first-stage decisions and exogenous uncertainty as proposed by~\citet{bazotte2025solving}. For each realization $u^w$ of $\boldsymbol{u}$ (with $w \in \mathcal{W}$), we obtain a realization ${\succ_{\mathcal{S},w}^{a} \left(\boldsymbol{\bar{q}}_x, u^w, K\right) = \left\{ \succ_{s,w}^{a} \left(\boldsymbol{\bar{q}}_x, u^w_s, K\right) \right\}_{s \in \mathcal{S}}}$ of the reported preference order. Note that students’ endogenous preference orders have their own (large) set of realizations, i.e., all ranked subsets of $K$ schools, and ${\succ_{\mathcal{S},w}^{a} \left(\boldsymbol{\bar{q}}_x, u^w, K\right)}$ belongs to this set. To simplify notation, we henceforth omit the explicit arguments $(\boldsymbol{\bar{q}}_x, \boldsymbol{u}, K)$ and write only the function names.} We denote \changed{$\mathcal{C}^{a}_{s,w}\left(\boldsymbol{\bar{q}}_x, u^w_s, K\right)$ or $\mathcal{C}^{a}_{s,w}$} as the set of schools in the reported preference $\succ_{s,w}^{a}$, satisfying $|\mathcal{C}_{s,w}^{a}| \leq K$ \changed{and $\mathcal{C}_{s,w}^{a} \subseteq \mathcal{C}_{s,w}$}. We let \changed{$r^{a}_{s,c,w} \left(\boldsymbol{\bar{q}}_x, u^w_s, K\right)$ or $r^{a}_{s,c,w}$} be the rank of school $c \in \mathcal{C}_{s,w}^{a}$ in the reported preference list $\succ_{s,w}^{a}$ ($r^{a}_{s,c,w}=1$ implies that $c$ is student $s$’s top preference in scenario $w$), and $r_{s,\varnothing,w}^{a} = |\mathcal{C}_{s,w}^{a}|+1$ be the penalty for student \changed{$s$'s unassignment} in scenario $w$. Table~\ref{tab:notation} illustrates our conventions for notation.
\begin{table}[t]
    \centering
    \caption{Notation convention for preferences for behavior $a \in \mathcal{A}$} \label{tab:notation}
    \resizebox{0.8\textwidth}{0.08\textwidth}{%
    \begin{tabular}{cccc}
        \hline
        Random True Preference & Realized True Preference & Random Reported Preference & Realized Reported Preference \\
        \hline
        $\boldsymbol{\succ}_{\mathcal{S}} = \left\{ \boldsymbol{\succ}_{s} \right\}_{s \in \mathcal{S}} $ & $\succ_{\mathcal{S},w} = \left\{ \succ_{s,w} \right\}_{s \in \mathcal{S}}$ & $\boldsymbol{\succ}_{\mathcal{S}}^{a} = \left\{ \boldsymbol{\succ}_{s}^a \right\}_{s \in \mathcal{S}}$ & $\succ_{\mathcal{S},w}^{a}   = \{ \succ^a_{s,w}  \}_{s \in \mathcal{S}}$ \\
        & \changed{$ \mathcal{C}_{s,w} \subseteq \mathcal{C}, \, \forall s \in \mathcal{S}$} & & \changed{$\mathcal{C}_{s,w}^a    \subseteq \mathcal{C}_{s,w}, \, \forall s \in \mathcal{S} $}\\
        & \changed{$r_{s,c,w}, \forall (s,c) \in \mathcal{S} \times \mathcal{C}_{s,w}$} & & \changed{$r_{s,c,w}^a  , \, \forall (s,c) \in \mathcal{S} \times \mathcal{C}_{s,w}^a   $}\\
        \hline
        \\
    \end{tabular}}
    \begin{minipage}{0.9\textwidth}
        \scriptsize 
         \changed{Subscripts: $\mathcal{S}$ (set of students); $s$ (a particular student); $w$ (a particular scenario). Superscript: $a$ (student behavior). We use $\succ_{s,w}$ and $\succ^a_{s,w}  $ to denote complete rankings of schools. Sets $\mathcal{C}_{s,w}$ and $\mathcal{C}^a_{s,w}  $ represent the schools preferred over the unassigned option $\varnothing$ in rankings $\succ_{s,w}$ and $\succ^a_{s,w}  $. Parameters $r_{s,c,w}$ and $r_{s,c,w}^a $ indicate the rank of school $c$ in rankings $\succ_{s,w}$ and $\succ^a_{s,w} $.}
    \end{minipage}
\end{table}

\subsection{Mathematical Formulation}\label{subsec:mathematical-form}

\changed{For a capacity expansion $x$, behavior $a \in \mathcal{A}$ and scenario $w \in \mathcal{W}$, let ${\Gamma_{x,w}^{a} = \langle \mathcal{S}, \mathcal{C}, \succ_{\mathcal{C}}, \succ^{a}_{\mathcal{S},w}\left(\boldsymbol{\bar{q}}_x, u^w,K\right), \boldsymbol{\bar{q}}_x \rangle}$, characterizing any instance with expanded capacities and \emph{reported} preferences to define a stable matching (Definition~\ref{def:stablematching}). The {\STMMP} (Problem~\ref{prob:stmmp}) seeks the expanded capacity $x$ that respects the budget and minimizes the expected value of student-optimal stable matchings in instances $\Gamma_{x,w}^{a}$ for scenarios $w \in \mathcal{W}$.}
\begin{definition}[\changed{Stable matching}]\label{def:stablematching}
     \changed{A matching ${\mu_w^a \subseteq \mathcal{S} \times \left( \mathcal{C}^a_{s,w}  \cup \varnothing \right)}$ for instance $\Gamma_{x,w}^a$} maps each student $s \in \mathcal{S}$ to a school \changed{$\mu^a_w(s) \in \mathcal{C}_{s,w}^{a}  \cup \{\varnothing\}$}, ensuring that no school $c \in \mathcal{C}$ is assigned more students than its \changed{expanded capacity, i.e., $\mu^a_w(c) \subseteq \mathcal{S}$ with $ |\mu^a_w(c) | \le q_c + \sum_{j \in \mathcal{B}}x_c^j$.}  
     A matching $\mu_w^a$ is \emph{stable} if it does not admit any pairs $(s, c)$ such that:  
     (i)~\changed{$c$ is preferred by student $s$ over $\mu^a_w(s)$, i.e.,} $c \succ_{s,w}^{a} \mu^a_w(s)$, and  
     (ii)~$ | \mu^a_w(c) | $ is strictly inferior to the expanded capacity of school $c$ \changed{(i.e., $ | \mu^a_w(c) | < q_c + \sum_{j \in \mathcal{B}} x_c^j$)} or there exists $ s' \in \mu^a_w(c) $ \changed{such that $s$ is preferred over $s'$ by school $c$, i.e., $s \succ_c s'$}. In other words, $\mu^a_w $ is stable if it has no \emph{blocking pairs}. 
\end{definition}
\begin{problem} \label{prob:stmmp}
The {\STMMP} under behavior $a \in \mathcal{A}$ ({\STMMP}-$a$) is defined as
\begin{equation}\label{prog:stmmp-1st}
    \min_{x} \left\{ \mathbb{E}_{\boldsymbol{u}} \left[ Q^a(x,\boldsymbol{u}) \right]: x \in \mathcal{X} \right\},
\end{equation}
where $\mathcal{X} = \left\{\sum_{c \in \mathcal{C}} \sum_{j \in \mathcal{B}} x_c^j \leq B, \, x_c^{j} \geq x_c^{j+1}, \forall (c,j) \in \mathcal{C} \times \mathcal{B}\backslash\{B\}, \, x \in \{0,1\}^{\mathcal{C} \times \mathcal{B}} \right\}$. Given an instance $\Gamma_{x,w}^{a}$ for a realization $u^w$ of $\boldsymbol{u}$, the second-stage program $Q^a(x,u^w)$ under behavior $a \in \mathcal{A}$ is:
\begin{equation}\label{prog:stmmp-2st}
   \min_{\mu_w^a} \left\{ \sum_{(s,c) \in \mu_w^a} r^{a}_{s,c,w}: \mu_w^a \mbox{\small is a stable matching in \changed{instance} $\Gamma_{x,w}^{a}$ \changed{as in Definition~\ref{def:stablematching}}}   \right\}.
\end{equation}
\end{problem}
In $\mathcal{X}$, the first constraint ensures that the capacity expansion $x$ satisfies the budget, while the second guarantees the correct definition of variables $x$, and the third enforces their binary nature. The objective is to minimize the expected value of the second-stage Program~\eqref{prog:stmmp-2st} over $\boldsymbol{u}$. The constraint in Program~\eqref{prog:stmmp-2st} ensures that $\mu^a_w$ is a stable matching in $\Gamma_{x,w}^{a}$, while the objective minimizes students' assignment preferences and the penalty for \changed{unassignment}, ensuring that a student-optimal stable matching is obtained. We note that $B$ can be interpreted as the number of extra-seat groups (e.g., additional classrooms), with binary variables $x$ indicating how many groups are installed at each school, allowing direct computation of final capacities. Extra seat allocation may involve operational costs, such as the construction of new infrastructure or hiring of \changed{extra} professionals, which can be incorporated as a linear function on $x$ in the objective. School-specific expansion limits, reflecting factors such as location, building size, growth potential\changed{, or the need to balance under- and over-subscribed schools,} can be included by limiting the maximum number of extra seats per school or fixing $x_c^j = 0$ for $j$ exceeding the school’s individual capacity. \changed{In other settings, capacity reductions can be modeled by extending variables $x$, i.e., defining indexes $x_c^{-j}$ to represent the removal of $j$ seats.}

Program~\eqref{prog:stmmp-1st} has complete recourse: for any $x \in \mathcal{X}$, Program~\eqref{prog:stmmp-2st} is feasible for every $u^w$. For fixed $x$ and $\boldsymbol{\bar{q}}_x$, the reported preference order $\succ_{\mathcal{S},w}^{a} (\boldsymbol{\bar{q}}_x, u^w,K)$ can be precomputed, and thus the {\DAA} solve Program~\eqref{prog:stmmp-2st}. Program~\eqref{prog:stmmp-1st} uses the \emph{exogenous true preferences} $\boldsymbol{u}$, which are transformed into \emph{endogenous reported preferences} in instances $\Gamma_{x,w}^{a}$ for strategic cases $a \in \{{\CEUM},{\IEUM}\}$, \changed{following the method of~\citet{bazotte2025solving}.}

\subsection{Reported Students' Preferences}\label{subsec:students-behavior}

\changed{We model students’ reported preferences, $\boldsymbol{\succ}^{a}_{\mathcal{S}}$, as a function of their true preferences, the list size limit $K$, and, for strategic behaviors, their perceived admission chances, and thus on capacity decisions.} The considered utility-maximizing behaviors are: truthful Utility Maximization ({\UM}), strategic Conjoint Expected Utility Maximization ({\CEUM}), and strategic Individual Expected Utility Maximization ({\IEUM}). These behavioral models build on prior empirical work estimating the distribution of student preferences from rich observational data~\citep{agarwal2020revealed}. \changed{In real-world problems}, empirical data on reported preferences can be used to estimate the parameters of these models \changed{and recover the} true preference distribution.

We denote by $p^w_{s,c}(x_c)$, with \changed{$x_c = \left\{x_c^j\right\}_{j \in \mathcal{B} \cup \{0,B+1\}}$}, the admission chance of student $s$ to school $c$ under scenario $w \in \mathcal{W}$ for \changed{$(s,c) \in \mathcal{S} \times \mathcal{C}_{s,w}$}, given the extra capacity $x_c$. The scenarios correspond to \emph{true} utilities $\boldsymbol{u}$ and are thus decision-independent. We introduce parameters $x_c^0 = 1$ and $x_c^{B+1} = 0$ for every school $c \in \mathcal{C}$; their purpose is explained below. We make the following assumption regarding this admission chance:
\begin{assumption}\label{assum:belief}
The perceived admission chance in scenario $w \in \mathcal{W}$ of student-school \changed{$(s,c) \in \mathcal{S} \times \mathcal{C}_{s,w}$} is:
    \begin{equation}
          p_{s,c}^w(x_c) = \sum_{j \in \text{\changed{$\mathcal{B} \cup \{0\}$}}} \overline{p}_{s,c}^j \left( x_c^j - x_c^{j+1} \right),
    \end{equation}
    where $\overline{p}_{s,c}^j \in (0,1]$ for \changed{$j \in \mathcal{B} \cup \{0\}$} denotes the perceived admission chance of student $s$ when school $c$ has capacity $(q_c + j)$, with $0 < \overline{p}_{s,c}^0 \leq \overline{p}_{s,c}^1 \leq \ldots \leq \overline{p}_{s,c}^B$. Thus, we define $\boldsymbol{\overline{p}}_s = \left\{p_{s,c}^j\right\}_{c \in \mathcal{C}, j \in \mathcal{B}}$ and $\boldsymbol{\overline{p}} = \left\{\boldsymbol{\overline{p}}_s\right\}_{s \in \mathcal{S}}$.
\end{assumption}
As school capacities increase, perceived admission chances improve, consistent with~\citet{agarwal2018demand}, \changed{being modeled by schools’ minimal acceptance requirements~\citep{kapor2020heterogeneous}.} The \emph{perceived} admission chance $p_{s,c}^w(x_c)$ may differ from students' \emph{actual} chances, as biases are common in practice~\citep{arteaga2022smart}. \changed{We next} detail the student behaviors and exemplify their differences in \changed{~\ref{app:example-behaviors}}.

\subsubsection{Utility Maximization ($a ={\UM}$)} 

In this case, we assume that students truthfully report their preferences as proposed in Assumption~\ref{assum:true}. Reported preferences are exogenous, independent of first-stage decisions $x$. Program~\eqref{prog:stmmp-2st} is solved for the instance $\Gamma_{x,w}^{{\UM}} = \langle \mathcal{S}, \mathcal{C}, \succ_{\mathcal{C}}, \succ_{\mathcal{S},w}^{\UM} \left(\boldsymbol{\bar{q}}_x, u^w, K\right), \boldsymbol{\bar{q}}_x \rangle$, where \changed{${\succ^{{\UM}}_{\mathcal{S},w}}\left(\boldsymbol{\bar{q}}_x, u^w, K\right)$ is equal} for all $x \in \mathcal{X}$.
\begin{assumption}\label{assum:true}
    Every student $s \in \mathcal{S}$ in scenario $w \in \mathcal{W}$ behaves truthfully by reporting their top $K$ truly preferred schools: \changed{$\mathcal{C}^{{\UM}}_{s,w} \in$} $\arg \max_{|C| \leq K, C \subseteq \mathcal{C}_{s,w}} \left\{\overline{{\UM}}(C) = \sum_{c \in C} u_{s,c}^w \right\}$. The reported preference \changed{$\succ_{s,w}^{{\UM}}$} is equal to the true preference order $\succ_{s,w}$ limited to the schools in \changed{$\mathcal{C}_{s,w}^{\UM}$}.
\end{assumption} 

\subsubsection{Conjoint Expected Utility Maximization ($a = {\CEUM}$)} 

\changed{We consider Assumption~\ref{assum:strategy}, which models} students' behavior via Program~\eqref{eq:max-exp-utility}, \changed{selecting} the subset of schools $\mathcal{C}^{{\CEUM}}_{s,w}$ that maximizes student $s$'s conjoint expected utility $\overline{{\CEUM}}_s^w(\cdot)$ based on perceived admission chances. This utility accounts for both the chance of admission to each school in $\mathcal{C}^{{\CEUM}}_{s,w}$ and the probability of not being admitted to higher-ranked schools according to true preferences. Hence, $\succ_{s,w}^{{\CEUM}}$ orders schools in $\mathcal{C}^{{\CEUM}}_{s,w}$ by decreasing true utilities $\boldsymbol{u}_s^w$, i.e., any deviation from this ordering reduces the conjoint expected utility. Note that true preferences $\succ_{s,w}$ follow the same ordering: $c \succ_{s,w} c'$ iff $u_{s,c}^w > u_{s,c'}^w$. Program~\eqref{eq:max-exp-utility} was introduced by~\citet{chade2006simultaneous} and has been widely used in prior research to model strategic student behavior with fixed admission chances (see Section~\ref{sec:literature}). In our model, however, perceived admission chances depend on the capacity decisions $x$. Program~\eqref{eq:max-exp-utility} maps first-stage decisions $x$ and exogenous random vector $\boldsymbol{u}$ to the endogenous random variables $\boldsymbol{\succ}^{{\CEUM}}_{\mathcal{S}}$, acting as a transformation~\citep{bazotte2025solving}.
\changed{While the transformation is well-defined, the distribution of $\boldsymbol{\succ}_\mathcal{S}^{\CEUM}$ does not admit a straighforward closed form.}
\begin{assumption}\label{assum:strategy}
    Every student $s \in \mathcal{S}$ in scenario $w \in \mathcal{W}$ behaves strategically by reporting a preference order $\succ^{{\CEUM}}_{s,w}$ over a set of schools $\mathcal{C}^{{\CEUM}}_{s,w}$ modeled by the following mathematical program:
    \begin{equation} \label{eq:max-exp-utility} 
        \mathcal{C}^{{\CEUM}}_{s,w} \in \arg \max\limits_{\substack{ |C| \leq K \\ C \subseteq \mathcal{C}_{s,w}}} \left\{ \overline{{\CEUM}}_s^w(x,C) = \sum_{c \in C} \varrho_{s,c}^w(x,C) u_{s,c}^w \right\},
    \end{equation}
    where $\varrho_{s,c}^w(x,C) =  p_{s,c}^w\left(x_c\right) \prod_{c' \in C : u_{s,c'}^w  > u_{s,c}^w} \left[ 1 -p_{s,c'}^w(x_{c'}) \right]$. The preference order $\succ^{{\CEUM}}_{s,w}$ ranks the schools in $\mathcal{C}^{{\CEUM}}_{s,w}$ according to $u_s^w$. Thus, for $c, c' \in \mathcal{C}^{{\CEUM}}_{s,w}$, we have $c \succ^{{\CEUM}}_{s,w} c'$ iff $u_{s,c}^w > u_{s,c'}^w$.
\end{assumption}
\changed{Program~\eqref{eq:max-exp-utility} represents rational behavior.} In a student-optimal stable matching, each student is assigned to their highest-ranked school in $\succ_{s,w}^{{\CEUM}}$ for which they have a sufficient score based on schools' rankings. It is thus rational to order schools in $\succ_{s,w}^{{\CEUM}}$ according to their true preferences, and Assumption~\ref{assum:strategy} ensures that students understand the mechanism. Program~\eqref{eq:max-exp-utility} may admit multiple optimal solutions when ${p_{s,c}^w(x_c) = 1}$ for some schools $c \in C' \subseteq \mathcal{C}_{s,w}$. In such cases, we consider students adopt the strategy in Assumption~\ref{assump:safety-strat}. Moreover, when the number of schools that a student $s$ \emph{truly} prefers over being unassigned in scenario $w$ satisfies $|\mathcal{C}_{s,w}| \leq K $, $ \mathcal{C}_{s,w}^{{\CEUM}} $ and $ \succ^{{\CEUM}}_{s,w} $ reduces to $ \mathcal{C}_{s,w} $ and $ \succ_{s,w} $, respectively, for all capacity decisions $x \in \mathcal{X}$.
\begin{assumption}\label{assump:safety-strat}
    Students adopt a safety strategy by listing the maximum possible number of acceptable schools, i.e., $ |\mathcal{C}_{s,w}^{{\CEUM}}| = \min\{|\mathcal{C}_{s,w}|, K\} $, and selecting the optimal solution $ \mathcal{C}_{s,w}^{{\CEUM}} $ that includes the subset of schools with the highest true utilities.\footnote{Consider a fixed $ x_c $ and an instance with three schools where $ u_{s,c_1}^w > u_{s,c_2}^w > u_{s,c_3}^w $, $ p_{s,c_2}^w(x_c) = 1 $, $ p_{s,c_1}^w(x_c) < 1 $, $ p_{s,c_3}^w(x_c) < 1 $, and $ K = 2 $. For example, if $ \mathcal{C}_{s,w}^{{\CEUM}} \in \{\{c_1, c_2\}, \{c_1, c_3\}\} $, which occurs when $ p_{s,c_1}^w(x_{c_1}) u_{s,c_1}^w > p_{s,c_2}^w(x_c) u_{s,c_2}^w $, we select $ \mathcal{C}_{s,w}^{{\CEUM}} = \{c_1, c_2\} $.}  
\end{assumption}

\begin{algorithm}[t]
    \caption{Marginal Improvement  Algorithm ({\MIA})} \label{alg:mia} 
    \resizebox{0.8\textwidth}{!}{%
    \begin{minipage}{\textwidth} %
    \small 
    \Input{$x \in \mathcal{X}$, $s \in \mathcal{S}$, $w \in \mathcal{W}$, $\boldsymbol{\overline{p}}_s$, $u_s^w $, $\mathcal{C}_{s,w}$;}\\
    \Output{$\succ_{s,w}^{{\CEUM}}$, $\mathcal{C}_{s,w}^{{\CEUM}}$;} \\
    \Stepone{Initialize $C=\emptyset$ and $k = 1$;} \\
    \Steptwo{Select  $c_k \in \arg \max_{ c \in \mathcal{C}_{s,w} \backslash C}${\scriptsize$ \left\{ {{\MI}_s^w(x,C,c) = \overline{{\CEUM}}_s^w \left(x, C \cup \{c\} \right) - \overline{{\CEUM}}_s^w \left(x, C \right) } \right\} $} with the highest $u_{s,c_k}^w$;} \label{alg:correct-select-schools} \\
    \Stepthree{Set $C = C \cup \{c_k\}$, $k=k+1$ and return to Step 2 if $k \leq K$;} \\
    \Stepfour{Set $\mathcal{C}_{s,w}^{{\CEUM}} = C$ and $\succ_{s,w}^{{\CEUM}}$ as its ordering from $u_s^w$ \tcp*{$\succ_{s,w}^{{\CEUM}} = f_{u_{s}^w}(C)$}} 
    \end{minipage}
    }
\end{algorithm}

~\citet{chade2006simultaneous} introduced the Marginal Improvement algorithm~({\MIA}), a greedy procedure that yields an optimal solution to Program~\eqref{eq:max-exp-utility} for a fixed capacity expansion $x$. Algorithm~\ref{alg:mia} details this procedure, where $f_{u_{s}^w}(C)$ ranks any subset $C \subseteq \mathcal{C}_{s,w}$ according to the true preferences $u_{s}^w$. At each iteration, the algorithm adds a school $c_k \in \mathcal{C}_{s,w} \setminus C$ to $\mathcal{C}_{s,w}^{{\CEUM}}$ that maximizes the marginal increase in conjoint expected utility. The marginal increase  of including school $c \in \mathcal{C}_{s,w} \setminus C$ is:
\begin{equation}\label{eq:marginal-increase}
    {\MI}_{s}^w(x,C,c) = \varrho_{s, c}^w(x, C)u_{s, c}^w - p_{s, c}^w(x_{c}) \sum\limits_{\substack{c' \in C: \\ u_{s,c'}^w < u_{s,c}^w 
}}  \varrho_{s, c'}^w(x, C) u_{s, c'}^w.
\end{equation}
We adapted Steps~2 and~3 of Algorithm~\ref{alg:mia} to incorporate Assumption~\ref{assump:safety-strat}. In Step~3, the algorithm continues to select schools even when the marginal improvement in adding $c_k$ is zero, stopping only upon reaching the list size limit $K$. In Step~2, if multiple optimal choices exist, we select $c_k$ with the highest true utility.

\subsubsection{Individual Expected Utility Maximization ($a ={\IEUM}$)} 

\changed{We consider Assumption~\ref{assum:strategy_simple}, which} models students' behavior with Program~\eqref{eq:max-exp-utility-simple}, \changed{selecting} the subset of schools $\mathcal{C}_{s,w}^{{\IEUM}}$ that maximizes the sum of individual expected utilities $\overline{{\IEUM}}_s^w\left(\cdot\right)$. Unlike the {\CEUM}, students do not consider the rejection chance from higher-ranked schools in their true preferences when computing the expected utilities. As with the {\CEUM}, the reported preferences $\succ_{s,w}^{{\IEUM}}$ orders schools in $\mathcal{C}_{s,w}^{{\IEUM}}$ by decreasing true preferences $u_{s}^w$, reflecting rational behavior as explained earlier. If $|\mathcal{C}_{s,w}| \leq K$, then $\mathcal{C}_{s,w}^{{\IEUM}} $ and $ \succ^{{\IEUM}}_{s,w} $ are also equal to $ \mathcal{C}_{s,w} $ and $ \succ_{s,w} $, respectively, for all capacity decisions $x \in \mathcal{X}$. We adopt this model for two reasons: (i) it captures bounded rationality, students approximate complex strategies with simple heuristics, and (ii) it allows \changed{analysis} on how behavioral variations impact optimal capacity decisions and outcomes.
\begin{assumption}\label{assum:strategy_simple}
    Every student $s \in \mathcal{S}$ in scenario $w \in \mathcal{W}$ behaves strategically by reporting a preference order $\succ^{{\IEUM}}_{s,w}$ over a set of schools $\mathcal{C}^{{\IEUM}}_{s,w}$ modeled by the following mathematical program:
    \begin{equation} \label{eq:max-exp-utility-simple} 
        \resizebox{0.5\linewidth}{!}{$
        \displaystyle 
        \mathcal{C}^{{\IEUM}}_{s,w} \in \arg\max\limits_{\substack{ |C| \leq K \\ C \subseteq \mathcal{C}_{s,w}}} \left\{ \overline{{\IEUM}}_s^w\left(x,C\right) = \sum_{c\ \in\ C} p_{s,c}^w\left(x_c\right) u_{s,c}^w  \right\}. $}
    \end{equation}
    The preference order $ \succ^{{\IEUM}}_{s,w} $ is defined as the ordering of schools in $ \mathcal{C}^{{\IEUM}}_{s,w} $ based on the true preference~$u_{s}^w$. Thus, for $ c, c' \in \mathcal{C}^{{\IEUM}}_{s,w} $, we have $c \succ^{{\IEUM}}_{s,w} c'$ iff $u_{s,c}^w > u_{s,c'}^w$.
\end{assumption}
Given a capacity decision $x$, the Individual Utility Ordering algorithm ({\IOA}) selects the $K$-th school with the highest individual expected utilities, yielding an optimal solution for Program~\eqref{eq:max-exp-utility-simple}. In the rare case of ties, the school with the highest true utility is chosen. Program~\eqref{eq:max-exp-utility-simple} acts as a transformation~\citep{bazotte2025solving}, mapping first-stage decisions $x$ and exogenous true preferences $\boldsymbol{u}$ to the endogenous reported preferences $\boldsymbol{\succ}_{\mathcal{S}}^{{\IEUM}}$. \changed{However, the distribution of $\boldsymbol{\succ}_{\mathcal{S}}^{{\IEUM}}$ lacks a simple closed-form.}

\section{Sampling Based Methodology}\label{sec:methodology}

We apply the {\SAA} method to solve the {\STMMP}, as the continuity of random utilities \(\boldsymbol{u}\) makes enumerating the set \(\mathcal{W}\) infeasible. The {\SAA} program approximating Program~\eqref{prog:stmmp-1st} under behavior \(a \in \mathcal{A}\) is:
\begin{equation}\label{prog:saa}
    \min \left\{ \hat{g}^a_N(x) = \frac{1}{|\mathcal{W_N}|} \sum_{w \in \mathcal{W_N}} Q^a(x,u^w) : x \in \mathcal{X} \right\},
\end{equation}
where $\hat{g}^a_N(x)$ is the {\SAA} unbiased estimator of $\mathbb{E}_{\boldsymbol{u}} \left[ Q^a \boldsymbol{(}x, \boldsymbol{u} \boldsymbol{)} \right]$ and $\hat{\sigma}_{a,N}^2(x)$ is the variance estimator of $\hat{g}^a_{N}(x)$, \changed{both computed using the set} $\mathcal{W}_N = \{w_1,\dots,w_N\}$ of $N$ independent and identically distributed samples of $\boldsymbol{u}$. The {\SAA} method consists of two steps, detailed in Algorithm~\ref{alg:saa}. Multiple instances of the {\SAA} Program~\eqref{prog:saa} with distinct sets $\mathcal{W_N}$ are usually solved in Step~1 to compute a lower bound estimator. However, due to the program's NP-hardness, we solve only one instance.
\begin{algorithm}[t]
    \resizebox{0.8\textwidth}{!}{%
    \begin{minipage}{\textwidth} %
    \caption{{\SAA} Method for the {\STMMP}}\label{alg:saa}  
    \small \Input{$a \in \mathcal{A}$, $\mathcal{S},\mathcal{C},\mathcal{B},\mathcal{K},\succ_{\mathcal{C}},\boldsymbol{q},\boldsymbol{\overline{p}},\mathcal{W_N},\mathcal{W_{N'}}$;}\\
    \Output{\changed{$\overline{x}_N^a,\hat{g}^a_{N'}(\overline{x}_N^a),\hat{\sigma}_{a,N'}^2(\overline{x}_N^a)$;}} \\
    \Stepone{\small Solve {\SAA} Program~\eqref{prog:saa} (exactly or heuristically) with behavior $a$ and set $\mathcal{W_N}$ to \changed{obtain $\overline{x}_N^a \in \mathcal{X}$;}} \\
    \Steptwo{Compute $\hat{g}^a_{N'}(\overline{x}_N^a)$ and $\hat{\sigma}_{a,N'}^2(\overline{x}_N^a)$ using a sufficiently large set of scenarios $\mathcal{W_{N'}}$ ($|\mathcal{W_{N'}}| \gg |\mathcal{W_{N}}|$):}\\
    \hspace{1.1em}\StepA{Solve Program~\eqref{eq:max-exp-utility} with {\MIA} if $a = {\CEUM}$, or Program~\eqref{eq:max-exp-utility-simple} with {\IOA} if $a = {\IEUM}$, using $\overline{x}_N^a$, for each $s \in \mathcal{S}$ and $w \in \mathcal{W_{N'}}$, to obtain $\mathcal{C}^{a}_{s,w}$ and $\succ^{a}_{s,w}$; Otherwise, if $a = {\UM}$, use \changed{precomputed $\mathcal{C}_{s,w}^{\UM}$ and $\succ_{s,w}^{\UM}$}};\\
    \hspace{1.1em}\StepB{Solve $Q^a(\overline{x}_N^a,u^w)$ with instance ${\Gamma_{\overline{x}_N^a,w}^{a} = \langle \mathcal{S}, \mathcal{C}, \succ_{\mathcal{C}}, \succ^{a}_{\mathcal{S},w}, \boldsymbol{\bar{q}}_{\overline{x}_N^a} \rangle}$ for every $w \in \mathcal{W_{N'}}$ using the {\DAA}; }
    \end{minipage}
    }
\end{algorithm}
The convergence of the optimal solution of the {\SAA} Program~\eqref{prog:saa} to the true optimum of the {\STMMP} depends on students’ exogenous true preferences, rather than on their endogenous reported preferences in strategic cases~\citep{bazotte2025solving}. \changed{We present the exact formulation of the {\SAA}~Program~\eqref{prog:saa} by defining the second-stage Program~\eqref{prog:stmmp-2st} [$Q^a(c,u^w)$] that composes it (Section~\ref{subsec:exact-approches}) and heuristic approaches to solve it (Sections~\ref{subsec:heuristic-lagrangian} and~\ref{subsec:heuristic-localsearch}).}

\subsection{Mathematical Formulations}\label{subsec:exact-approches}

\subsubsection{Truthful Behavior}\label{subsubsec:nonstrategic-behavior}

For the {\UM} behavior, we propose a modified version of the $L$-constraints formulation of~\citet{bobbio2025capacity} to Program~\eqref{prog:stmmp-2st} [$Q^{\UM}(x,u^w)$] for each scenario $w \in \mathcal{W_N}$. We introduce variables $y_{s,c}^w \in \{0,1\}$, which equal one if student $s$ is assigned to school $c$ and zero otherwise, for \changed{$(s,c) \in \mathcal{S} \times \left( \mathcal{C}_{s,w}^{\UM} \cup \varnothing \right)$.} The formulation of Program~\eqref{prog:stmmp-2st} under {\UM} behavior is:
\begin{subequations}\label{Prog:mod-lconstr-second-stage-exo}
    \setlength{\jot}{0.5pt}
    \setlength{\arraycolsep}{0.2pt}
    \begin{align}
        \min\mbox{: } \label{constr:exo-objective} & \text{\changed{$\sum_{s \in \mathcal{S}} \sum_{c \in \left( \mathcal{C}_{s,w}^{\UM} \cup \varnothing \right) }$}} r_{s,c,w} y_{s,c}^w  \\
        \label{constr:exo-assign}\mbox{s.t.:} & \text{\changed{$\sum_{c \in \left(\mathcal{C}_{s,w}^{\UM} \cup \varnothing \right)}$}} y_{s,c}^w = 1, && \forall s \in \mathcal{S} \\
        \label{constr:exo-capa}& \text{\changed{$\sum_{s \in \mathcal{S}: c \in \mathcal{C}_{s,w}^{\UM}}$}} y_{s,c}^w \leq q_c +  \sum_{j \in \mathcal{B}} x_c^j, && \forall c \in \mathcal{C} \\
        \label{constr:mod-lconstr-stab-exo} & \sum_{c' \succeq_{s,w}^K c } y_{s,c'}^w \geq y_{s',c}^w,  && \forall \text{\changed{$(s,c) \in \mathcal{S} \times \left( \mathcal{C}_{s,w}^{\UM} \cup \varnothing \right)$}}, s' \prec_c s \\
        \label{constr:exo-domain-y}& y_{s,c}^w \in \{0,1\}, && \forall \text{\changed{$(s,c) \in \mathcal{S} \times \left( \mathcal{C}_{s,w}^{\UM} \cup \varnothing \right)$}},
    \end{align}
\end{subequations}
where $r_{s,\varnothing,w} = \min(|\mathcal{C}_{s,w}|,K)+1$. Constraints~\eqref{constr:exo-assign} and~\eqref{constr:exo-capa} ensure that each student is assigned to a school or remains unassigned, and that schools' expanded capacities are respected. The modified $L$-Constraints~\eqref{constr:mod-lconstr-stab-exo} enforce stability: if \changed{$(s,c) \in \mathcal{S} \times \left( \mathcal{C}_{s,w}^{\UM} \cup \varnothing \right)$}, then any less-preferred student $s' \prec_c s$ may only be assigned to $c$ if $s$ is assigned either to $c$ or to a more preferred school $c' \succ^{\UM}_{s,w} c$. We have a constraint for each $s' \prec_c s$, while the $L$-constraints formulation uses a single constraint over the complement of such set of students. Although more numerous, our constraints tighten the linear relaxation and often reduce solution time.

\subsubsection{Strategic Behaviors}\label{subsec:strategic-behavior}

When $|\mathcal{C}_{s,w}| > K$, reported preferences depend on capacity decisions $x$; in this case, we provide formulations to Programs~\eqref{eq:max-exp-utility} and~\eqref{eq:max-exp-utility-simple} to define $\mathcal{C}_{s,w}^{a}$ and $\succ_{s,w}^{a}$ under {\CEUM} and {\IEUM} behaviors ($a \in \{\CEUM, \IEUM\}$). We introduce the indicator variables $\xi^{a}_{s,c,w} \in \{0,1\}$ for $w \in \mathcal{W_N}$ and \changed{$(s,c) \in \mathcal{S} \times \mathcal{C}_{s,w} $}, which equals one if $c \in \mathcal{C}_{s,w}^{a}$ and zero otherwise. If $|\mathcal{C}_{s,w}| \leq K$, reported preferences coincide with the true ones. We then present the complete formulation of Program~\eqref{prog:stmmp-2st} under these strategic behaviors. 

\paragraph{Conjoint Expected Utility Maximization ($a={\CEUM}$)} We follow the {\MIA} to formulate Program~\eqref{eq:max-exp-utility} for each \changed{student-scenario $(s,w) \in \mathcal{S} \times \mathcal{W_N}$} when ${|\mathcal{C}_{s,w}| > K}$. Let $\overline{\xi}_{s,c}^{w,k}$ for $c \in \mathcal{C}_{s,w}$ and \changed{$k \in \mathcal{K} \cup \{0\}$} be binary variables such that $\overline{\xi}_{s,c}^{w,k} = 1$ if school $c$ is selected on or before iteration $k$ of the {\MIA} to compose $\mathcal{C}_{s,w}^{{\CEUM}}$, with ${\overline{\xi}_{s,c}^{w,0} = 0}$. Thus, ${\xi_{s,c,w}^{\CEUM} = \overline{\xi}_{s,c}^{w,K} = 1}$ if and only if ${c \in \mathcal{C}_{s,w}^{{\CEUM}}}$. We introduce variables $\phi_{s,c}^{w,k} \geq 0$ and $\upsilon_{s,c}^{w,k} \geq 0$, where ${\phi_{s,c}^{w,k} = \prod_{c' \succ_{s,w} c} \left( 1 - p^w_{s,c}(x_c) \xi_{s,c'}^{w,k} \right)}$ represents the probability that student $s$ is rejected by all schools ranked higher than $c$ among the first $k$ selected schools, with ${\phi_{s,c}^{w,0} = 1}$, and $\upsilon_{s,c}^{w,k} =  p_{s,c}^w(x_c)  u_{s,c}^w \phi_{s,c}^{w,k} \xi_{s,c}^{w,k}$ denotes the expected utility of school $c$ in this preference order, with $\upsilon_{s,c}^{w,0} = 0$. Let $L_{s,w}^k \geq 0$ be the total expected utility at iteration $k$, with $L_{s,w} = \left\{L_{s,w}^k\right\}_{\text{\changed{$k \in \mathcal{K} \cup \{0\}$}}}$ and $L_{s,w}^0 = 0$. The following constraints model Program~\eqref{eq:max-exp-utility}:
\begin{align}
    \label{constr:select-school-it}& \sum_{c \in \mathcal{C}_{s,w}} \big( \overline{\xi}_{s,c}^{w,k} - \overline{\xi}_{s,c}^{w,k-1} \big) = 1, && \forall k \in \mathcal{K} \\
    \label{constr:restrict-schools}& \overline{\xi}_{s,c}^{w,k-1} \leq \overline{\xi}_{s,c}^{w,k}, \quad \overline{\xi}_{s,c}^{w,k} \in \{0,1\}, &&  \forall k \in \mathcal{K}, c \in \mathcal{C}_{s,w} \\
    \label{constr:def-var-L}& L_{s,w}^k - L_{s,w}^{k-1} = \sum_{c \in \mathcal{C}_{s,w}}  \big(\upsilon_{s,c}^{w,k} - \upsilon_{s,c}^{w,k-1} \big), && \forall k \in \mathcal{K} \\
    \label{constr:correct-select-schools}
    &\text{\changed{$L_{s,w}^k - L_{s,w}^{k-1} \geq p_{s,c}^w(x_c) \bigg( u_{s,c}^w \phi_{s,c}^{w,k-1} - \sum_{c' \preceq_{s,w} c} \upsilon_{s,c'}^{w,k-1} \bigg),$}} && \forall k \in \mathcal{K}, c \in \mathcal{C}_{s,w}.
\end{align}
Constraints~\eqref{constr:select-school-it} and~\eqref{constr:restrict-schools} ensure that each position in the reported preference order is filled and that no school appears more than once. Constraints~\eqref{constr:def-var-L} define $L_{s,w}^k$ based on variables $\upsilon_{s,c}^{w,k}$. \changed{Constraints~\eqref{constr:correct-select-schools} implement the {\MIA} by selecting, at each (iteration) $k \in \mathcal{K}$, the unchosen school in \( \mathcal{C}_{s,w} \) with the highest marginal increase in conjoint expected utility (Equation~\eqref{eq:marginal-increase}) while accounting the extra capacity $x$.}
   
The variables \( \phi_{s,c}^{w,k} \) and \( \upsilon_{s,c}^{w,k} \) are defined by nonlinear terms. To linearize them, we apply the probability chain technique of~\citet{o2013probability} \changed{(see~\ref{app:lin-ceum})}, which introduces \( |\mathcal{C}_{s,w}| K (B+2) \) continuous variables and \( |\mathcal{C}_{s,w}| K (B+4) \) constraints. Constraints~\eqref{constr:correct-select-schools} also contain nonlinear terms, whose linearization, based on the function \( p_{s,c}^w(x_c) \) defined over binary variables, is provided in the appendix. By Lemma~\ref{lemma:safety-strategy} (\changed{see~\ref{app:proofs}} for the proof), when Program~\eqref{eq:max-exp-utility} has multiple solutions, we add the following constraints to ensure Assumption~\ref{assump:safety-strat} is respected for \changed{$\hat{j} \in \mathcal{B} \cup \{0\}$} such that $\overline{p}_{s,c}^{\hat{j}} = 1$ and $\overline{p}_{s,c}^{\hat{j}-1} < 1$:
\begin{align}\label{constr:safety-strategy}
    \overline{\xi}_{s,c'}^{w,K} \geq x_c^{\hat{j}}, && \forall c \in \mathcal{C}, c' \prec_{s,w} c: r_{s,c',w} \leq K.
\end{align}
\begin{lemma}\label{lemma:safety-strategy}
    Constraints~\eqref{constr:safety-strategy} ensure that Assumption~\ref{assump:safety-strat} is satisfied by the formulation.
\end{lemma}
Program~\eqref{eq:max-exp-utility} is modeled by $\overline{\Xi}^{s,w}_{{\CEUM}}(x) = \left\{ \xi_{s,w}^{\CEUM} \in \{0,1\}^{\mathcal{C}_{s,w}} : \eqref{constr:select-school-it}-\eqref{constr:safety-strategy}, \, \xi^{{\CEUM}}_{s,c,w} = \overline{\xi}_{s,c}^{w,K} \, \forall c \in \mathcal{C}_{s,w}  \right\}$, with $\xi_{s,w}^{\CEUM} = \{\xi_{s,c,w}^{\CEUM}\}_{c \in \mathcal{C}_{s,w}}$. The set $\Xi^{s,w}_{{\CEUM}}(x)$ extends this formulation to include the linearized expressions of $\phi_{s,c}^{w,k}$, $\upsilon_{s,c}^{w,k}$, and Constraints~\eqref{constr:correct-select-schools} (\changed{see~\ref{app:lin-ceum}}). Students’ reported preferences are defined by ${\xi^{{\CEUM}}_{s,w} \in \Xi^{s,w}_{{\CEUM}}(x)}$.

\paragraph{Individual Expected Utility Maximization ($a={\IEUM}$)} 
We formulate Program~\eqref{eq:max-exp-utility-simple} for each \changed{student-scenario $(s,w) \in \mathcal{S} \times \mathcal{W_N}$} when ${|\mathcal{C}_{s,w}| > K}$. With slight abuse of notation, we let the binary variables $\overline{\xi}_{s,c,j}^{w,k}$ for $c \in \mathcal{C}_{s,w}$, \changed{$k \in \mathcal{K} \cup \{0\}$}, and \changed{$j \in \mathcal{B} \cup \{0\}$} be equal to one if school $c$ has the $k$-th highest individual expected utility for student $s$, considering the allocation of extra seats $x$, and if $j$ extra seats are allocated to $c$ (i.e., $x_c^j - x_c^{j+1}=1$), and zero otherwise. We set $\overline{\xi}_{s,c,j}^{w,0} = 0$, and $\xi_{s,c,w}^{{\IEUM}} = \sum_{k \in \mathcal{K}}\sum_{j \in \text{\changed{$\mathcal{B} \cup \{0\}$}}} \overline{\xi}_{s,c,j}^{w,k} =1$ iff $c \in \mathcal{C}^{{\IEUM}}_{s,w}$. Let variables $L_{s,w}^k \geq 0$ for $k \in \mathcal{K}$ be the $k$-th highest individual expected utility of schools. \changed{Program~\eqref{eq:max-exp-utility-simple} is:}
\begin{align}
    \label{constr:strat-simple-select}& \sum_{c \in \mathcal{C}_{s,w}} \sum_{j \in \text{\changed{$\mathcal{B} \cup \{0\}$}}}\overline{\xi}_{s,c,j}^{w,k} = 1, && \forall k \in \mathcal{K} \\
    \label{constr:strat-simple-restric-schools}& \sum_{k \in \mathcal{K}} \sum_{j \in \text{\changed{$\mathcal{B} \cup \{0\}$}}}\overline{\xi}_{s,c,j}^{w,k} \leq 1, && \forall c \in \mathcal{C}_{s,w} \\
    \label{constr:strat-simple-link-xi-x}& \sum_{k \in \mathcal{K}}  {\overline{\xi}}_{s,c,j}^{w,k} \leq x_c^j - x_c^{j+1}, && \forall c \in \mathcal{C}_{s,w}, \text{\changed{$j \in \mathcal{B} \cup \{0\}$}} \\
    \label{constr:strat-simple-def-utility}& L_{s,w}^k = \sum_{c \in \mathcal{C}_{s,w}} \sum_{j \in \text{\changed{$\mathcal{B} \cup \{0\}$}}} \overline{p}_{s,c}^j u_{s,c}^w {\overline{\xi}_{s,c,j}^{w,k}}, && \forall k \in \mathcal{K}\\
    \label{constr:strat-simple-max-utility} & \text{\changed{$L_{s,w}^k \geq \sum_{j \in \text{\changed{$\mathcal{B} \cup \{0\}$}}} \overline{p}_{s,c}^j u_{s,c}^w \bigg( x_c^j -x_c^{j+1} - \sum_{l < k} \overline{\xi}_{s,c,j}^{w,l} \bigg)$}} && \forall k \in \mathcal{K}, c \in \mathcal{C}_{s,w}.
\end{align}
Constraints~\eqref{constr:strat-simple-select} and~\eqref{constr:strat-simple-restric-schools} ensure that all positions in the reported preference order are filled and that each school is assigned to at most one position in the order, respectively. Constraints~\eqref{constr:strat-simple-link-xi-x} enforce that \(\overline{\xi}_{s,c,j}^{w,k} = 1\) only if exactly \(j\) additional seats are allocated to school \(c\). \changed{Constraints~\eqref{constr:strat-simple-def-utility} and~\eqref{constr:strat-simple-max-utility} define \(L_{s,w}^k\) in terms of \(\overline{\xi}_{s,c,j}^{w,k}\) and ensure that \(\mathcal{C}_{s,w}^{\IEUM}\) has the \(K\) schools with the highest individual expected utilities, respectively.} We model Program~\eqref{eq:max-exp-utility-simple} by ${\Xi_{{\IEUM}}^{s,w}(x) = \left\{ \xi_{s,w}^{\IEUM} \in \{0,1\}^{\mathcal{C}_{s,w}}: \eqref{constr:strat-simple-select}-\eqref{constr:strat-simple-max-utility}, \, \xi_{s,c,w}^{{\IEUM}} = \sum_{k \in \mathcal{K}}\sum_{j \in \text{\changed{$\mathcal{B} \cup \{0\}$}}} \xi_{s,c,j}^{w,k} \, \forall c \in \mathcal{C}_{s,w} \right\}}$, with $\xi_{s,w}^{{\IEUM}} = \{\xi_{s,c,w}^{{\IEUM}}\}_{c \in \mathcal{C}_{s,w}}$. The set $\Xi_{{\IEUM}}^{s,w}(x)$ has integral vertices in $ \xi_{s,w}^{\IEUM}$, ensured by the integrality of $x$ for $\overline{\xi}_{s,c,j}^{w,1}$ and of both $x$ and $\overline{\xi}_{s,c,j}^{w,k-1}$ for $\overline{\xi}_{s,c,j}^{w,k}$, for $ k \in \mathcal{K} \backslash \{1\}$.

\paragraph{Complete Formulation} 
The modified $L$-Constraints~\eqref{constr:mod-lconstr-stab-exo} can be adapted to {\CEUM} and {\IEUM} (\changed{see~\ref{app:mod-lconstr-strategic}}), but their linear relaxation is loose for strategic behavior. We propose an alternative formulation for Program~\eqref{prog:stmmp-2st} [$Q^{\CEUM}(x,u^w)$, $Q^{\IEUM}(x,u^w)$] for each $w \in \mathcal{W_N}$, extending the cutoff score model of~\citet{agoston2016integer}. Preliminary results show it outperforms the modified $L$-constraints under strategic behavior (the opposite holds for {\UM} behavior; \changed{see~\ref{app:cutoff-nonstrategic}}). As in the nonstrategic case, we define variables $y_{s,c}^w \in \{0,1\}$, however, for \changed{$(s,c) \in \mathcal{S} \times \left( \mathcal{C}_{s,w}  \cup \varnothing \right)$}. We let $\varphi_s^w \geq 0$ be students' reported preference ranking of their assigned school. We let $z_c^w \geq 0$ be \changed{the cutoff score required for acceptance to school $c$}, and $f_c^w \in \{0,1\}$ be equal to one if school $c$ rejects any student, and zero otherwise. The formulation of Program~\eqref{prog:stmmp-2st} under $a \in \{{\CEUM},{\IEUM}\}$ is:
\begin{subequations}\label{Prog:cutoff-second-stage}
     \begin{align}
        \min\mbox{:} \label{constr:objective} & \sum_{s \in \mathcal{S}}  \varphi_{s}^{w}  \\
        \label{constr:assign}  \mbox{s.t.:} &  \text{\changed{$\sum_{c \in \left( \mathcal{C}_{s,w} \cup \varnothing \right)}$}} y_{s,c}^w = 1, && \forall s \in \mathcal{S} \\
        \label{constr:capa} & \text{\changed{$\sum_{s \in \mathcal{S}: c \in \mathcal{C}_{s,w}}$}} y_{s,c}^w \leq q_c +  \sum_{j \in \mathcal{B}} x_c^j, && \forall c \in \mathcal{C} \\ 
        \label{constr:cutoff-score1} & z_c^w \leq \left( 1 - y_{s,c}^w \right) \left( n+1 \right) + e_{s,c}, && \forall \text{\changed{$(s,c) \in \mathcal{S} \times \mathcal{C}_{s,w}$}} \\
        \label{constr:cutoff-score2} &  
        \left( e_{s,c} + \epsilon \right) \xi_{s,c,w}^{a} \leq z_c^w +  \left( n+1 \right)\sum_{c' \succeq_{s,w} c} y_{s,c'}^w, &&  \forall \text{\changed{$(s,c) \in \mathcal{S} \times \mathcal{C}_{s,w}$}}
        \\ 
        \label{constr:cutoff-waste1} & \sum_{j \in \mathcal{B}} x_c^j -  B + \hat{q}_c f_c^w  \leq \text{\changed{$\sum_{s \in \mathcal{S}: c \in  \mathcal{C}_{s,w}}$}} y_{s,c}^w, && \forall c \in \mathcal{C} \\
        \label{constr:cutoff-waste2} & z_c^w \leq f_c^w\big( n+1\big), && \forall c \in \mathcal{C}\\
         \label{constr:respect-report} & y_{s,c}^w \leq \xi_{s,c,w}^{a}, && \forall \text{\changed{$(s,c) \in \mathcal{S} \times \mathcal{C}_{s,w}$}} \\
        \label{constr:ranking-general} & 
        \varphi_{s}^{w} \geq \sum_{c' \succeq_{s,w} c} \xi_{s,c,w}^{a} - d_{s,c}^w \sum_{c' \succ_{s,w} c } y_{s,c'}^w , && \forall \text{\changed{$(s,c) \in \mathcal{S} \times \mathcal{C}_{s,w}$}} \\
        \label{constr:ranking-unassigned} & \varphi_{s}^{w} \geq d_{s,\varnothing}^w y_{s,\varnothing}^w, && \forall s \in \mathcal{S} \\
        \label{constr:def-endogenous} & \xi_{s,w}^{a} \in \Xi^{s,w}_a(x), && \forall s \in \mathcal{S} \\
        \label{constr:domain-y} & y_{s,c}^w \in \{0,1\}, && \forall \text{\changed{$(s,c) \in \mathcal{S} \times \left( \mathcal{C}_{s,w} \cup \varnothing \right)$}} \\
        \label{constr:domain-z}& f_c^w \in \{0,1\}, \, z_c^w \geq 0, && \forall c \in \mathcal{C},
     \end{align}
\end{subequations}
where $d_{s,c}^w = \min(r_{s,c,w},K)$, $d_{s,\varnothing}^w = \min(|\mathcal{C}_{s,w}|,K)+1$ and $\hat{q}_c= q_c + B$. Constraints~\eqref{constr:assign} and~\eqref{constr:capa} ensure valid student assignments and enforce school capacity limits. Constraints~\eqref{constr:cutoff-score1} and~\eqref{constr:cutoff-score2} ensure students meet the cutoff score of their assigned school and cannot be assigned to ranked schools for which they do not qualify. Constraints~\eqref{constr:cutoff-waste1} and~\eqref{constr:cutoff-waste2} prevent blocking assignments to schools with empty seats by setting their cutoff scores to zero. Constraints~\eqref{constr:respect-report} restrict students assignment to schools in their reported preferences. Constraints~\eqref{constr:ranking-general} and~\eqref{constr:ranking-unassigned} guarantee the correct definition of variables $\varphi_s^w$, i.e., $\varphi_s^w = \sum_{c \in \mathcal{C}^{a}_{s,w}} r_{s,c,w}^{a} y_{s,c}^w + d_{s,\varnothing}^w y_{s,\varnothing}^w$, with \( r^{a}_{s,c,w} = \sum_{c' \succeq_{s,w} c} \xi_{s,c',w}^{a} \) if \( \xi_{s,c,w}^{a} = 1 \), and 0 otherwise. Constraints~\eqref{constr:ranking-general} and~\eqref{constr:ranking-unassigned} define variables $\varphi_s^w$ when students are assigned to schools or remain unassigned. Constraints~\eqref{constr:def-endogenous} define the endogenous variables $\xi_{s,c,w}^{a}$ under {\CEUM} or {\IEUM}, so that $\Xi^{s,w}_a(x)$ corresponds to $\Xi_{{\CEUM}}^{s,w}(x)$ or $\Xi_{{\IEUM}}^{s,w}(x)$. The objective minimizes students' preference rankings over their assigned schools.

\subsection{\changed{Lagrangian Heuristics}}\label{subsec:heuristic-lagrangian}

\changed{We propose a Lagrangian ({\LR}) heuristic for the {\SAA} Program~\eqref{prog:saa} under {\UM} behavior; initial tests showed that it has poor performance and long runtimes for {\CEUM} and {\IEUM}.} It relaxes the stability Constraints~\eqref{constr:mod-lconstr-stab-exo} of the modified $L$-constraints formulation. Rearranging the objective terms yields the {\SAA} {\LR} subproblem:
\begin{equation}
     \text{\changed{$\texttt{LR} (\alpha)$}}  : \min \left\{ \frac{1}{|\mathcal{W_N}|} \sum_{w \in \mathcal{W_N}} Q^{{\UM}}_{{\LR}}(x,u^w,\alpha^w) : x \in \mathcal{X} \right\}, \nonumber
\end{equation}
where ${Q^{\UM}_{{\LR}}(x,u^w,\alpha^w) = \min \left\{ \text{\changed{$\sum_{(s,c) \in \mathcal{S} \times \left( \mathcal{C}_{s,w}^{\UM} \cup \varnothing\right)}$}} \left[ r_{s,c,w} + \overline{\alpha}_{s,c}^w \left( \alpha^w \right) \right] y_{s,c}^w: \eqref{constr:exo-assign},\eqref{constr:exo-capa},\eqref{constr:exo-domain-y} \right\}}$. \changed{Here, $\alpha = \{\alpha^w\}_{w \in \mathcal{W}_N}$ denotes the set of Lagrangian multipliers for all scenarios, 
where $\alpha^w \ge 0$ is the vector of multipliers associated with scenario $w$, and the aggregated multiplier term is ${\overline{\alpha}_{s,c}^w(\alpha^w) = \left( \sum_{s' \succ_c s} \alpha_{s',c,s}^w - \sum_{c' \preceq_{s,w}^K c} \sum_{s' \prec_{c'} s } \alpha_{s,c',s'} \right)}$.} The {\SAA} {\LR} subproblem does not satisfy the integrality property~\citep{geoffrion2009lagrangean}, as variables $x$ and $y$ may take fractional values when the integrality constraints are relaxed. Thus, this Lagrangian relaxation yields stronger bounds than the linear relaxation of the {\SAA}~Program~\eqref{prog:saa}. \changed{Solving the {\SAA} {\LR} subproblem yields a feasible capacity expansion $x \in \mathcal{X}$, though the matchings $y^w$ are not necessarily stable (Constraints~\eqref{constr:mod-lconstr-stab-exo} are relaxed). Moreover, it can be solved efficiently with standard {\MILP} solvers (see Section~\ref{sec:results}).}

\paragraph{\changed{Lagrangian Dual}}
For any Lagrangian multiplier $\alpha$, the {\SAA} {\LR} subproblem $\text{LR}(\alpha)$ finds a lower bound to the {\SAA} Program~\eqref{prog:saa}. The Lagrangian dual then maximizes the Lagrangian function: $\max_{\alpha \geq 0} \text{LR}(\alpha)$. Although $\text{LR}(\alpha) $ is nondifferentiable, we can easily compute a subgradient. Thus, we apply the subgradient method, a fast and widely used approach to solve the Lagrangian dual. At each iteration of this method, we solve the subproblem $\text{LR}(\hat{\alpha})$ with the current multiplier $\hat{\alpha}$ and obtain its optimal solution $(\hat{x}, \hat{y})$, where $\hat{x} \in \mathcal{X}$. The subgradient for $w \in \mathcal{W_N}$, \changed{$(s,c) \in \mathcal{S} \times \left( \mathcal{C}_{s,w}^{\UM} \cup \varnothing\right)$}, and $s' \prec_c s$ at the current iteration is:
\begin{equation}
    \hat{\gamma}_{s,c,s'}^w = - \hat{y}_{s',c}^w + \sum_{c' \succeq_{s,w}^K c} \hat{y}_{s,c'}^w,
\end{equation}
representing the violation of the relaxed constraints by $(\hat{x},\hat{y})$. The step size at each iteration follows~\citet{held1974validation}: ${\hat{\lambda} = \hat{\tau}\left( \frac{\overline{z} - \text{LR}(\hat{\alpha})}{ \| \hat{\gamma} \|^2}\right)}$, where $ \hat{\gamma} $ is the vector of all $\hat{\gamma}_{s,c,s'}^w$, $\hat{\tau} \in [0,2] $ is a scalar that tends toward zero, and $ \overline{z} $ is the value of the best feasible solution found so far. The Lagrangian multipliers are updated by:
\begin{equation}
    \hat{\alpha}_{s,c,s'}^w = \left[\hat{\alpha}_{s,c,s'}^w - \hat{\lambda}\hat{\gamma}_{s,c,s'}^w \right]^{+},
\end{equation}
for $w \in \mathcal{W_N}$, \changed{$(s,c) \in \mathcal{S} \times \left( \mathcal{C}_{s,w}^{\UM} \cup \varnothing\right)$}, and $s' \prec_c s$ for the next iteration, where $[v]^{+}$ denotes $\max\{0,v\}$. 

\paragraph{Upper Bound Computation} At each iteration of the subgradient method, we compute the upper bound $\hat{g}^{{\UM}}_{N}(\hat{x})$ to {\SAA} Program~\ref{prog:saa} using the current optimal solution $\hat{x} \in \mathcal{X}$ of the {\SAA} {\LR} subproblem, as in Step~2 of the {\SAA} method \changed{for {\UM} behavior: for each $w \in \mathcal{W_N}$, $Q^{\UM}(\hat{x}, u^w)$ is solved by the {\DAA} with preference $\succ_{\mathcal{S},w}^{\UM}$.}

\paragraph{Lagrangian Heuristic} 
We solve the Lagrangian dual via the subgradient method, evaluating the solutions $\hat{x} \in \mathcal{X}$ obtained at each iteration. The heuristic returns: (i) the best Lagrangian lower bound across all subproblems $LR(\hat{\alpha})$; (ii) the best upper bound $\hat{g}_N^{\UM}(\overline{x}_N^{\UM})$ from evaluating current solutions; and (iii) the corresponding solution $\overline{x}_N^{\UM} \in \mathcal{X}$. The algorithm terminates when the step size $\hat{\lambda}$ becomes sufficiently small, the lower bound stagnates for a fixed number of iterations, or the maximum iteration limit is reached.

\paragraph{\changed{Assignment Heuristic}} \changed{We propose a special case of the {\LR} heuristic, the Assignment ({\ASG}) heuristic, which performs a single subgradient iteration with $\alpha = 0$, reducing the {\LR} {\SAA} subproblem to:}
\begin{equation} \label{prog:asg-saa}
    \min \left\{ \frac{1}{|\mathcal{W_N}|} \sum_{w \in \mathcal{W_N}} Q^{\UM}_{{\ASG}}(x,u^w) : x \in \mathcal{X} \right\}, 
\end{equation}
where $Q^{\UM}_{ASG}(x,u^w) = \min \{\eqref{constr:exo-objective}: \eqref{constr:exo-assign},\eqref{constr:exo-capa},\eqref{constr:exo-domain-y} \}$. \changed{The {\ASG} heuristic reduces to two steps: Step~1 solves the {\ASG} {\SAA} Program~\eqref{prog:asg-saa} to obtain $\overline{x}_N^{\UM}$ and Step~2 computes the upper bound $\hat{g}^{\UM}_N(\overline{x}_N^{\UM})$} (\changed{see~\ref{app:heur-algo} for details}). This heuristic generalizes the linear programming-based approach of~\citet{bobbio2025capacity} for the deterministic nonstrategic version of the {\STMMP}. In our stochastic setting, multiple scenarios prevent integral vertices for capacity decisions, so integrality is enforced in both $x$ and $y$.

\subsection{\changed{Local Search Heuristics}}\label{subsec:heuristic-localsearch}

We propose two local search heuristics for the {\SAA} Program~\eqref{prog:saa} under distinct behaviors: a Greedy Local Search ({\LS}) and a probabilistic Simulated Annealing ({\SA}).  Both methods explore the neighborhood of a current solution in an attempt to iteratively improve it, but differ in how new solutions are accepted and how they balance exploration and exploitation. The neighborhood $\text{NG}(x)$ of a feasible solution $x \in \mathcal{X}$ is defined by: (i) increasing a school’s capacity by one if budget allows (one neighbor per school); (ii) transferring one unit of capacity between two schools in either direction (two neighbors per pair); and (iii) swapping the entire extra capacity of two schools (one neighbor per pair). These operations ensure the feasibility of the neighbors while encouraging diversity in the search.

The {\LS} heuristic implements a hill-climbing strategy (\changed{see~\ref{app:heur-algo} for details}). It initially sets $\overline{x}_N^a$ to the feasible solution without additional seats. Then, it explores the {\LS} neighborhood $\text{NG}_{\LS}(\overline{x}_N^a)$, using only operations (i) and (iii) to keep the neighborhood manageable. For each neighbor $x \in \text{NG}_{\LS}(\overline{x}_N^a)$, the objective $\hat{g}_N^a(x)$ is computed as in Step~2 of the {\SAA} method using scenario set $\mathcal{W_N}$. The incumbent solution is thus iteratively updated if a better neighbor is found and terminates when no further improvement is possible. In contrast, the {\SA} heuristic is a probabilistic local search allowing occasional uphill moves to escape local optima (\changed{see~\ref{app:heur-algo} for details}). It maintains both the best solution $\overline{x}_N^a$ and the current solution $\widetilde{x}_N^a$, initialized as the feasible solution without additional seats. The objective $\hat{g}_N^a(\hat{x})$ of a randomly selected neighbor $\hat{x} \in \text{NG}(\widetilde{x}_N^a)$ is then computed as in Step~2 of the {\SAA} method using $\mathcal{W_N}$. In this case, worse neighbors may be accepted as current solutions probabilistically via the Metropolis criterion, based on the current temperature $T$. The heuristic terminates according to a predefined cooling schedule for $T$.

\section{Experimental Results}\label{sec:results}

The goals of our experimental analysis are (i) to estimate the value of explicitly modeling stochastic preferences versus using an average scenario ({\EV} problem); (ii) to examine how capacity allocation decisions vary under different student behaviors; and (iii) to assess the performance of our {\SAA}-based methods for the {\STMMP}. The {\SAA} program, the {\SAA} {\LR} subproblems, and the {\EV} problem were implemented in C++ and solved with Gurobi 12.0.0. The {\DAA}, {\MIA}, and {\IOA} were also coded in C++. These algorithms are used to compute $\hat{g}_N^a(x)$ for any $x \in \mathcal{X}$, enabling evaluation of first-stage decisions in the Step~2 of the {\SAA} method, the upper-bound procedure in the {\LR} heuristic, the neighborhood evaluations of both {\LS} and {\SA}\changed{, as well as the} expected result of using the {\EV} solution ({\EEV}). All experiments were run single-threaded on an Intel Gold 6148 Skylake CPU @ 2.4 GHz. All instances, results, and code are publicly available.\footnote{\url{https://github.com/mariabazotte/endog-school-choice}\label{fn:github}}.

\changed{We generate small instances with $|\mathcal{S}| \in \{50,100\}$, $|\mathcal{C}| = 6$, $K \in \{2,3,4\}$, and $B \in \{1,3,5,10,15\}$, and large instances with $|\mathcal{S}| \in \{500,1000\}$, $|\mathcal{C}| \in \{20,40,60\}$, $K \in \{2,3,4\}$, and $B \in \{1,5,15,30,45\}$. We consider five random seeds per configuration, yielding 150 small and 450 large instances in total.}
\changed{~\ref{app:instance-generation}} details the random generation of school preferences ($\succ_{\mathcal{C}}$), capacities ($q_c$), and student utilities ($\boldsymbol{u}_{s,c}$), modeled as a standard random utility model~\citep{agarwal2018demand}, combining deterministic student-school characteristics with a Gumbel-distributed term.~\ref{app:instance-generation} points the adapted bootstrapping procedure of~\citet{agarwal2018demand} for estimating students’ perceived admission chances ($\boldsymbol{\overline{p}}$) and \changed{~\ref{app:preprocess}} details preprocessing conditions.

\subsection{Value of the Stochastic Solution}\label{subsec:vss}

This section analyzes the value of the stochastic solution ({\VSS})\changed{, a standard metric in stochastic programming~\citep[see][]{birge2011introduction} for comparing the recourse program (Program~\eqref{prog:stmmp-1st}) with the approximate {\EV} approach,} for the {\STMMP} under all student behaviors, i.e., whether accounting for scenario realizations adds value. The relative ${\textrm{{\VSS}}^a}$ under behavior $a \in \mathcal{A}$ is ${(\textrm{{\EEV}}^a - v^{a}_{opt} )/v^{a}_{opt}}$, where $v^{a}_{opt}$ is the optimal value of Program~\eqref{prog:stmmp-1st} and $\textrm{{\EEV}}^a$ is the objective value of $x^{a}_{ev}$ in Program~\eqref{prog:stmmp-1st}, with $x^{a}_{ev}$ being the optimal solution of the {\EV} problem. \changed{Our} {\EV} problem is the deterministic version of Program~\eqref{prog:stmmp-1st}, with the expected value of $\boldsymbol{u}$. \changed{It is thus} an approximation (not the conventional stochastic programming version), as it relies on \changed{expected utilities} rather than on the expected value of preferences $\boldsymbol{\succ}^{a}_{\mathcal{S}}$ (\changed{see~\ref{app:vss}} for details). 

We compute the estimator ${\overline{\textrm{{\VSS}}}^a= 100\% \cdot [\overline{\textrm{{\EEV}}}^a - \hat{g}^a_{N'}(\overline{x}_N^a)]}/{\hat{g}^a_{N'}(\overline{x}_N^a)}$ for $a \in \mathcal{A}$, where ${\overline{\textrm{{\EEV}}}^a = \hat{g}^a_{N'}(x^a_{ev})}$. We also evaluate the effect of capacity expansion decisions on the average number of students who 
(i) enter the matching, i.e., previously unassigned students with \(B = 0\) who are assigned to one of their preferences under expanded capacities, and;
(ii) improve their assignments, i.e., previously assigned students with $B = 0$ who are reassigned to more preferred schools under expanded capacities. Let \(\hat{e}^a_{N'}(x)\) and \(\hat{i}^a_{N'}(x)\) denote the estimators of the average number of entering and improving students, respectively, under behavior $a \in \mathcal{A}$ for a given allocation $x$ over $N'$ scenarios. We compute ${\overline{\textrm{{\VSSENTER}}}^a= 100\% \cdot [\hat{e}^a_{N'}(\overline{x}_N^a) - \hat{e}^a_{N'}(x_{ev}^a)]}/{\hat{e}^a_{N'}(\overline{x}_N^a)}$ and ${\overline{\textrm{{\VSSIMPROV}}}^a= 100\% \cdot [ \hat{i}^a_{N'}(\overline{x}_N^a) - \hat{i}^a_{N'}(x_{ev}^a) ]}/{\hat{i}^a_{N'}(\overline{x}_N^a)}$ for all behaviors. Since larger values for the number of students entering or improving are desirable, the numerator’s sign is reversed compared to $\overline{\textrm{{\VSS}}}^a$. We focus our analysis on large instances and use the feasible solution $\overline{x}_N^a$ from the {\SAA} method with $N = 100$ to compute the estimators for all behaviors. We set $N' = 5 \times 10^3$, as the standard deviation of $\hat{g}^a_{N'}(\overline{x}_N^a)$ and $\hat{g}^a_{N'}(x_{ev}^a)$ were below 0.125\% of their mean across all large instances, indicating low variance.

\begin{figure}[t]
    \caption{Average $\overline{{\VSS}}$, $\overline{{\VSSENTER}}$, and $\overline{{\VSSIMPROV}}$ per budget $B$ for large instances} \label{fig:vss}
    \centering
    \begin{subfigure}{0.32\textwidth}
        \centering
        \includegraphics[width=\linewidth]{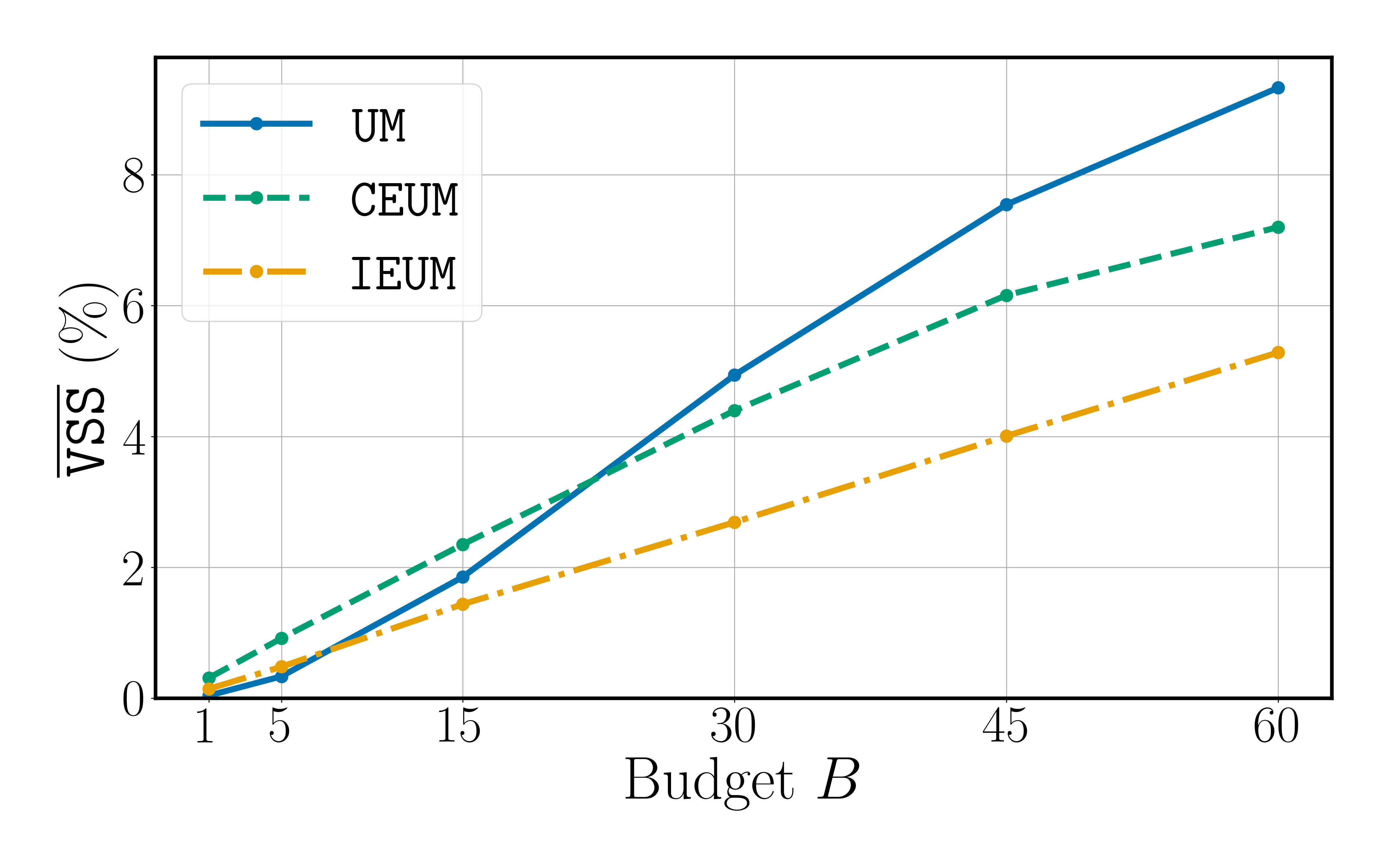}
        \vspace{0.5em}     
        \includegraphics[width=\linewidth]{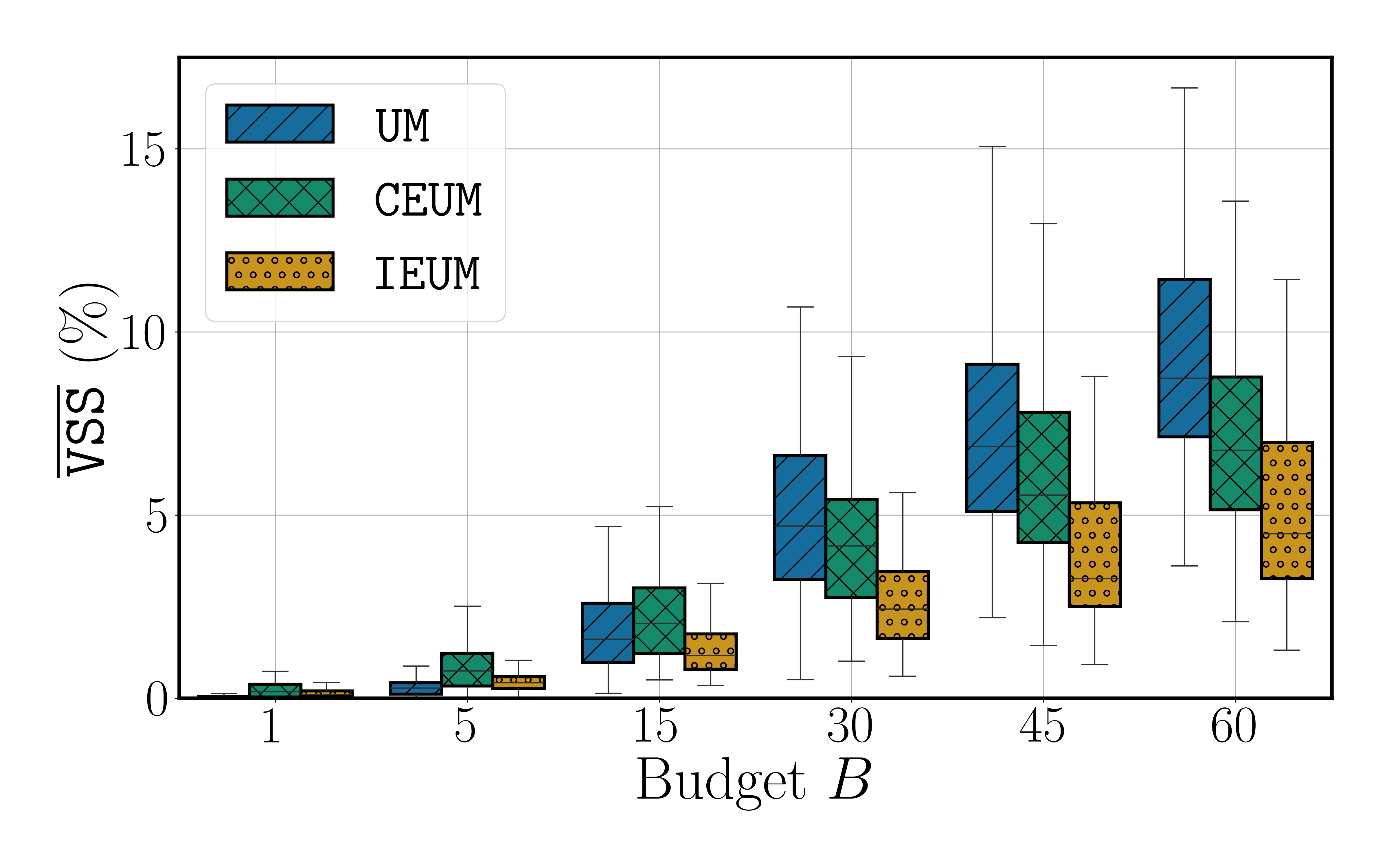}
    \end{subfigure}
    \hfill
    \begin{subfigure}{0.32\textwidth}
        \centering
        \includegraphics[width=\linewidth]{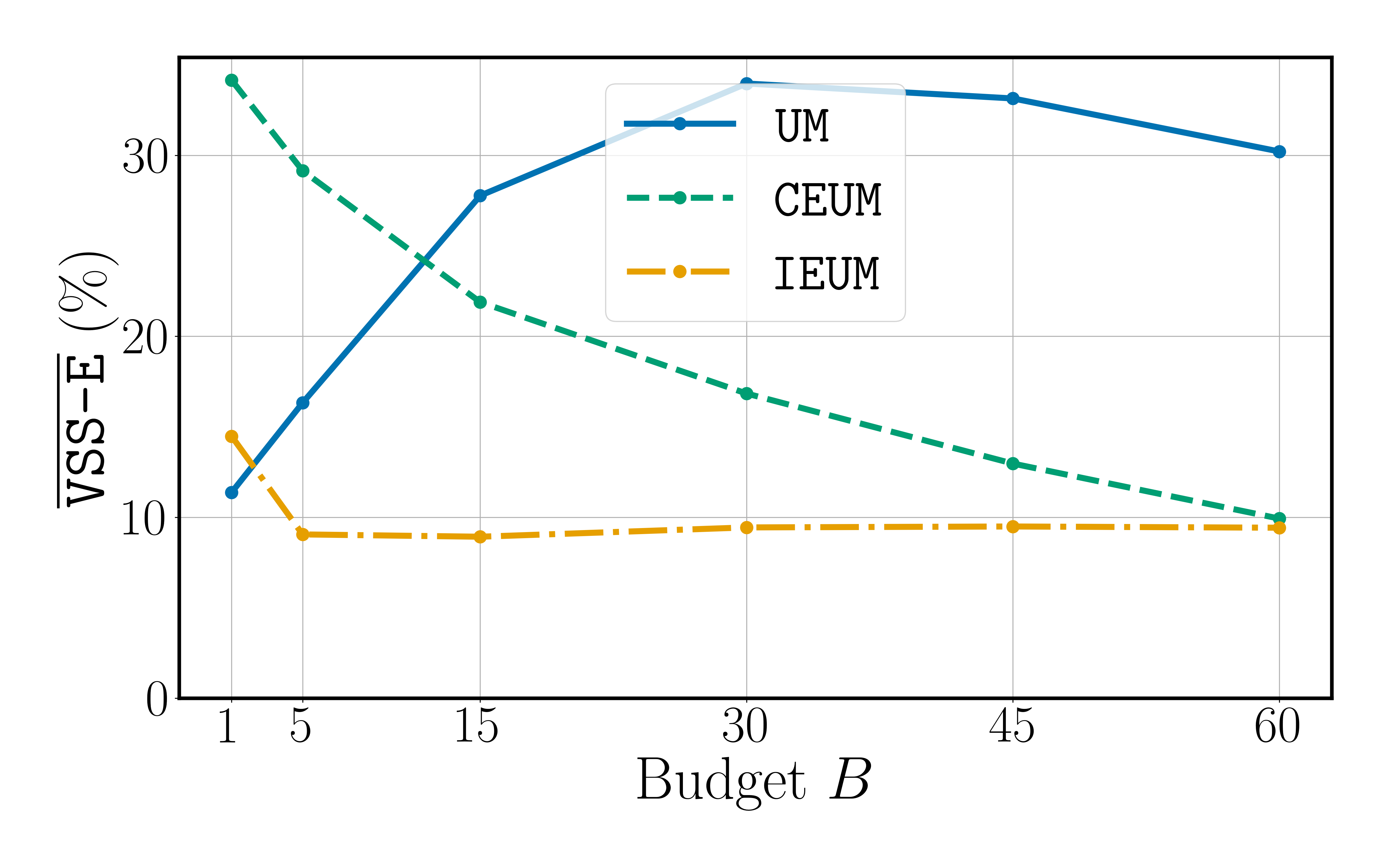}
        \vspace{0.5em}
        \includegraphics[width=\linewidth]{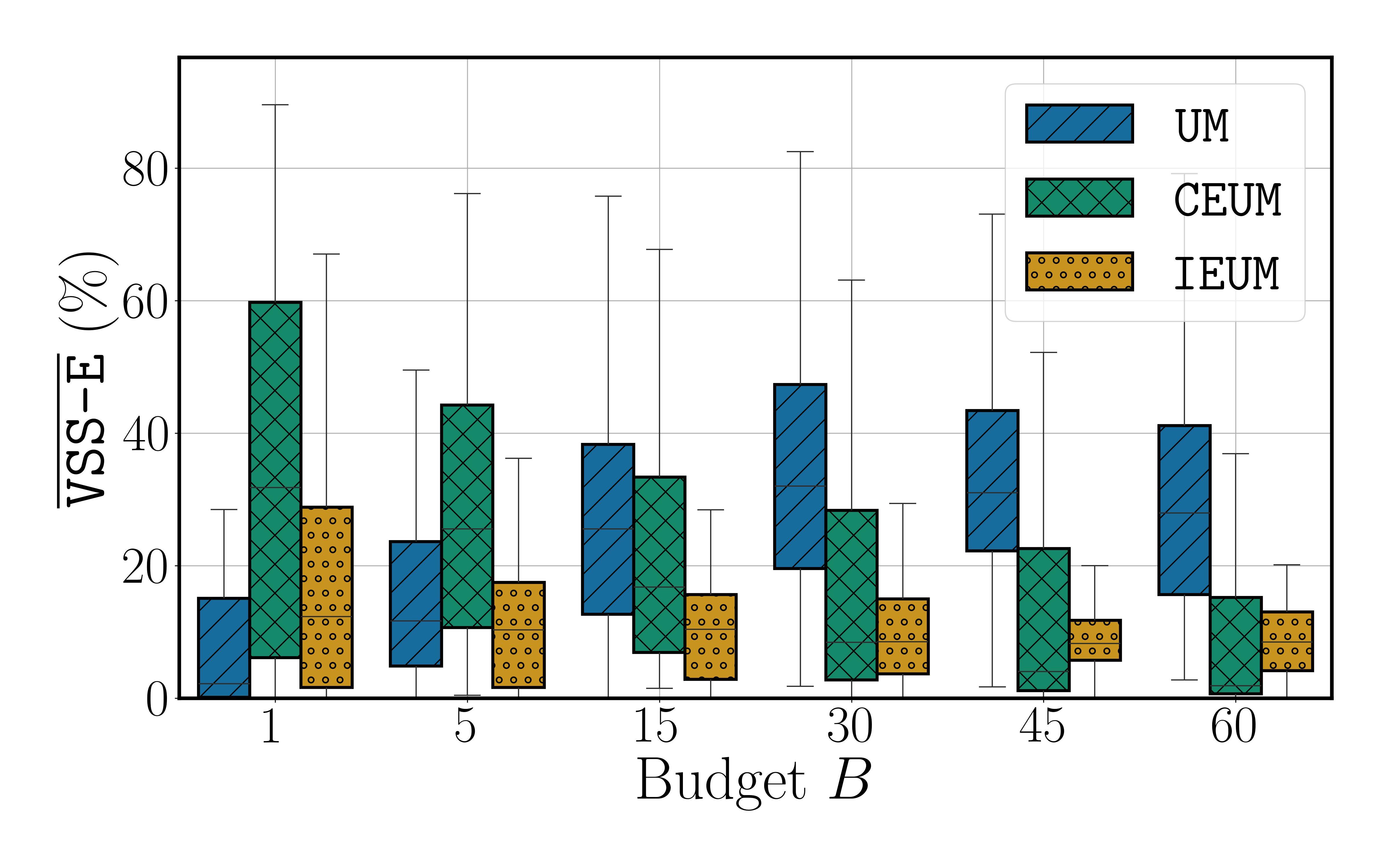}
    \end{subfigure}
    \hfill
    \begin{subfigure}{0.32\textwidth}
        \centering
        \includegraphics[width=\linewidth]{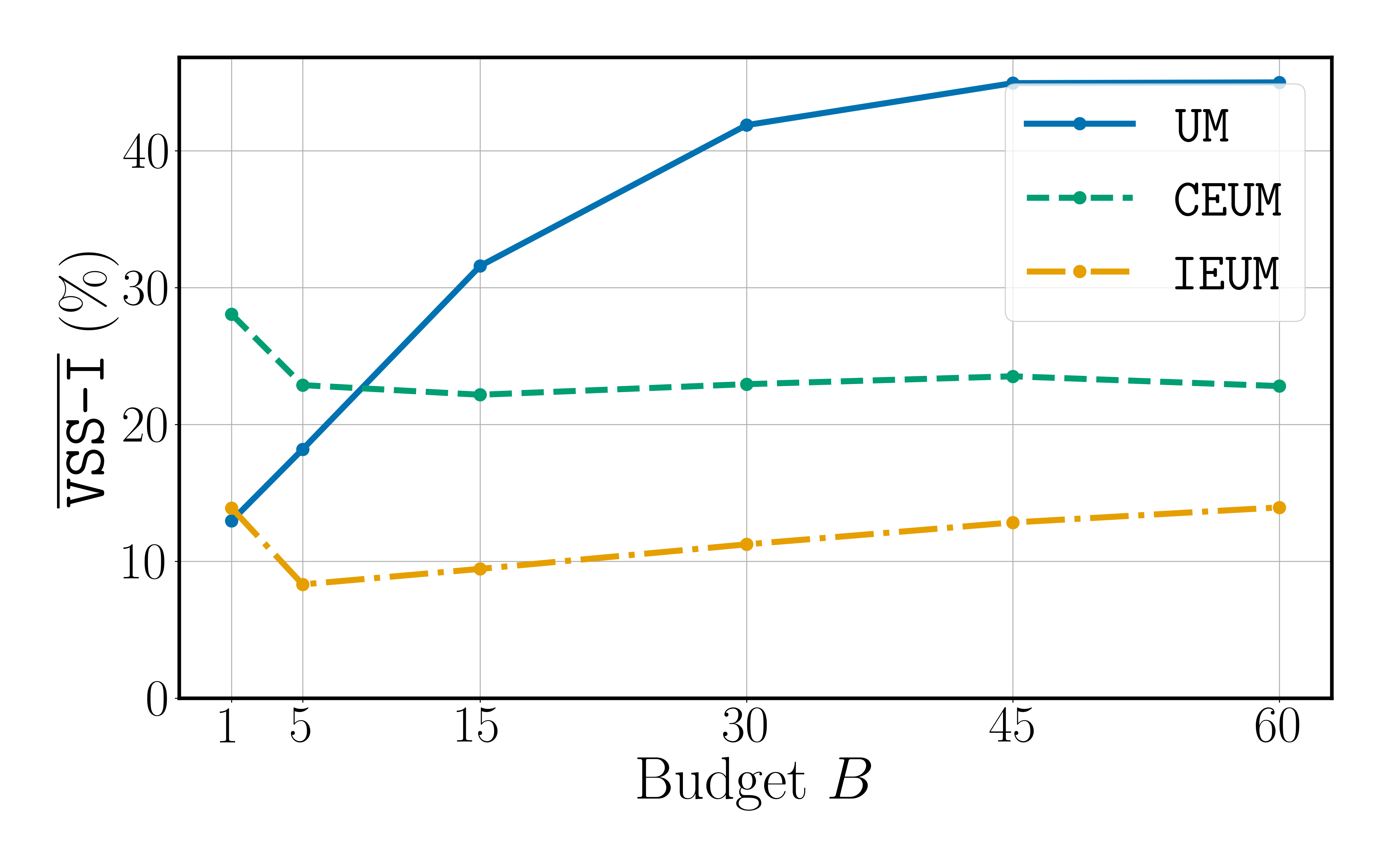}
        \vspace{0.5em}
        \includegraphics[width=\linewidth]{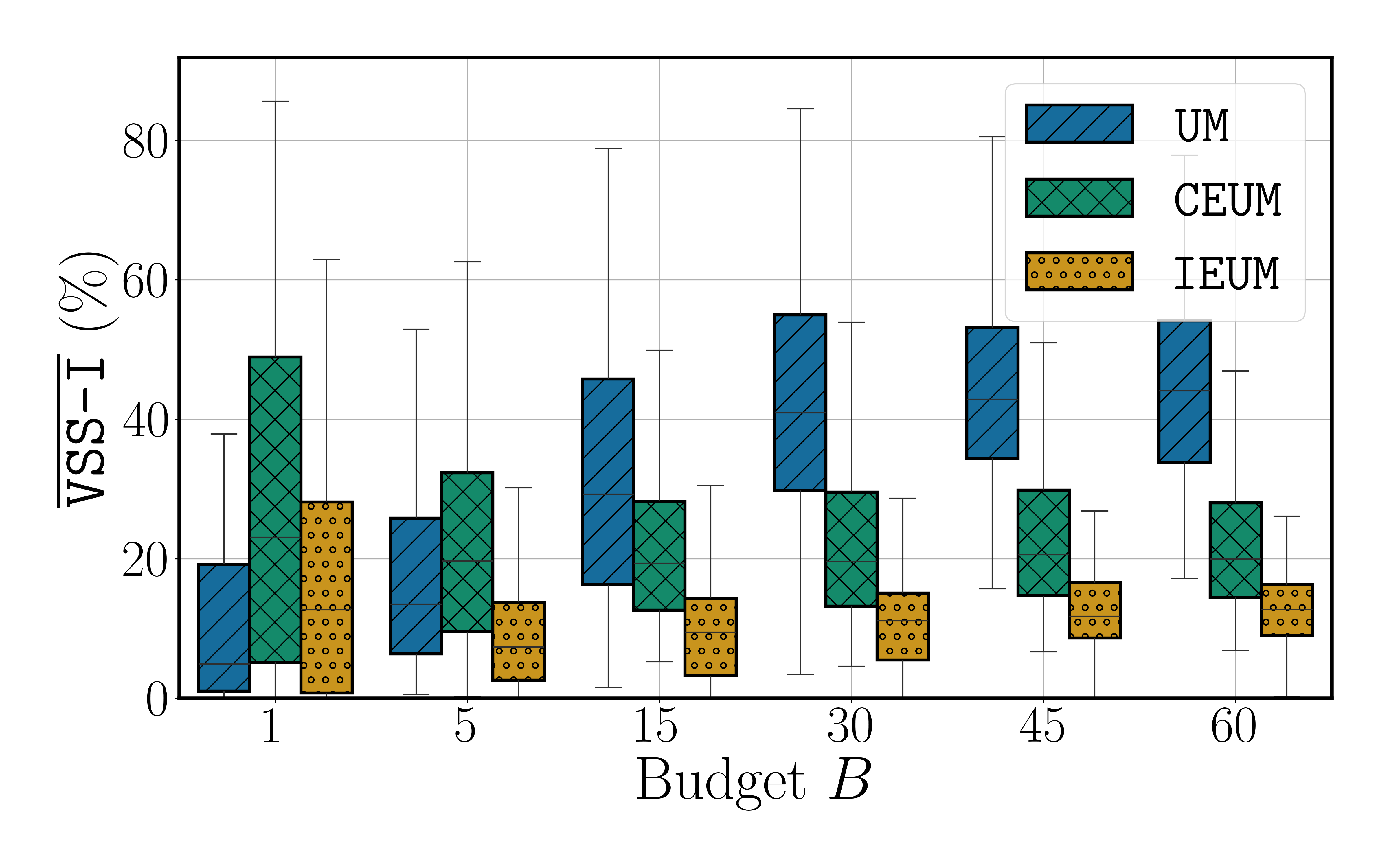}
    \end{subfigure}
\end{figure}

Figure~\ref{fig:vss} shows the mean and quantiles of $\overline{{\VSS}}^a$, $\overline{{\VSSENTER}}^a$, and $\overline{{\VSSIMPROV}}^a$, averaged over large instances with same budgets, for each behavior $a \in \mathcal{A}$ (\changed{see~\ref{app:vss}} for detailed results by budget $B$ and list size limit $K$). Based on the analysis in Section~\ref{subsec:comparison-performance}, the {\SAA} program was solved to termination using the {\LR} heuristic under {\UM}, and the {\SA} heuristic under {\CEUM} and {\IEUM}, with hyperparameters detailed in \changed{~\ref{app:detailed-results}}. For the {\EV} problem, instances under {\UM} were solved to optimality with Gurobi using the modified $L$-constraints formulation. For {\CEUM} and {\IEUM}, exact methods were inefficient, and the {\SA} heuristic was solved (hyperparameters in \changed{~\ref{app:detailed-results}}) until termination. The best feasible solution found was used in place of the optimum $x_{ev}^a$.

Figure~\ref{fig:vss} shows that capacity expansion decisions obtained via the {\SAA} method consistently outperform those from the {\EV} problem across all instances, behaviors, and metrics (students’ matching preferences and the number of entering and improving students). This confirms that solving the {\SAA} program instead of the {\EV} problem is beneficial for all behaviors, whether with endogenous or exogenous uncertainty. \changed{Thus, the $\overline{{\VSS}}^{\UM}$, $\overline{{\VSSIMPROV}}^{\UM}$, and $\overline{{\VSSENTER}}^{\UM}$ clearly demonstrate that our method yields substantially higher-quality matchings than the approaches proposed in prior work, which assume deterministic student preferences.}

Across behaviors, $\overline{{\VSS}}^a$ increases with the budget $B$, showing that stochasticity has a greater impact on students’ matching preferences as the first-stage feasibility set expands. A larger set of possible capacity decisions exacerbates the drawbacks of optimizing on an average scenario, leading to capacity allocations that diverge more from those obtained under explicit uncertainty. For example, at $B = 60$, the average $\overline{{\VSS}}^a$ reaches 9.332\% for {\UM}, 7.201\% for {\CEUM}, and 5.285\% for {\IEUM}. The variance in $\overline{{\VSS}}^a$ increases with $B$, further emphasizing the importance of accounting for uncertainty in preferences.

The benefits of accounting for uncertainty are more pronounced when evaluating the number of students entering or improving in the matching. At $B = 60$, $\overline{{\VSSIMPROV}}^a$ averages 44.985\% for {\UM}, 22.800\% for {\CEUM}, and 13.936\% for {\IEUM}, while $\overline{{\VSSENTER}}^a$ reaches 30.216\%, 9.925\%, and 9.421\%, respectively. Notably, $\overline{{\VSSENTER}}^a$ tends to decrease as $B$ increases, especially under {\CEUM} and {\IEUM}, while $\overline{{\VSSIMPROV}}^a$ stabilizes at higher budgets. These trends occur because, beyond a threshold, available seats exceed initially unassigned students, reducing the marginal benefit of additional capacity for both matching entry and assignment improvements. This diminishing marginal effect is amplified for $\overline{{\VSSENTER}}^a$ by the penalty for unassigned students (set to $|\mathcal{C}_{s,w}^a| + 1$), which encourages the clearinghouse to prioritize improving match quality over reducing unassignments. Additionally, both $\overline{{\VSSIMPROV}}^a$ and $\overline{{\VSSENTER}}^a$ decline as the list size limit $K$ increases, though they remain substantial (see \changed{~\ref{app:vss}}), since a larger $K$ accommodates more students in the initial matching, reducing the marginal impact of extra seats on entry and assignment improvements.

The $\overline{{\VSS}}^a$ is consistently higher in the nonstrategic exogenous case ($\overline{{\VSS}}^{\UM}$) than in the strategic endogenous cases ($\overline{{\VSS}}^{\CEUM}$ and $\overline{{\VSS}}^{\IEUM}$). This is because, under exogenous preferences, the {\EV} problem and the {\SAA} program rely on the same fixed distribution. In contrast, under endogenous uncertainty, the distribution of reported preferences depends on the decisions, so the distribution defined by $x_{ev}^a$ may differ from that defined by $\overline{x}_{N}^a$. The {\EV} problem benefits from additional information embedded in the decision-dependent distribution, which shapes its solution and reduces the perceived impact of uncertainty due to bias. 

\begin{figure}
    \caption{Average $\overline{{\texttt{DIFF-RANK}}}^k$ per list size limit $K$ for large instances with budget $B=60$}\label{fig:diff-rank}
    \centering
    \includegraphics[width=0.67\textwidth]{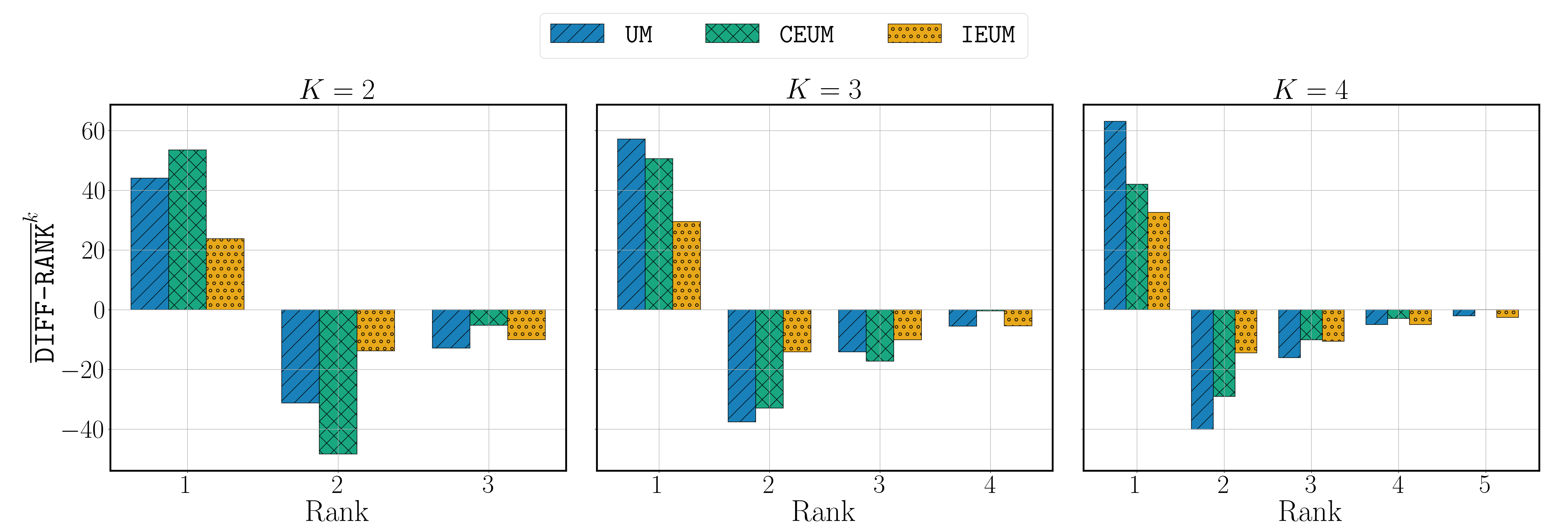}
\end{figure}

We further analyze the impact of stochasticity in capacity expansion by examining differences in the average number of students assigned to each rank. For instances with a maximum of $K$ ranked schools, we define ${\overline{{\texttt{DIFF-RANK}}}^k = \hat{n}^{k,a}_{N'}(\overline{x}^a_N) - \hat{n}^{k,a}_{N'}(x_{ev}^a)}$, where $\hat{n}^{k,a}_{N'}(x)$ is the estimated average number of students assigned to their $k$th preferred school under decision $x$, for $k \in \mathcal{K} \cup \{K+1\}$, with $K+1$ representing unassigned students. Figure~\ref{fig:diff-rank} shows the value of $\overline{{\texttt{DIFF-RANK}}}^k$, averaged over all instances with same list size limit $K$ and budget $B = 60$; results for other budgets (see \changed{~\ref{app:vss}}) are qualitatively similar. The {\EV} and {\SAA} problems are solved using the same algorithms and $N, N'$ values as in the previous analysis.

Figure~\ref{fig:diff-rank} shows that incorporating stochasticity in capacity expansion decisions increases the number of students assigned to their top-ranked schools across all behaviors. Compared to capacity designs based on a single average scenario, solutions from the {\SAA} method consistently reduce lower-ranked assignments and increase matchings to first choices. This highlights that modeling uncertainty enhances student satisfaction, leading to capacity decisions that achieve better matchings on average. The results also indicate that, under the considered unassignment penalty, the clearinghouse implicitly prioritizes match quality over minimizing the number of unmatched students. Overall, solving the {\SAA} program with $N=100$ scenarios yields better feasible solutions than relying on deterministic approximations across all behaviors.

\subsection{Comparison of Students' Behavior} \label{subsec:comparison-strategic}

In this section, we evaluate how capacity decisions derived under one assumed student behavior perform when applied to other behaviors, assessing their robustness to behavioral misspecification. Given a feasible solution $\overline{x}_N^{a'}$ from the {\SAA} method under behavior $a' \in \mathcal{A}$, its performance under a different behavior $a \in \mathcal{A}\backslash\{a'\}$ is measured by $\hat{g}_{N'}^{a}(\overline{x}_N^{a'})$, which computes the expected students' matching preferences under behavior $a$ but using capacity decisions planned for $a'$. The stochastic ($a'$--$a$) behavior gap is ${\overline{\texttt{gap}}^{a'-a}_{N'} = 100\% \cdot [\hat{g}_{N'}^a(\overline{x}_N^{a'}) - \hat{g}_{N'}^a(\overline{x}_N^{a}) ] / \hat{g}_{N'}^a(\overline{x}_N^{a})}$, capturing the relative loss in expected students' matching preferences under behavior $a$ when implementing capacity decisions optimized for $a'$. Similarly, for deterministic approximations, the optimal {\EV} solution $x^{a'}_{ev}$ under $a'$ behavior is evaluated via $\hat{g}_{N'}^a(x^{a'}_{ev})$ for a distinct behavior $a$, and the deterministic ($a'$--$a$) behavior gap is ${\overline{\texttt{gap}}_{ev,N'}^{a'-a} = 100\% \cdot [\hat{g}_{N'}^a(x^{a'}_{ev}) - \hat{g}_{N'}^a(x^{a}_{N}) ] / \hat{g}_{N'}^a(x^{a}_{N})}$, measuring the relative loss in expected student matching preference under behavior $a$ when using the average-scenario capacity decision obtained under $a'$ instead of the stochastic solution under behavior $a$.

As discussed, the relatively low penalty for unassigned students leads the clearinghouse to prioritize improving assignment quality over minimizing the number of unmatched students. We thus evaluate the average number of students benefiting from expanded capacities compared to the case of no extra capacities ($B=0$), which includes both students who newly enter the matching and those who improve their assigned school, \changed{i.e., the benefit estimator is ${\hat{b}_{N'}^a(x) = \hat{i}_{N'}^a(x) + \hat{e}_{N'}^a(x)}$.} Using this metric, we compute the ($a'$--$a$) stochastic and deterministic behavior gaps: ${\overline{\texttt{gap-b}}_{N'}^{a'-a} = 100\% \cdot [ \hat{b}_{N'}^a(\overline{x}_N^{a}) -\hat{b}_{N'}^a(\overline{x}_N^{a'}) ] / \hat{b}_{N'}^a(\overline{x}_N^{a})}$ and ${\overline{\texttt{gap-b}}_{ev,N'}^{a'-a} = 100\% \cdot [ \hat{b}_{N'}^a(x^{a}_{N}) - \hat{b}_{N'}^a(x^{a'}_{ev})] / \hat{b}_{N'}^a(x^{a}_{N})}$. The sign of the numerator is reversed relative to the original objective-based gaps as the benefit metric is maximized, while the original objective is minimized.

\begin{figure}[t]
    \caption{Stochastic behavior gaps per list size limit $K$ and budget $B$ for large instances.}
    \label{fig:stoch_behavior_gaps}
    \centering
    \begin{subfigure}[b]{\textwidth}
        \centering
        \includegraphics[width=0.85\textwidth]{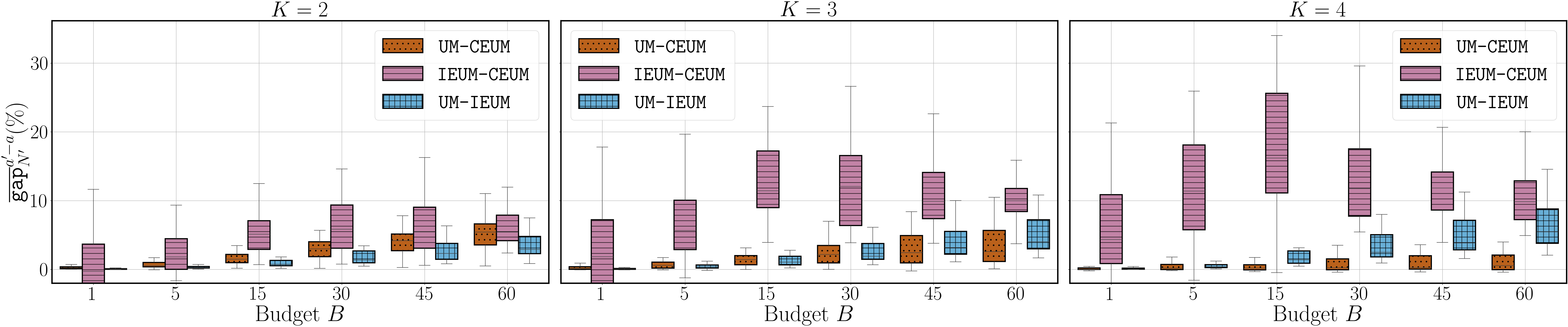}
    \end{subfigure}
    \hfill
    \begin{subfigure}[b]{\textwidth}
        \centering
        \includegraphics[width=0.85\textwidth]{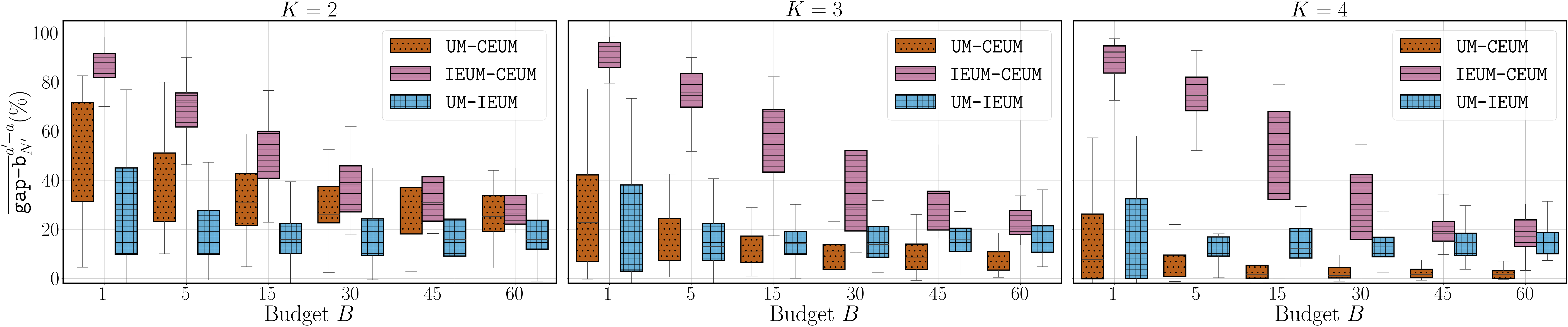}
    \end{subfigure}
\end{figure}

\begin{figure}[t]
    \caption{Deterministic behavior gaps per list size limit $K$ and budget $B$ for large instances.}
    \label{fig:det_behavior_gaps}
    \centering
    \begin{subfigure}[b]{\textwidth}
        \centering
        \includegraphics[width=0.85\textwidth]{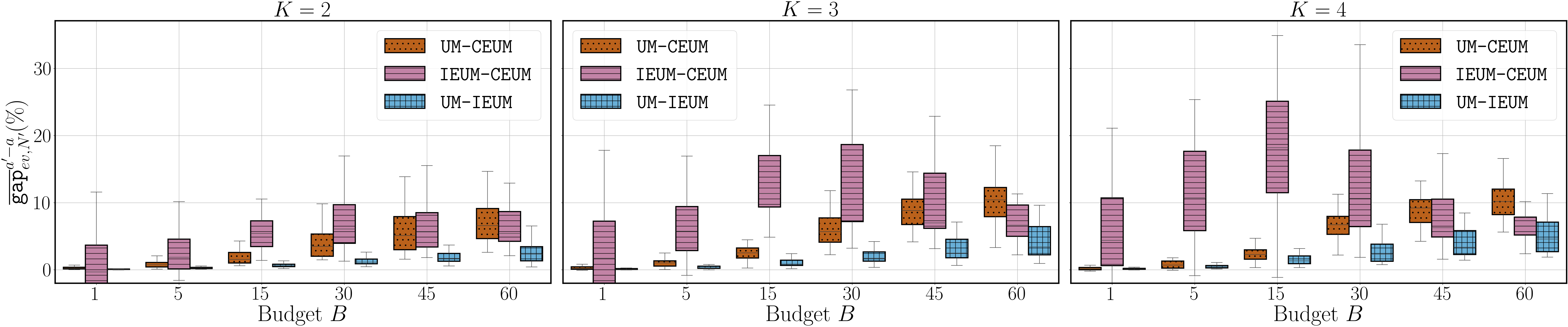}
    \end{subfigure}
    \hfill
    \begin{subfigure}[b]{\textwidth}
        \centering
        \includegraphics[width=0.85\textwidth]{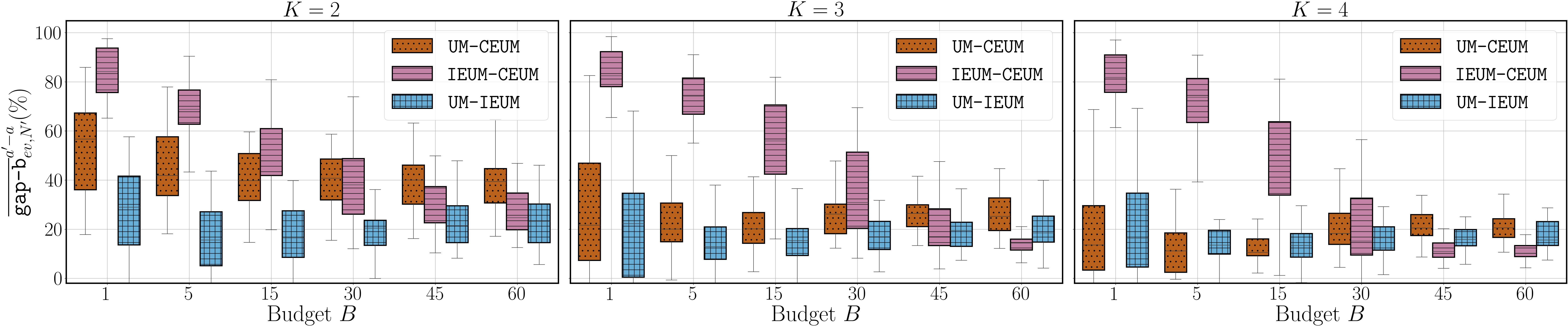}
    \end{subfigure}
\end{figure}

We analyze the effect of assuming truthful behavior when students act strategically, \changed{and of} misrepresenting strategic behavior, on capacity decisions \changed{by computing} stochastic and deterministic behavior gaps ({\UM}--{\CEUM}), ({\UM}--{\IEUM}), and ({\IEUM}--{\CEUM}) for both metrics. Figures~\ref{fig:stoch_behavior_gaps} and~\ref{fig:det_behavior_gaps} show the gaps averaged over instances with the same list size limit $K$ and budget $B$ (see \changed{~\ref{app:comparison-strategic}} for detailed results). We also set $N' = 5 \times 10^3$, solve the {\SAA} program with $N = 100$ to obtain $\overline{x}_N^a$, and follow the solution procedures described earlier.

Figures~\ref{fig:stoch_behavior_gaps} and~\ref{fig:det_behavior_gaps} show that capacity decisions optimized for truthful {\UM} behavior perform worse across both metrics when students act strategically according to {\CEUM}, compared to decisions optimized for the correct behavior. For the students’ matching preferences metric, both stochastic and deterministic ({\UM}--{\CEUM}) gaps increase with budget $B$. The trends diverge with respect to the list size limit $K$: the stochastic gap decreases as $K$ increases, averaging 5.477\%, 3.430\%, and 1.106\% for $K = 2$, 3, and 4 at $B = 60$, while the deterministic gap rises to 7.434\%, 10.253\%, and 10.534\%, respectively.

The trend in stochastic gaps arises from how students report preferences under each behavior. When $K$ is small, students cannot list all truly preferred schools, so reported preferences under truthful {\UM} and strategic {\CEUM} behaviors diverge more. Under {\CEUM}, students account for rejection risk in their utility evaluations, so the probability of attending a given school reflects both the admission chance to that school and the rejection risk from more preferred schools, captured by the function $\varrho_{s,c}^w(\cdot)$. With a larger $K$, the expected utility of truly less preferred schools is more heavily penalized by rejection risk, reducing their likelihood of being reported. In contrast, truly preferred schools are less affected by this risk, making them more likely to be in the reported preferences. Thus, as $K$ increases, reported preferences under {\CEUM} behavior become increasingly aligned with those under {\UM}, enhancing the quality of capacity expansion decisions based on {\UM} behavior when students actually follow {\CEUM}. Alternatively, deterministic gaps increase with $K$ because relying on the average scenario without explicitly modeling uncertainty further misrepresents reported preferences, amplifying divergence between decisions under the assumed {\UM} versus true {\CEUM} behavior. As the feasibility set for capacity decisions expands with budget $B$, the impact of misrepresented behavior grows, explaining the lower quality of decisions based on incorrect {\UM} behavior at higher budgets.

For the number of benefiting students, both stochastic and deterministic ({\UM}--{\CEUM}) behavior gaps decrease as $B$ and $K$ increase, though they remain substantial. At $B = 60$, average stochastic gaps are 25.634\%, 7.998\%, and 1.919\% for $K = 2$, 3, and 4, respectively, while deterministic gaps are higher at 37.853\%, 26.065\%, and 21.709\%. These reflect two effects: (i) when $B$ exceeds student demand, extra capacity has diminishing returns, and (ii) when $K$ is large, more students are accommodated without extra seats. Both reduce the marginal benefit of capacity expansion, moderating the impact of behavior misspecification.


Figures~\ref{fig:stoch_behavior_gaps} and~\ref{fig:det_behavior_gaps} also show that capacity decisions based on nonstrategic {\UM} behavior yield lower-quality outcomes when students act strategically under {\IEUM}, for both evaluation metrics, compared to decisions based on the correct behavior. For the students’ matching preferences, both stochastic and deterministic ({\UM}--{\IEUM}) behavior gaps increase with the list size limit $K$ and budget $B$. At $B = 60$, average stochastic gaps are 3.888\%, 5.461\%, and 6.572\%, while deterministic gaps are 2.784\%, 4.281\%, and 5.297\% for $K = 2$, 3, and 4, respectively. Under {\IEUM}, students’ reported preferences reflect their true preferences and each school’s admission chance, but (unlike {\CEUM}) they ignore rejection risk from more preferred schools. This increases the discrepancy between reported preferences under {\UM} and {\IEUM}, especially as $K$ grows and students face more strategic options, and reduces the quality of capacity decisions based on {\UM} behavior when students actually follow {\IEUM. Moreover, the rise in behavior gap with larger $B$ reflects the expanded and more variable capacity decision space, as observed in the ({\UM}--{\CEUM}) case.

For the number of benefiting students metric, both stochastic and deterministic ({\UM}--{\IEUM}) behavior gaps remain substantial and relatively stable across budgets $B$ and list size limits $K$, with slightly higher values at smaller $B$ and $K$. At $B=60$, the average stochastic gaps are 18.417\%, 16.984\%, and 15.413\%, while the deterministic gaps are slightly higher at 23.156\%, 20.198\%, and 17.590\% for $K=2$, 3, and 4. As in the ({\UM}--{\CEUM}) case, larger $B$ or $K$ reduces the marginal impact of strategic misalignment, since more students can be accommodated even under suboptimal capacity allocations. \changed{The deterministic ({\UM}--{\CEUM}) and ({\UM}--{\IEUM}) behavioral gaps clearly indicate that our method yields substantially higher-quality matchings than the approaches proposed in prior work, which assume deterministic and truthful student preferences.}


Capacity decisions based on misrepresenting strategic behavior (assuming {\IEUM} instead of {\CEUM}) exhibit substantially lower quality across all budgets $B$, list size limits $K$, and evaluation metrics compared to decisions based on the correct behavior. For the students’ matching preferences metric, the stochastic and deterministic ({\IEUM}–{\CEUM}) behavior gaps increase with $K$ and $B$, reaching averages of 6.387\%, 10.045\%, and 10.425\% for $K = 2,3,4$ at $B = 60$ (stochastic) and 6.898\%, 7.668\%, and 6.845\% (deterministic). In contrast, for the number of benefiting students metric, both stochastic and deterministic gaps decrease as $K$ and $B$ increase, yet remain significant: at $B = 60$, stochastic gaps are 29.012\%, 21.886\%, and 17.837\%, and deterministic gaps are 27.689\%, 15.627\%, and 10.432\% for $K = 2,3,4$. These results indicate that recognizing strategic behavior is insufficient if its form is misassumed; in particular, whether students account for rejection probabilities from more preferred schools, as in {\CEUM} versus {\IEUM}, has a substantial effect on outcomes. It is therefore essential to recognize strategic behavior and accurately model its form.

Misrepresenting student behavior results in lower-quality capacity decisions. However, when $K$ is large, decisions based on {\UM} behavior closely approximate those under the strategic {\CEUM} behavior, whereas for small $K$, they resemble decisions under the strategic {\IEUM} behavior. This motivates solving the {\STMMP} under {\UM} in such cases, as it produces a simpler optimization problem with exogenous uncertainty while still yielding accurate decisions in the relevant $K$ regimes. Nevertheless, even when the impact on students’ matching preferences is limited, the loss in the number of students benefiting from extra capacity can be substantial (a critical metric for effective policy design). The similarity between stochastic and deterministic behavior gaps highlights that accurately modeling student behavior is as important as accounting for uncertainty: failing to capture accurate behavior can undermine otherwise well-structured capacity planning.

\subsection{Computational Performance} \label{subsec:comparison-performance}

A sufficiently large $\mathcal{W_N}$ is required for convergence of the {\SAA} program and its optimal solution to the true expectation and optimum of the {\STMMP}. However, even the {\STMMP} on a single scenario is NP-hard~\citep{bobbio2022capacity}. Thus, we evaluate the computational performance of our methodologies.

\subsubsection{Truthful Behavior}\label{subsubsec:perf-um}
We compare the solution quality and runtime of the proposed methods for the {\SAA} Program~\eqref{prog:saa} of the {\STMMP}-{\UM} using $N=100$ scenarios: the modified $L$-constraints formulation solved with Gurobi, as well as the {\ASG}, {\LR}, {\LS}, and {\SA} heuristics (see \changed{~\ref{app:perf-nonstrat}} for details on hyperparameter values for the {\LR} and {\SA} heuristics). The warm-start procedure for all approaches is defined in \changed{~\ref{app:warm-start}}. Figure~\ref{fig:perf-um} reports the results for large instances, with all runs limited to a time limit of two hours. Figure~\ref{subfig:time-um} shows the runtime in seconds, and Figure~\ref{subfig:optgap-um} the optimality gap [\({\text{gap}_{opt} = 100\% \cdot (ub - lb) / ub}\)], where \(ub\) and \(lb\) denote the best upper and lower bounds obtained by the approach on the optimal value of {\SAA} Program~\eqref{prog:saa}, per percentage of instances for each approach. The {\LS} and {\SA} heuristics do not provide optimality gaps. Figure~\ref{subfig:ubgap-um} reports the upper bound gap [\({\text{gap}_{ub^*} = 100\% \cdot (ub - ub^*) / ub}\)], where \(ub\) is defined as before, and \(ub^*\) is the best upper bound found across all approaches within the time limit. We refer to \changed{~\ref{app:detailed-results}} for detailed results.

\begin{figure}
    \begin{subfigure}[c]{0.32\linewidth}
        \includegraphics[width=\linewidth]{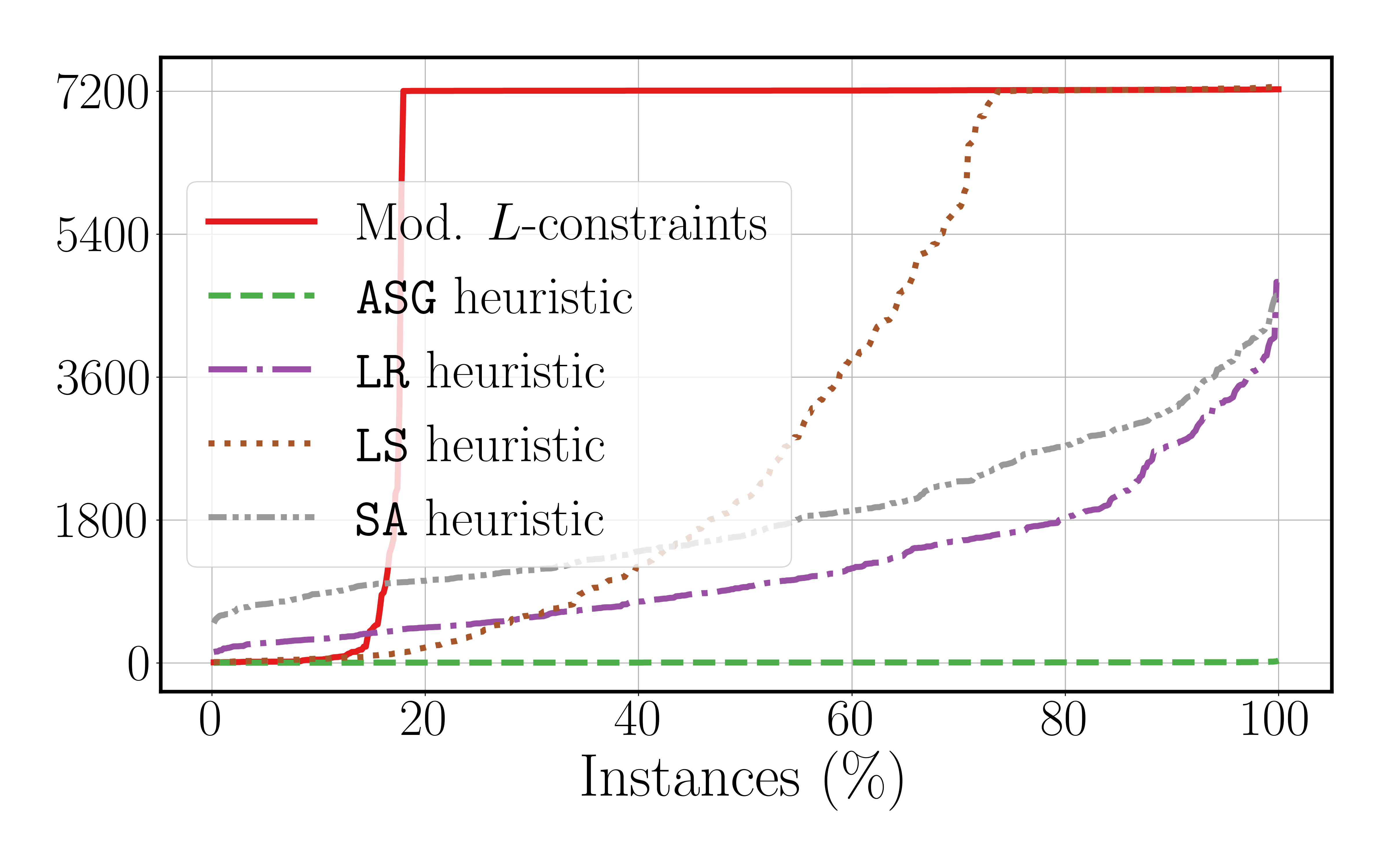}
        \caption{Time (seconds)}
        \label{subfig:time-um}
    \end{subfigure}
    \hfill
    \begin{subfigure}[c]{0.32\linewidth}
        \includegraphics[width=\linewidth]{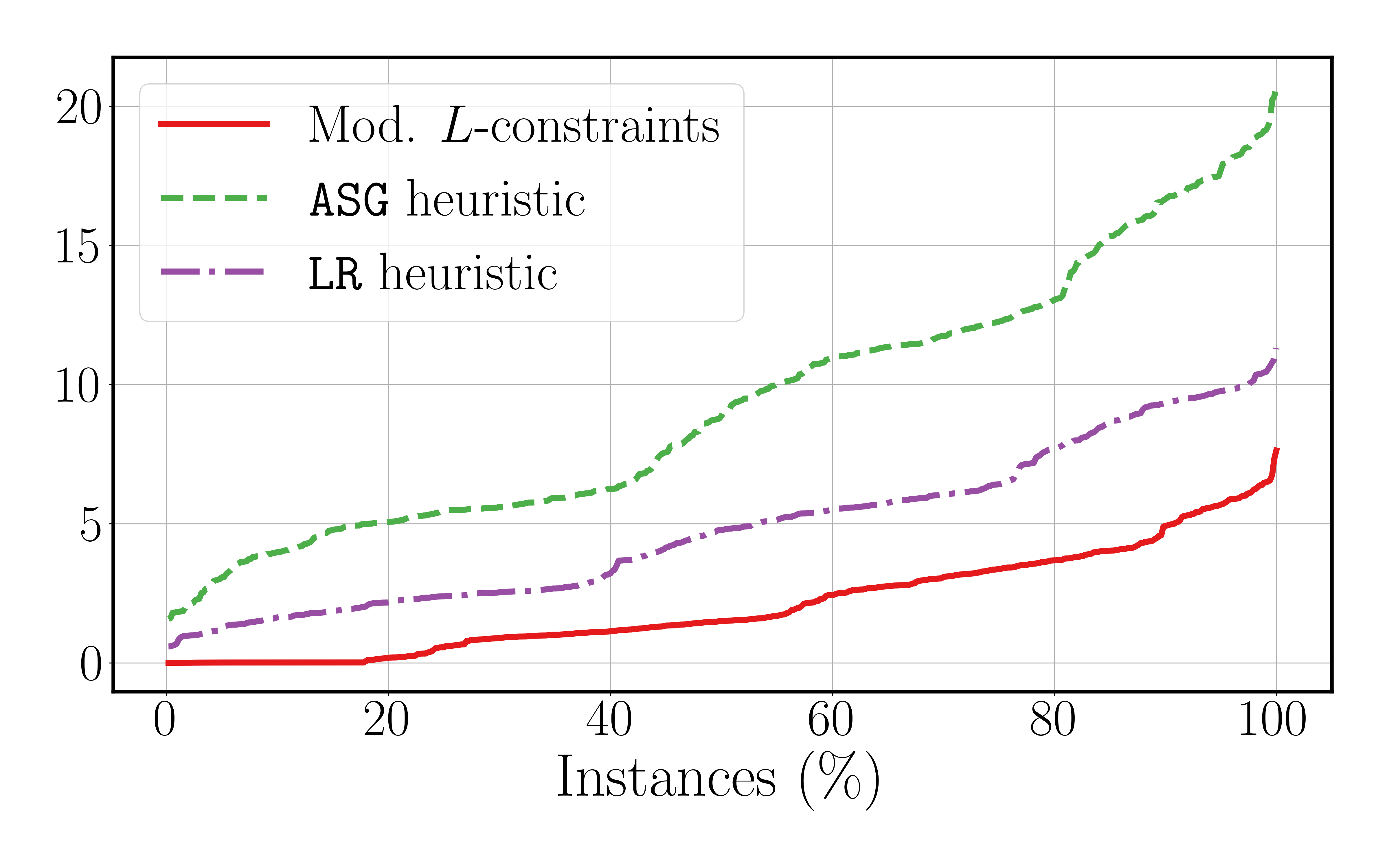}
        \caption{Optimality gap [$\text{gap}_{opt}$]  (\%)}
        \label{subfig:optgap-um}
    \end{subfigure}
    \hfill
    \begin{subfigure}[c]{0.32\linewidth}
        \includegraphics[width=\linewidth,height=0.23\textheight]{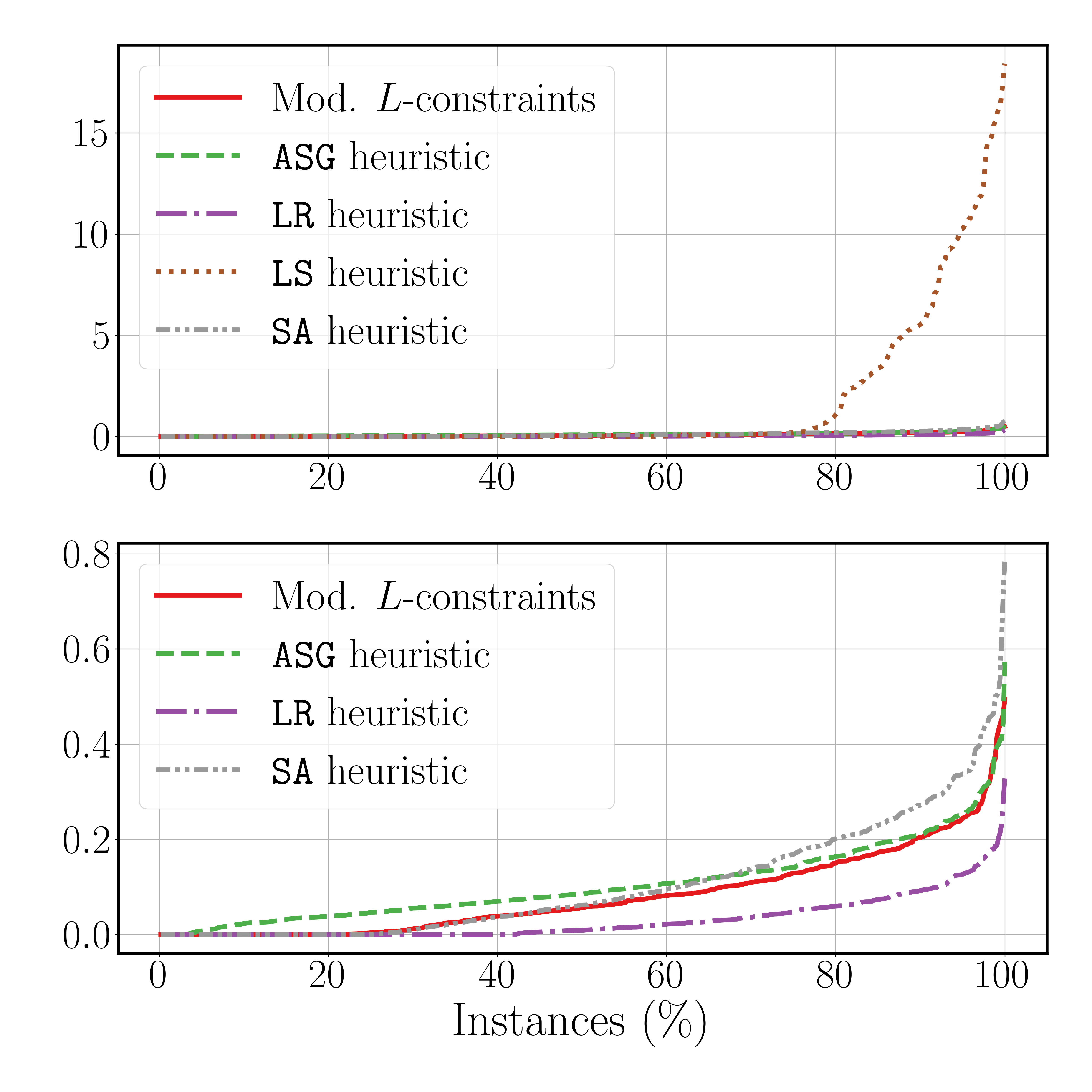}
        \caption{Upper bound gap [$\text{gap}_{ub^{*}}$] (\%)}
        \label{subfig:ubgap-um}
    \end{subfigure}
    \caption{Computational performance for large instances of approaches for {\SAA} Program~\eqref{prog:saa} under {\UM} behavior.}
    \label{fig:perf-um}
\end{figure}

Figure~\ref{fig:perf-um} shows that the {\SAA} Program~\eqref{prog:saa} under {\UM} behavior can be effectively solved with the modified $L$-constraints approach within a reasonable time. This approach solved 18\% of large instances to optimality, achieved optimality gaps below 5\% in 90\% of cases, and maintained upper bound gaps under 0.5\% across all instances (average of 0.082\%) within a two-hour limit, with an average runtime of 5980.1 seconds. This approach is particularly suitable when capacity decisions are made well in advance of the matching process, for example, during annual strategic planning, where computational time is less restrictive.

However, improving the quality of the {\SAA} method may require solving multiple {\SAA} Programs~\eqref{prog:saa} with different sets $\mathcal{W}_\mathcal{N}$ or increasing the number of scenarios $N$ to better capture uncertainty. In such cases, solving the {\SAA} program exactly becomes more challenging. This is particularly relevant in time-sensitive contexts (e.g., when public funding, such as government grants, must be allocated and implemented rapidly), where relying solely on exact methods may be impractical and efficient heuristics are necessary.

Figure~\ref{fig:perf-um} shows that the {\ASG} heuristic yields high-quality solutions; it produced feasible solutions with an average upper bound gap of 0.106\% under 27 seconds across all instances. The {\LR} heuristic performs even better, achieving the lowest average upper bound gap of 0.031\% and smaller optimality gaps than the {\ASG} heuristic (4.834\% vs.\ 9.323\%, see Figure~\ref{subfig:optgap-um}). While heuristics typically lack formal guarantees on solution quality, the {\LR} lower bounds provide performance guarantees. The runtime of both the {\ASG} and {\LR} heuristics is stable over instances (see Figure~\ref{subfig:time-um}), indicating their efficacy in handling instances with large first-stage feasibility sets. These methods are highly effective when speed is critical for the matching.

Figure~\ref{subfig:ubgap-um} shows that the {\LS} heuristic consistently yields higher upper bound gaps than the other approaches, averaging 1.475\%. Its performance is budget-sensitive, i.e., influenced by the size of the first-stage feasibility set. As shown in Figure~\ref{subfig:time-um}, the {\LS} heuristic reached the time limit in 26.5\% of the instances, primarily for the large budgets $B \in \{30, 45, 60\}$, resulting in lower solution quality, with an average upper bound gap of 2.94\% and runtime of 5380 seconds. For the small budgets $B \in \{1, 5, 15\}$, it produced high-quality solutions within the time limit, with an average gap of 0.011\% and runtime of 1054 seconds (see \changed{~\ref{app:perf-nonstrat}} for details). These trends stem from the neighborhood in the {\LS} heuristic, which directly affects runtime. For large-budget instances, the heuristic requires more iterations to converge, resulting in longer runtimes. It is then more effective for small-budget instances, where convergence occurs in fewer iterations.

In contrast, the {\SA} heuristic maintains low upper bound gaps across all budgets (averaging 0.105\%) and exhibits lower, more stable runtimes than the {\LS} heuristic (see Figure~\ref{subfig:time-um}). Unlike the {\LS} heuristic, the {\SA} heuristic performs a randomized search over the neighborhood rather than exhaustively exploring it, so its runtime is more closely tied to hyperparameters (iteration limits and temperature decay) than on the size of the feasibility set, making it more scalable to large-budget instances. Notably, the {\LR} heuristic achieves slightly lower runtimes and upper bound gaps than the {\SA} heuristic, confirming that it remains the most effective of the three in both solution quality and efficiency.

\subsubsection{Strategic Behaviors.}\label{subsubsec:perf-strat}

To solve the {\SAA} Program~\eqref{prog:saa} of the {\STMMP}-{\IEUM} and {\STMMP}-{\CEUM}, we compare the exact cutoff score formulation solved with Gurobi against the {\LS} and {\SA} heuristics. Initial experiments showed that, even when relaxing stability constraints (as in the {\ASG} and {\LR} heuristics), the problem remains computationally challenging due to the endogenous preferences. In particular, solving the {\SAA} program exactly under {\CEUM} or {\IEUM} is much harder than under {\UM}, with large instances often exhibiting optimality gaps near 100\% after eight hours. We explored integer and alternating L-shaped decomposition techniques~\citep{angulo2016improving}, but found them ineffective due to weak cuts, common for two-stage models with integer second-stage decisions, resulting in slow convergence and loose bounds. \changed{This limitation extends to the endogenous, distribution-specific L-shaped cuts by~\cite{pantuso2025shaped}, since each capacity design induces a distinct distribution, making the resulting cuts separate only a small subset of first-stage solutions.}

We thus structure our analysis in two parts: (i) evaluating the computational performance of the proposed methods on small instances, where optimality can be reliably assessed; and (ii) comparing the performance of the {\LS} and {\SA} heuristics on large instances, where exact methods become impractical (see \changed{~\ref{app:perf-strat}} for the {\SA} heuristic hyperparameters). \changed{~\ref{app:warm-start}} describes the warm-start strategy used in the exact approach.

Figures~\ref{fig:perf-ceum-small} and~\ref{fig:perf-ieum-small} present the results for small instances with $N = 50$ scenarios under the {\CEUM} and {\IEUM} behaviors, respectively. All runs had an eight-hour time limit. We report the runtime (Figures~\ref{subfig:time-ceum-small} and~\ref{subfig:time-ieum-small}), the optimality gap [$\text{gap}_{opt}$] (Figures~\ref{subfig:optgap-ceum-small} and~\ref{subfig:optgap-ieum-small}), and the upper bound gap [$\text{gap}_{ub^*}$] (Figures~\ref{subfig:ubgap-ceum-small} and~\ref{subfig:ubgap-ieum-small}) as functions of the percentage of instances, following definitions from the previous section. Figures~\ref{fig:perf-ceum-large} and~\ref{fig:perf-ieum-large} show the results for large instances with $N = 100$ scenarios for {\CEUM} and {\IEUM}, respectively, under the same time limit. For these large instances, we focus on local search-based heuristics, reporting the runtime (Figures~\ref{subfig:time-ceum-large} and~\ref{subfig:time-ieum-large}) and the upper bound gap (Figures~\ref{subfig:ubgap-ceum-large} and~\ref{subfig:ubgap-ieum-large}). \changed{~\ref{app:perf-strat}} has detailed results.

\begin{figure}[t]
    \begin{subfigure}[b]{0.32\linewidth}
        \includegraphics[width=\linewidth]{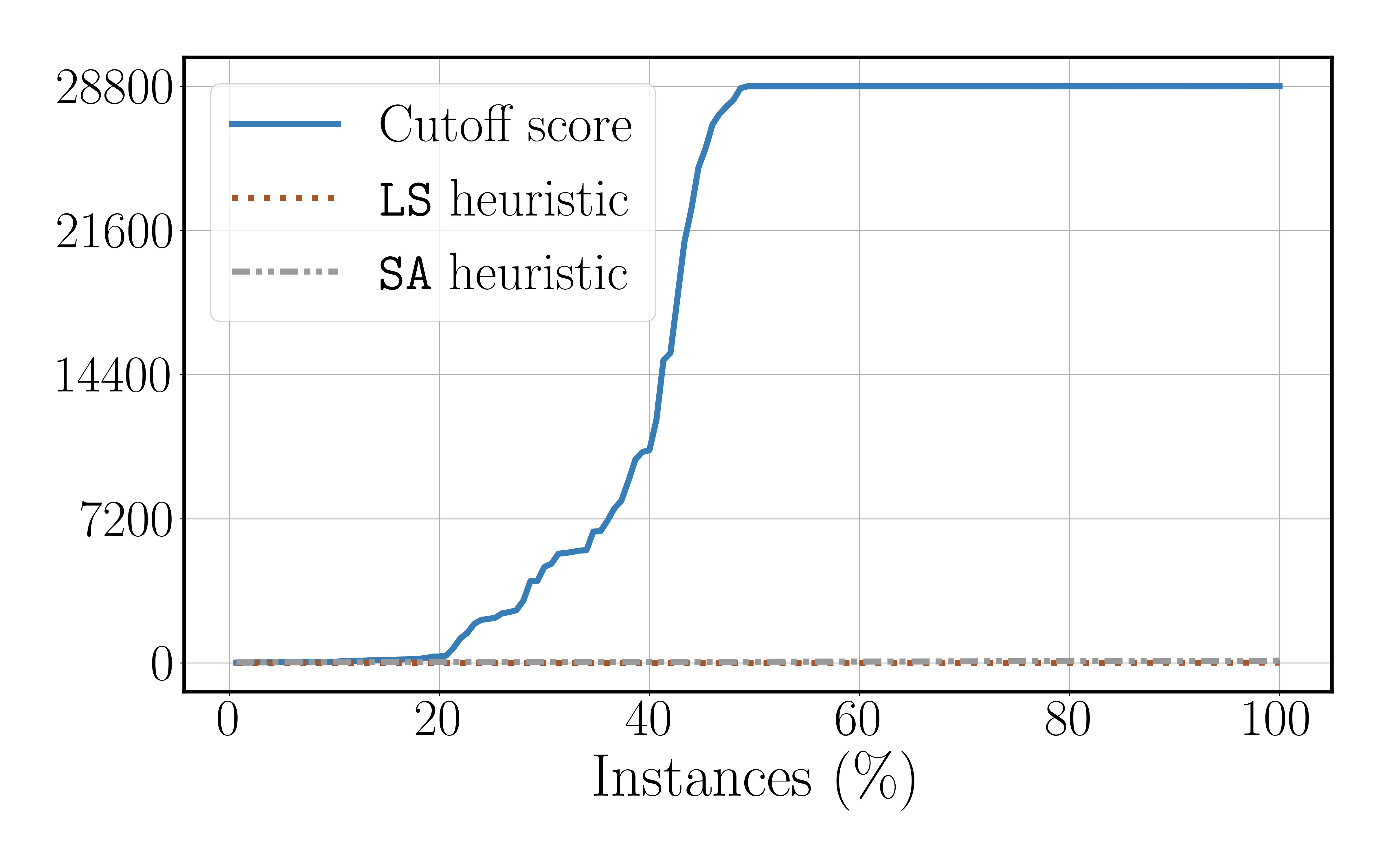}
        \caption{Time (seconds)}
        \label{subfig:time-ceum-small}
    \end{subfigure}
    \hfill
    \begin{subfigure}[b]{0.32\linewidth}
        \includegraphics[width=\linewidth]{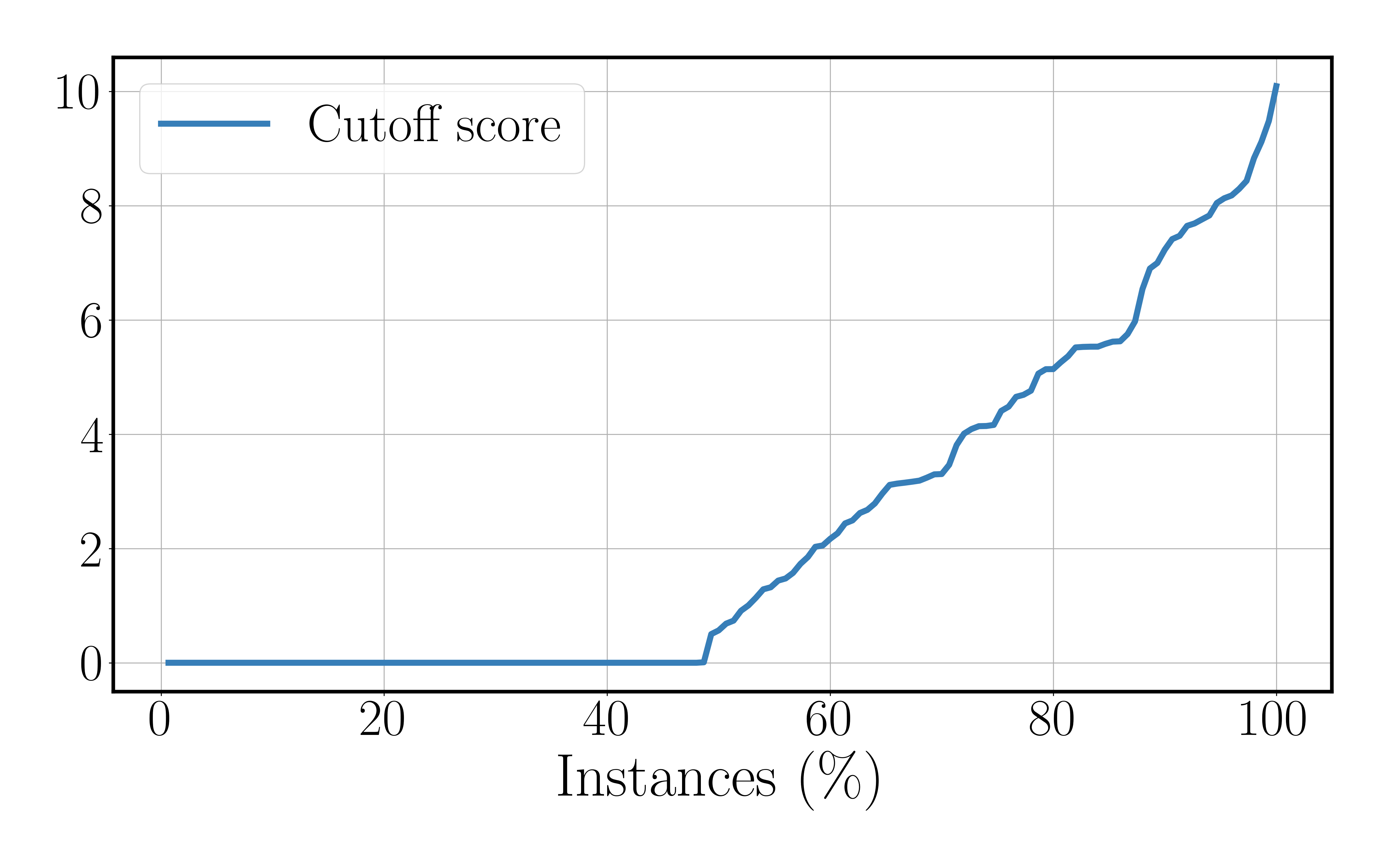}
        \caption{Optimality gap [$\text{gap}_{opt}$]  (\%)}
        \label{subfig:optgap-ceum-small}
    \end{subfigure}
    \hfill
    \begin{subfigure}[b]{0.32\linewidth}
        \includegraphics[width=\linewidth]{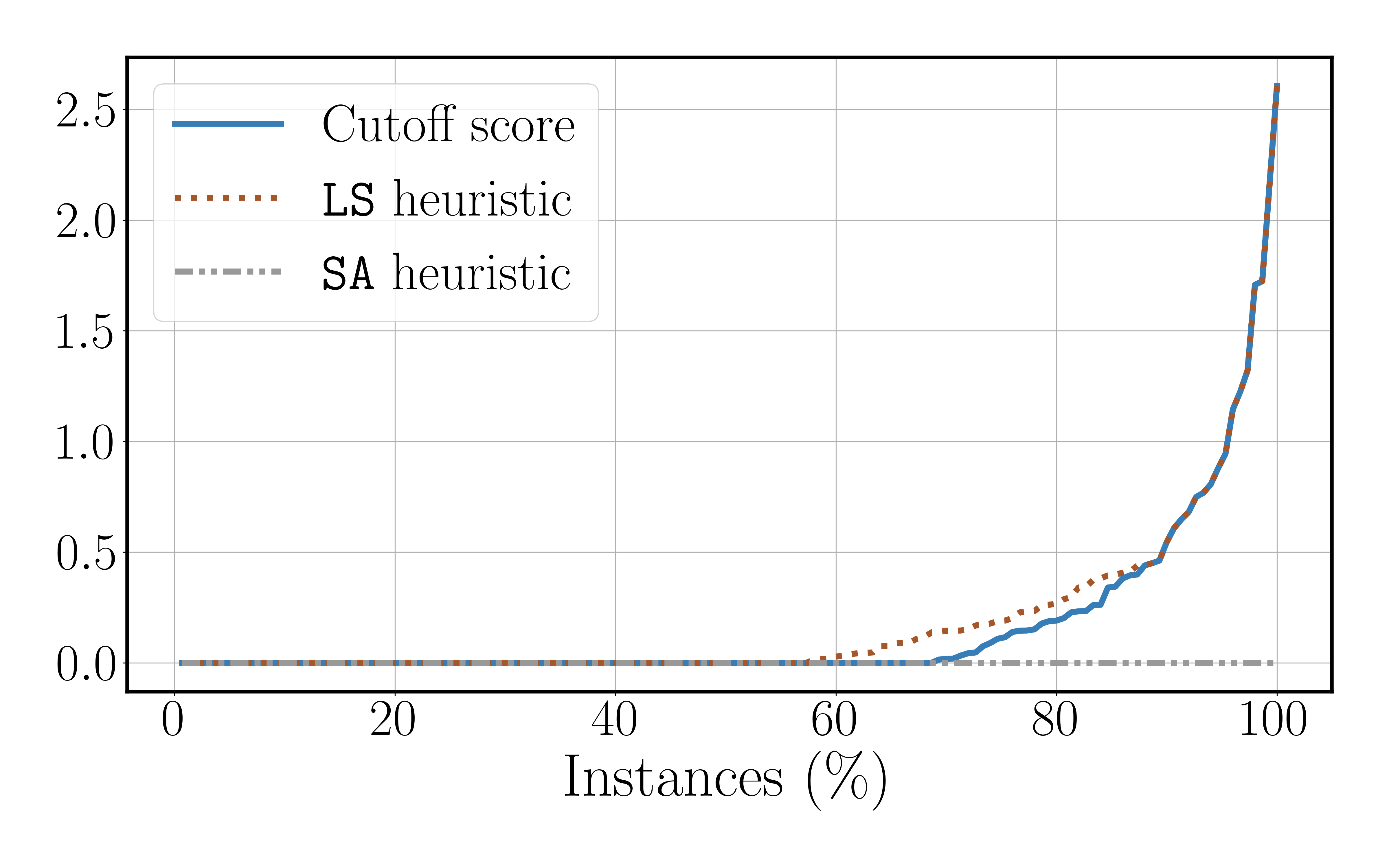}
        \caption{Upper bound gap [$\text{gap}_{ub^{*}}$] (\%)}
        \label{subfig:ubgap-ceum-small}
    \end{subfigure}
    \caption{Computational performance for small instances of approaches for {\SAA} Program~\eqref{prog:saa} under {\CEUM} behavior}
    \label{fig:perf-ceum-small}
\end{figure}
\begin{figure}[t]
    \begin{subfigure}[b]{0.32\linewidth}
        \includegraphics[width=\linewidth]{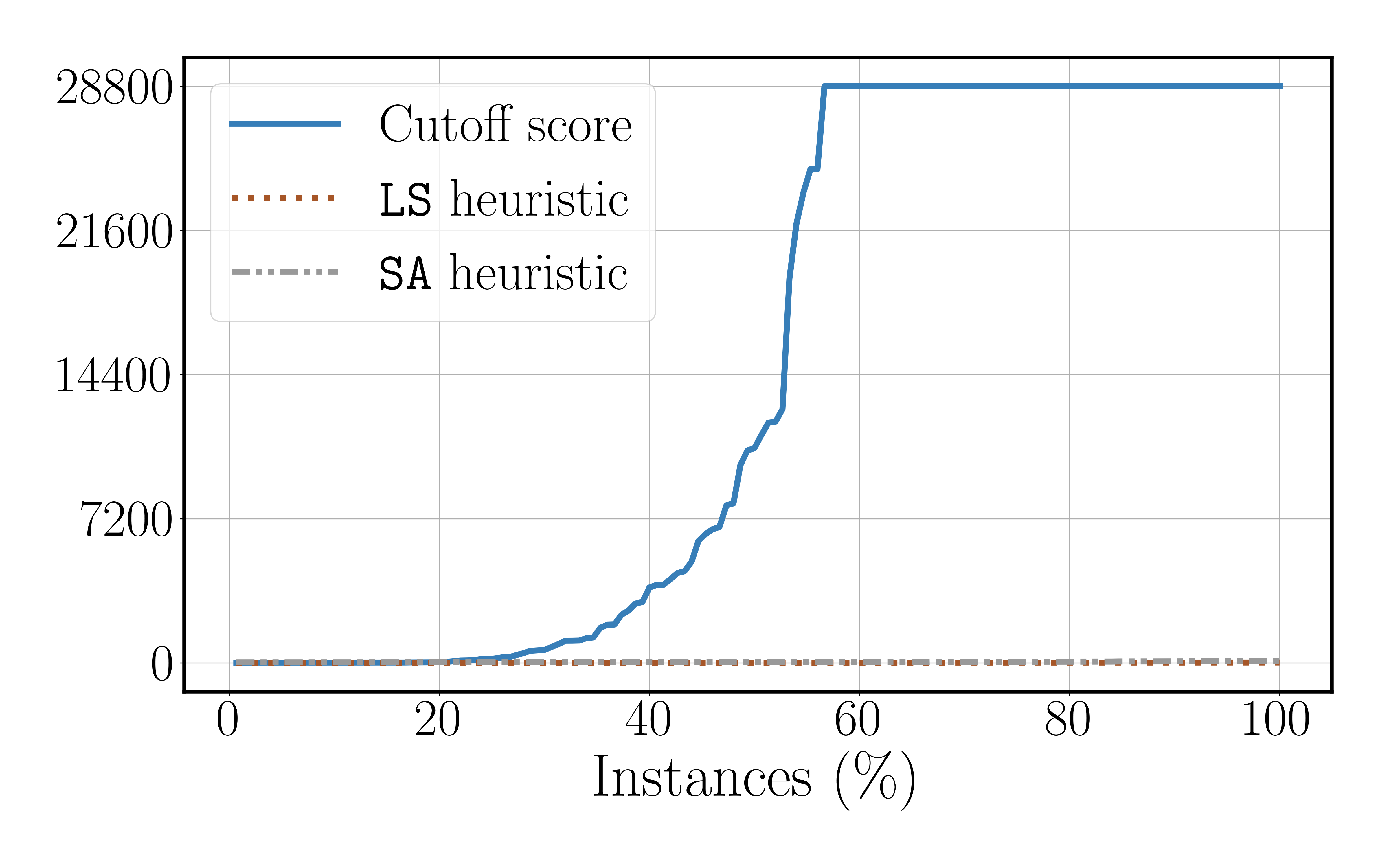}
        \caption{Time (seconds)}
        \label{subfig:time-ieum-small}
    \end{subfigure}
    \hfill
    \begin{subfigure}[b]{0.32\linewidth}
        \includegraphics[width=\linewidth]{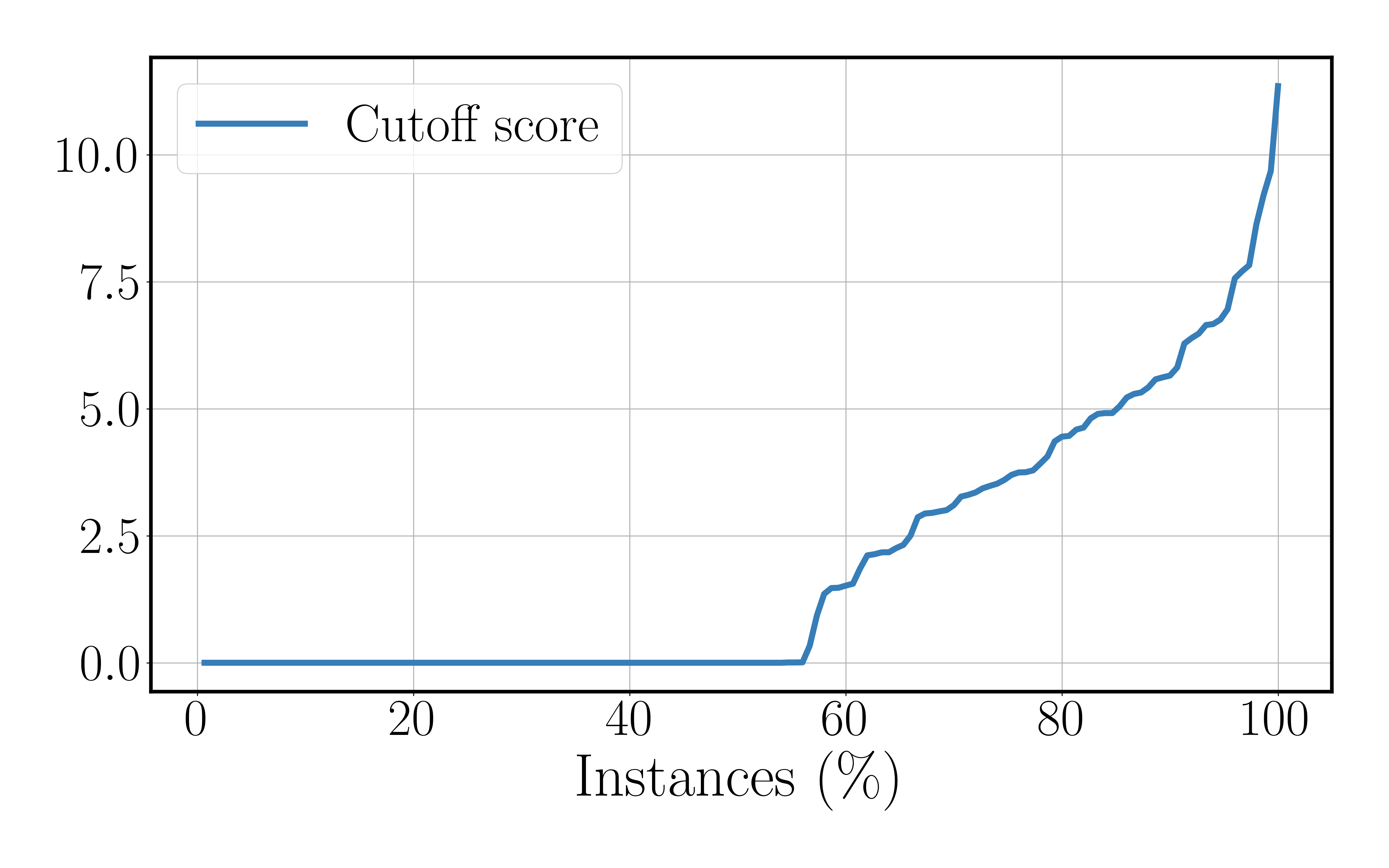}
        \caption{Optimality gap [$\text{gap}_{opt}$]  (\%)}
        \label{subfig:optgap-ieum-small}
    \end{subfigure}
    \hfill
    \begin{subfigure}[b]{0.32\linewidth}
        \includegraphics[width=\linewidth]{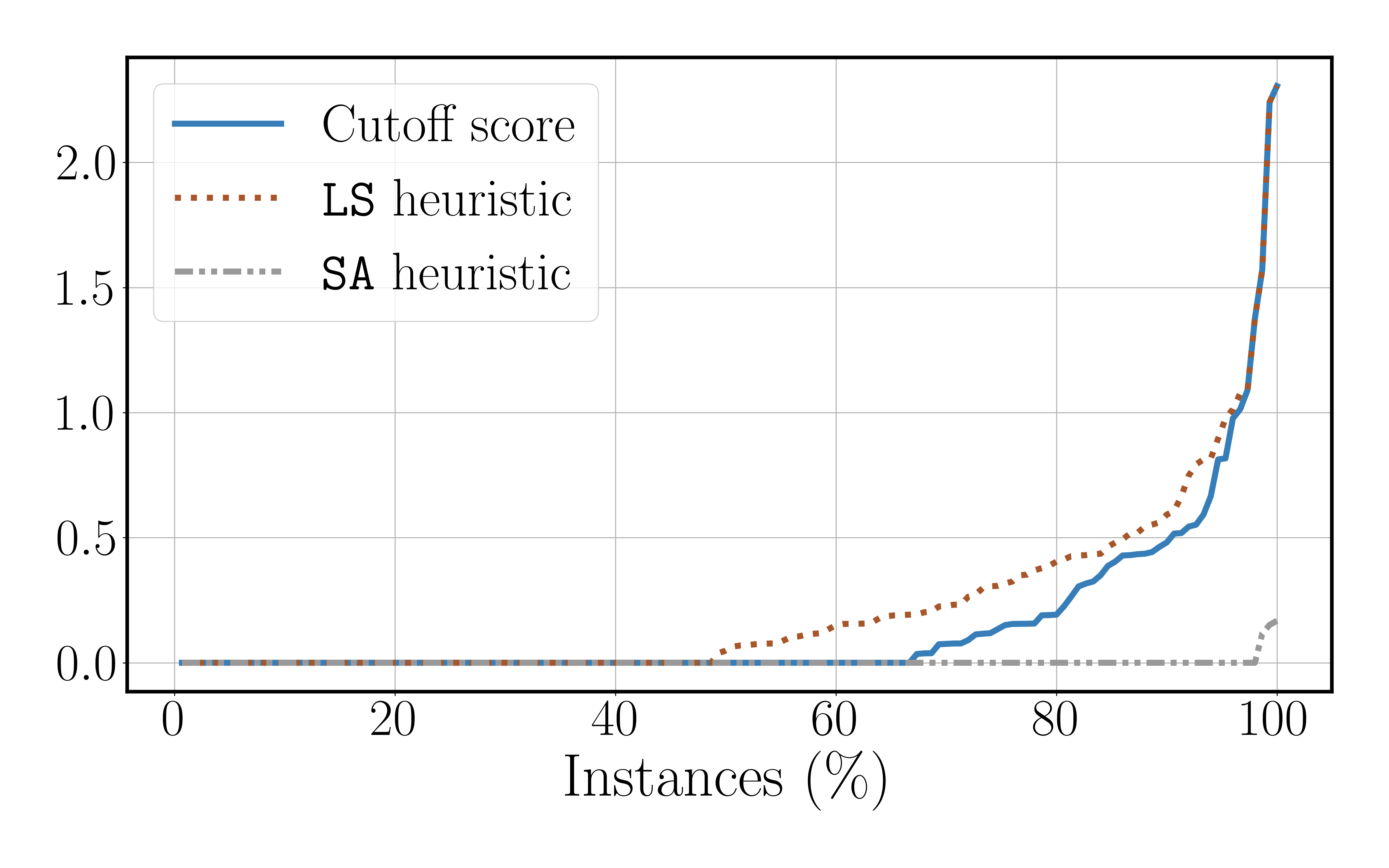}
        \caption{Upper bound gap [$\text{gap}_{ub^{*}}$] (\%)}
        \label{subfig:ubgap-ieum-small}
    \end{subfigure}
    \caption{Computational performance for small instances of approaches for {\SAA} Program~\eqref{prog:saa} under {\IEUM} behavior}
    \label{fig:perf-ieum-small}
\end{figure}
\begin{figure}[t]
    \centering
    \begin{subfigure}[b]{0.48\linewidth}
    \centering
        \includegraphics[width=0.65\linewidth]{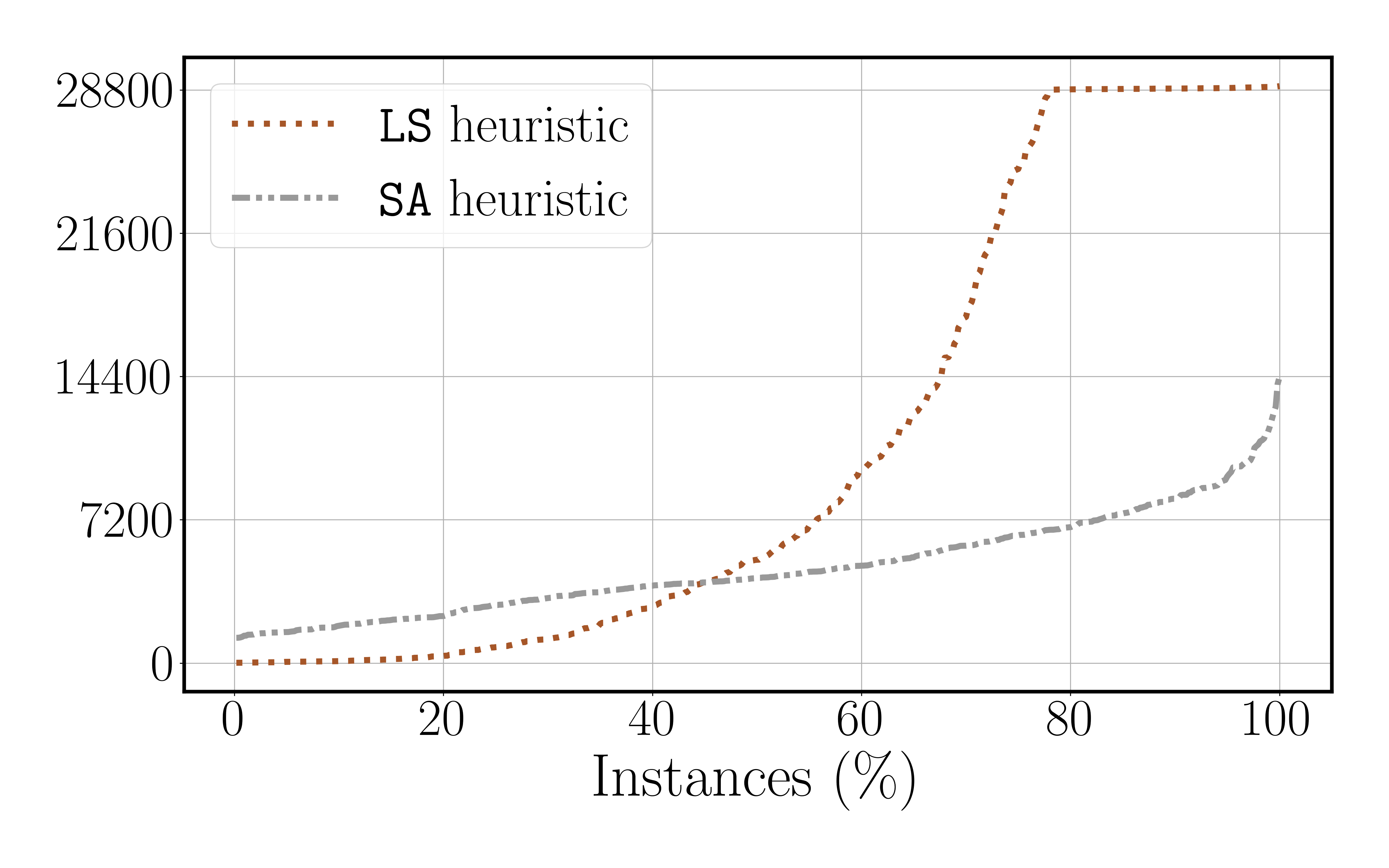}
        \caption{Time (seconds)}
        \label{subfig:time-ceum-large}
    \end{subfigure}
    \hfill
    \begin{subfigure}[b]{0.48\linewidth}
    \centering
        \includegraphics[width=0.65\linewidth]{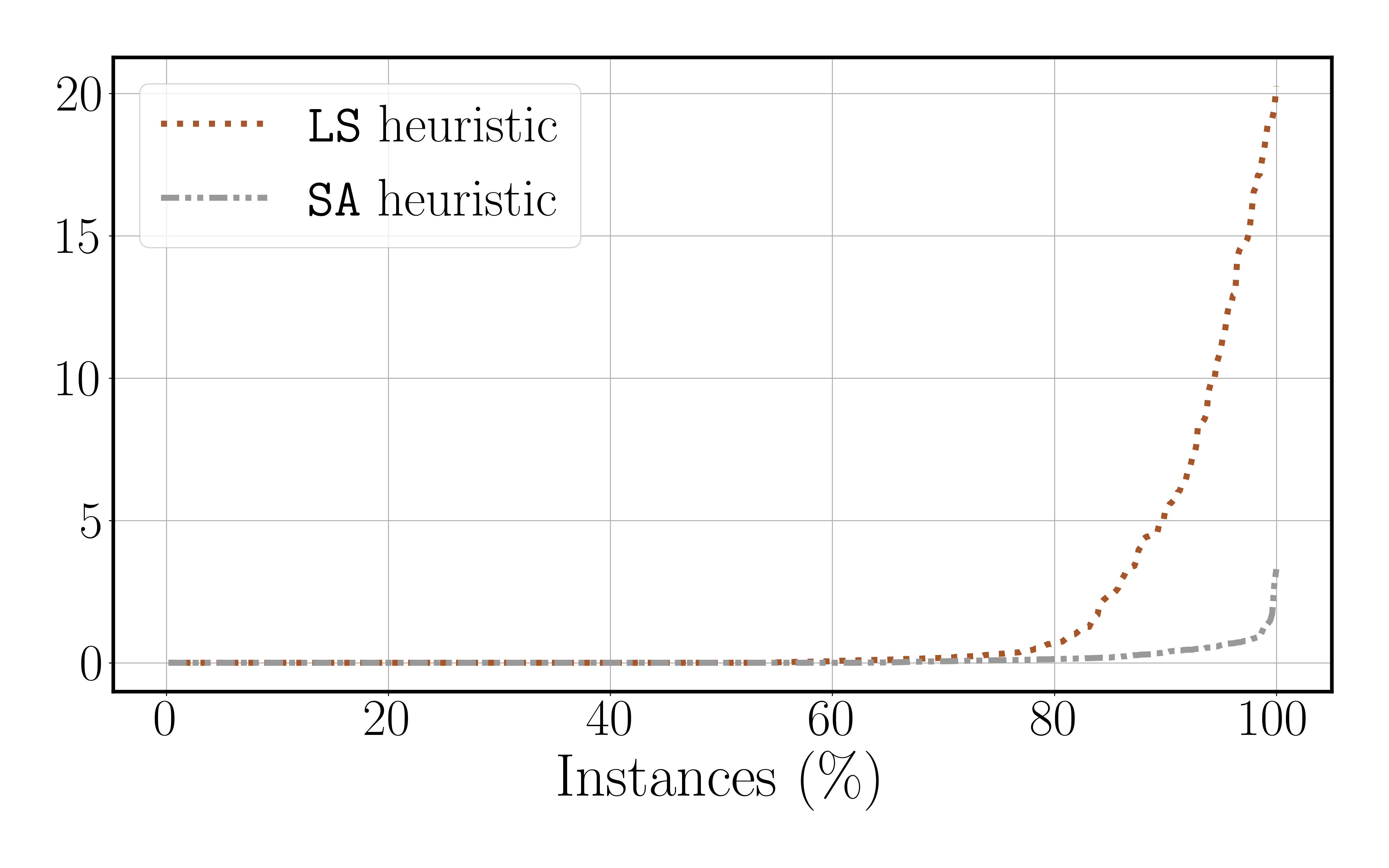}
        \caption{Upper bound gap [$\text{gap}_{ub^{*}}$] (\%)}
        \label{subfig:ubgap-ceum-large}
    \end{subfigure}
    \caption{Computational performance for large instances of approaches for {\SAA} Program~\eqref{prog:saa} under {\CEUM} behavior}
    \label{fig:perf-ceum-large}
\end{figure}
\begin{figure}[t]
    \centering
    \begin{subfigure}[b]{0.48\linewidth}
        \centering
        \includegraphics[width=0.65\linewidth]{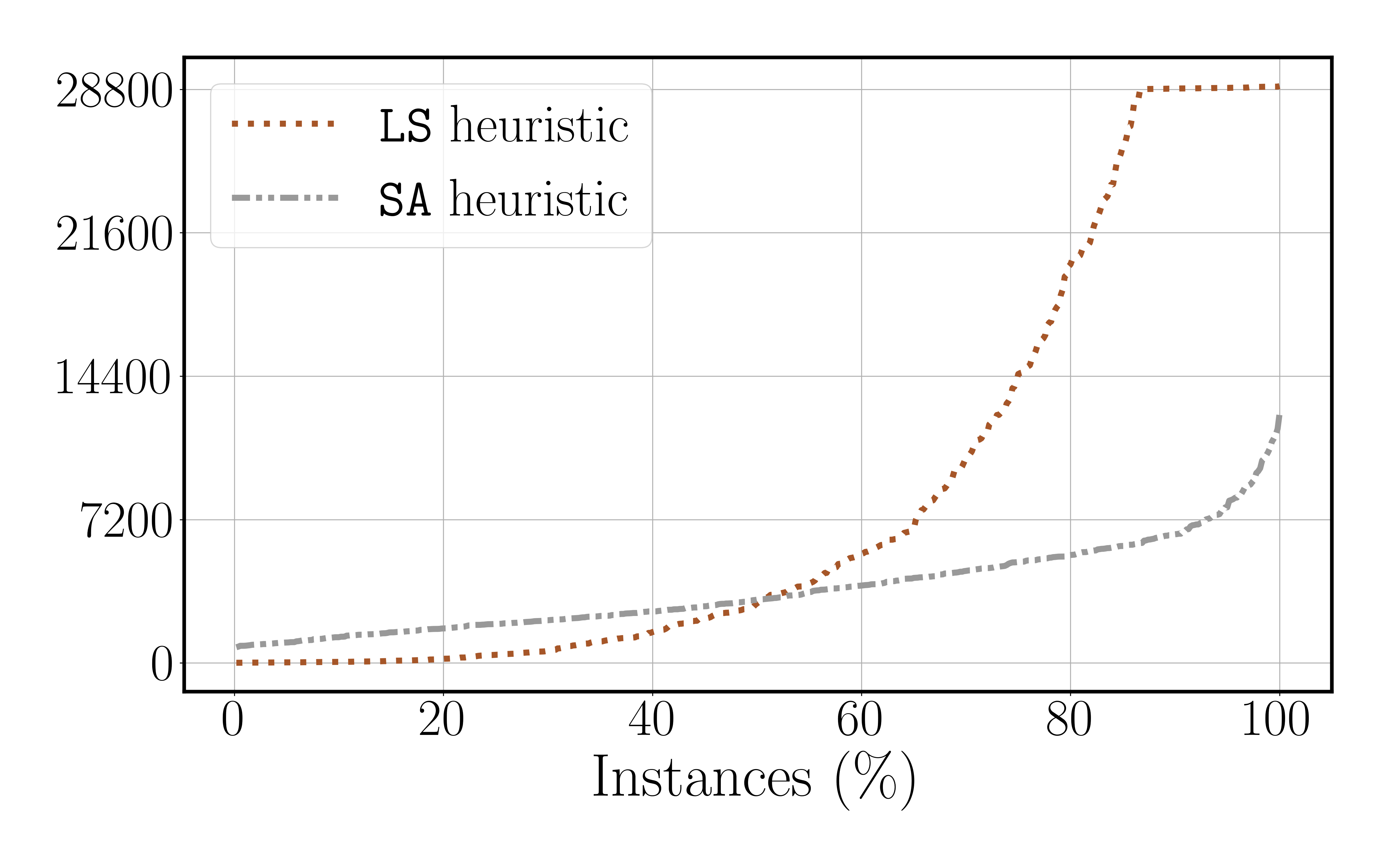}
        \caption{Time (seconds)}
        \label{subfig:time-ieum-large}
    \end{subfigure}
    \hfill
    \begin{subfigure}[b]{0.48\linewidth}
    \centering
        \includegraphics[width=0.65\linewidth]{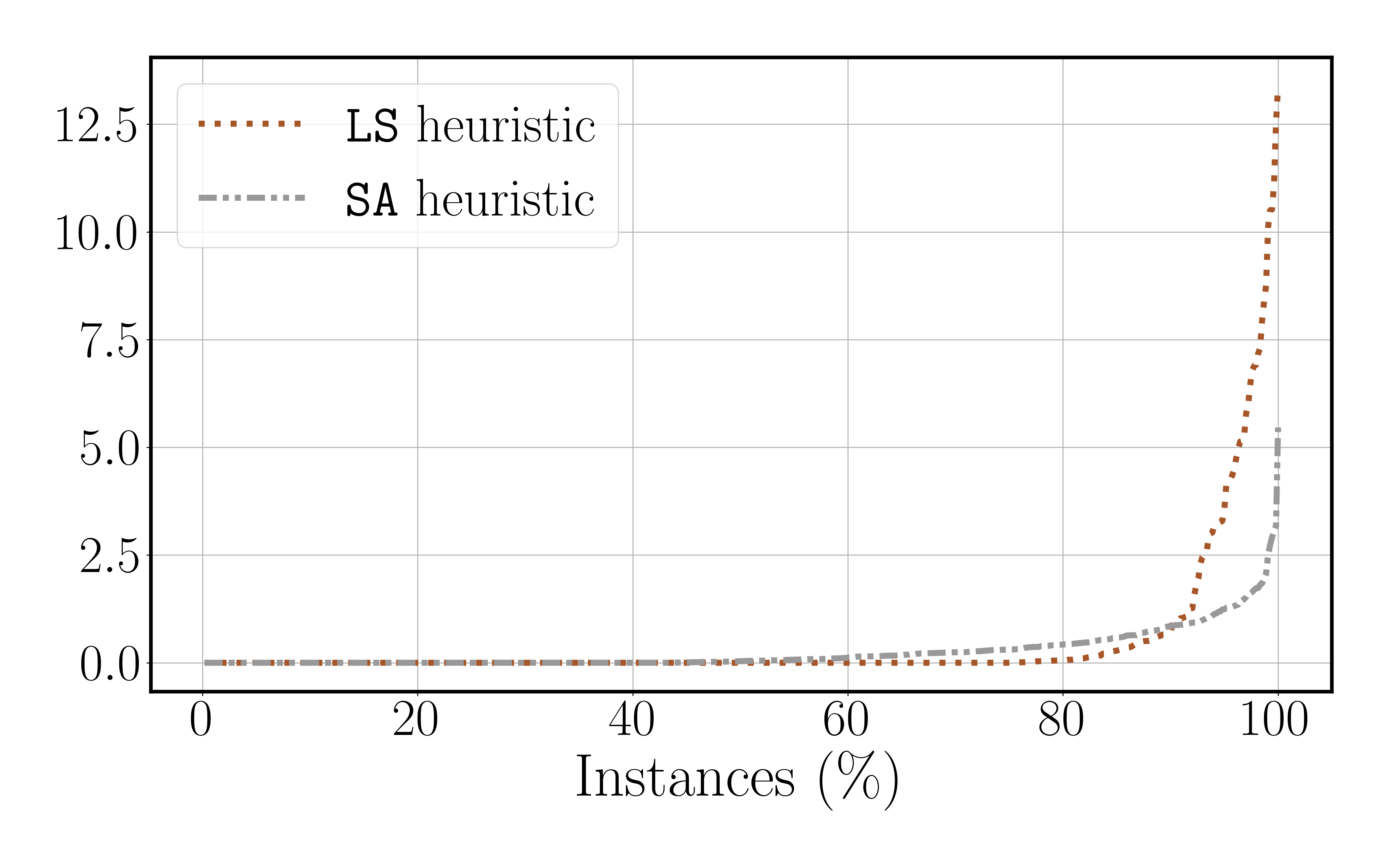}
        \caption{Upper bound gap [$\text{gap}_{ub^{*}}$] (\%)}
        \label{subfig:ubgap-ieum-large}
    \end{subfigure}
    \caption{Computational performance for large instances of approaches for {\SAA} Program~\eqref{prog:saa} under {\IEUM} behavior}
    \label{fig:perf-ieum-large}
\end{figure}

Figures~\ref{fig:perf-ceum-small} and~\ref{fig:perf-ieum-small} show that small instances of the {\SAA} Program~\eqref{prog:saa} under the strategic {\CEUM} and {\IEUM} behaviors are challenging to solve exactly, even within an eight-hour time limit. The exact approach solved approximately 48.7\% and 56\% of small instances under {\CEUM} and {\IEUM}, respectively, to optimality, achieved optimality gaps below 5\% for about 78\% and 85\% of the small instances, and below 10\% for 99\% of the small instances in both cases. The upper bound gaps were below 2.6\% and 2.3\%, respectively. These results illustrate the added complexity from endogenous preferences, as the same problem is considerably easier under {\UM}. Among the strategic settings, the problem under {\IEUM} is slightly easier to solve than under {\CEUM}.

This trend is expected: when students' reported preferences are endogenous, they must be explicitly defined for each scenario via functions that depend on capacity decisions and true preferences. These functions involve embedded optimization problems, Program~\eqref{eq:max-exp-utility} for {\CEUM} and Program~\eqref{eq:max-exp-utility-simple} for {\IEUM}, rendering the overall program bilevel. While single-level reformulations exist, they require numerous binary variables and loose constraints, resulting in weak linear relaxations and poor computational performance. Hence, even in contexts where capacity expansion is determined well in advance of the matching process, heuristic methods may be necessary to obtain high-quality feasible solutions, especially when a large scenario set $\mathcal{W}_{\mathcal{N}}$ is required or multiple {\SAA} Programs~\eqref{prog:saa} must be solved with distinct scenario sets. 

Both the {\SA} and {\LS} heuristics have extremely low runtimes for small instances (under 109 and 8 seconds, respectively). Although slightly slower, the {\SA} heuristic consistently produces higher-quality feasible solutions, with average upper bound gaps of 0.000\% ({\CEUM}) and 0.003\% ({\IEUM}), versus 0.189\% and 0.223\% of the {\LS} heuristic. The exact approach takes a significantly longer time to reach comparable or slightly worse solutions, notably when compared to the {\SA} heuristic. Hence, the {\SA} heuristic offers the best trade-off between runtime and solution quality, outperforming both the exact method and the {\LS} heuristic.

Figures~\ref{subfig:ubgap-ceum-large} and~\ref{subfig:ubgap-ieum-large} show that for large instances, the {\LS} heuristic yields higher upper bound gaps than the {\SA} heuristic, averaging 1.478\% ({\CEUM}) and 0.494\% ({\IEUM}). Its performance is budget-dependent and influenced by the size of the first-stage feasible region. Figures~\ref{subfig:time-ceum-large} and~\ref{subfig:time-ieum-large} reveal that its runtime reached the time limit in 22\% ({\CEUM}) and 13.5\% ({\IEUM}) of the large instances, mostly for the large budgets $B \in \{30, 45, 60\}$, for which average runtimes and upper bound gaps were 19,128s and 2.915\% ({\CEUM}) and 15,340.3s and 0.976\% ({\IEUM}). For the small budgets $B \in \{1, 5, 15\}$, runtimes and upper bound gaps dropped to 2,761s and 0.400\% ({\CEUM}) and 1,515s and 0.011\% ({\IEUM}). In contrast, the {\SA} heuristic maintained low upper bound gaps across all budgets (0.109\% for {\CEUM}, 0.266\% for {\IEUM}), and consistently exhibited lower, more stable runtimes across budget levels than the {\LS} heuristic. These trends reflect the influence of the {\LS} neighborhood on convergence and the role of hyperparameters in the {\SA} heuristic, as in the {\UM} case.

Both the {\LS} and {\SA} heuristics run faster under {\UM} behavior than under strategic {\IEUM} and {\CEUM} behaviors. On average, the {\SA} heuristic is 2.6 and 2 times slower, and the {\LS} heuristic 3.4 and 2.6 times slower for {\CEUM} and {\IEUM}, respectively, compared to {\UM}. This slowdown arises from the neighborhood evaluation: under {\UM}, the {\DAA} is used directly, whereas under {\IEUM} and {\CEUM}, reported preferences must be computed via the {\IOA} and {\MIA}, increasing computational cost.

\section{Conclusions} \label{sec:conclusions}

This work introduced and formalized the {\STMMP}, covering applications such as school choice and residency matching, where capacity expansions must be decided before students’ reported preferences are revealed. After reported preferences are observed, a student-optimal stable matching is implemented based on the expanded capacities. We considered truthful preference reporting ({\UM}) and two forms of strategic behavior ({\CEUM} and {\IEUM}). For each, we formulated the corresponding {\STMMP} and solved it using the {\SAA} method, supported by a toolkit of exact and heuristic approaches. Our results show that {\SAA} programs under strategic behaviors are substantially harder to solve to optimality than under truthful behavior due to endogenous preferences; nonetheless, our heuristics consistently produce feasible, high-quality solutions.
 
We showed that modeling stochasticity via the {\SAA} method consistently outperforms deterministic {\EV} solutions across all behaviors and metrics, including students' matching preferences, number of entering students, and those improving their assignments. Moreover, assuming incorrect behavioral models when making capacity decisions consistently leads to worse outcomes. These findings underscore the importance of both accounting for uncertainty and accurately modeling student behavior in capacity expansion planning.

Our work opens several research directions. Extensions could address alternative settings, including capacity reductions, tuition waiver allocations, secured enrollment mechanisms, lower and common quotas, couples matching, and paired applications, as well as alternative behaviors, such as reputation-based biases~\citep{kleinberg2024modeling}. Methodologically, promising directions include developing scenario clustering techniques to reduce scenario counts while preserving key stochastic features. Beyond school choice, the {\STMMP} framework applies to a wide range of applications where capacity must be allocated before preferences are revealed, such as refugee settlement expansion and medical resource allocation, contexts where uncertainty in preferences is a central challenge that our approach is well-suited to address.

\paragraph{\textbf{Acknowledgements}} This research was funded by Scale AI [the SCALE-AI Chair], the Fonds de recherche du Québec [the FRQ-IVADO Research Chair], IVADO [the FRQ-IVADO Research Chair], and the Natural Sciences and Engineering Research Council of Canada [Grant 2024-04051].


\bibliographystyle{apalike} 
\bibliography{myrefs}




%
%

\appendix

\section{Example for Students' Behaviors}\label{app:example-behaviors}

We illustrate the differences of the three considered behaviors using a simple instance for a student $s$ and scenario $w$. Let ${\mathcal{C} = \{c_1, \ldots, c_5\}}$, ${\mathcal{C}_{s,w} = \{c_1, \ldots, c_5\}}$, ${B = 1}$, and ${K = 3}$. The utilities are $u_{s,c_1}^w = 10$, $u_{s,c_2}^w = 8$, $u_{s,c_3}^w = 7$, $u_{s,c_4}^w = 6$, and $u_{s,c_5}^w = 5$, with true preference order $c_1 \succ_{s,w} c_2 \succ_{s,w} c_3 \succ_{s,w} c_4 \succ_{s,w} c_5$. The perceived admission chances are: $c_1$: $[0.3, 0.6]$, $c_2$: $[0.6, 0.8]$, $c_3$: $[0.5, 0.7]$, $c_4$: $[0.7, 0.9]$, and $c_5$: $[0.8, 0.9]$, where the first and second values represent $\overline{p}_{s,c}^0$ and $\overline{p}_{s,c}^1$, respectively. The first-stage decisions $\{x_c^1\}_{c \in \mathcal{C}}$ correspond to allocating one extra capacity to at most one school, with the feasible set ${\mathcal{X} = \{(0,0,0,0,0), (1,0,0,0,0), (0,1,0,0,0), (0,0,1,0,0), (0,0,0,1,0), (0,0,0,0,1)\}}$. The reported preferences under each behavior are:
\begin{itemize}
    \item \textbf{\UM} behavior: For all $x \in \mathcal{X}$, $\mathcal{C}^{\UM}_{s,w} = \{c_1,c_2,c_3\}$ and $c_1 \succ_{s,w}^{\UM} c_2 \succ_{s,w}^{\UM} c_3$.
    
    \item \textbf{\IEUM} behavior:
    \[
    \begin{aligned}
    x &\in \{(0,0,0,0,0),(0,1,0,0,0),(0,0,0,1,0),(0,0,0,0,1)\}: & \mathcal{C}_{s,w}^{\IEUM} &= \{c_2,c_4,c_5\},\ c_2 \succ_{s,w}^{\IEUM} c_4 \succ^{\IEUM}_{s,w} c_5,\\
    x &= (1,0,0,0,0): & \mathcal{C}_{s,w}^{\IEUM} &= \{c_1,c_2,c_4\},\ c_1 \succ^{\IEUM}_{s,w} c_2 \succ^{\IEUM}_{s,w} c_4,\\
    x &= (0,0,1,0,0): & \mathcal{C}_{s,w}^{\IEUM} &= \{c_2,c_3,c_4\},\ c_2 \succ^{\IEUM}_{s,w} c_3 \succ^{\IEUM}_{s,w} c_4.
    \end{aligned}
    \]
    For the feasible solution $(0,0,0,0,0)$, the individual expected utilities are 
    \[
    \left(p_{s,c_1}^w(x_{c_1}) u_{s,c_1}^w,\, p_{s,c_2}^w(x_{c_2}) u_{s,c_2}^w,\, p_{s,c_3}^w(x_{c_3}) u_{s,c_3}^w,\, p_{s,c_4}^w(x_{c_4}) u_{s,c_4}^w,\, p_{s,c_5}^w(x_{c_5}) u_{s,c_5}^w\right) = (3.0,\, 4.8,\, 3.5,\, 4.2,\, 4.0),
    \]
    leading to the selection of $\mathcal{C}_{s,w}^{\IEUM} = \{c_2, c_4, c_5\}$. When one extra capacity is allocated to $c_1$, i.e., for $(1,0,0,0,0)$, its admission chance increases to $0.6$, with an expected utility of $0.6 \times 10 = 6.0$. Consequently, the selected schools change to $\mathcal{C}_{s,w}^{\IEUM} = \{c_1, c_2, c_4\}$. For the feasible solution $(0,0,1,0,0)$, the admission chance of $c_3$ increases to $0.9$, resulting in an expected utility of $0.9 \times 7 = 6.3$. Hence, the selected schools become $\{c_2, c_3, c_4\}$.
    \item \textbf{\CEUM} behavior:
    \[
    \begin{aligned}
    x &\in \{(0,0,0,0,0),(1,0,0,0,0),(0,1,0,0,0),(0,0,0,1,0)\}: & \mathcal{C}_{s,w}^{\CEUM} &= \{c_1,c_2,c_4\},\ c_1 \succ_{s,w}^{\CEUM} c_2 \succ_{s,w}^{\CEUM} c_4.\\
    x &= (0,0,1,0,0): & \mathcal{C}_{s,w}^{\CEUM} &= \{c_1,c_2,c_3\},\ c_1 \succ_{s,w}^{\CEUM} c_2 \succ_{s,w}^{\CEUM} c_3\\
    x &= (0,0,0,0,1): & \mathcal{C}_{s,w}^{\CEUM} &= \{c_1,c_2,c_5\},\ c_1 \succ_{s,w}^{\CEUM} c_2 \succ_{s,w}^{\CEUM} c_5.
    \end{aligned}
    \]
    For the feasible solution $(0,0,0,0,0)$, we apply the {\MIA}. We first select school $c_2$, which has the highest expected utility, $p_{s,c_2}^w(x_{c_2})u_{s,c_2}^w = 0.6 \times 8 = 4.8$. In the next step, the largest marginal increase is obtained by selecting school $c_4$, with \({\MI}\left((0,0,0,0,0),\{c_2\},c_4\right) = (0.6\times8 + 0.4\times0.7\times6) - (0.6\times8)\). In the third iteration, school $c_1$ is chosen based on its marginal improvement: \({\MI} \left( (0,0,0,0,0),\{c_2,c_4\},c_1 \right) = (0.3\times10 + 0.7\times0.6\times8 + 0.7\times0.4\times0.7\times6) - (0.6\times8 + 0.4\times0.7\times6)\). For the feasible solution $(0,0,1,0,0)$, the process begins with school $c_3$, whose expected utility increases to $0.7\times7 = 4.9$. Next, school $c_2$ is selected, with \({\MI}\left((0,0,1,0,0),\{c_3\},c_2\right) = (0.6\times8 + 0.4\times0.7\times7) - (0.7\times7)\). Finally, school $c_1$ is chosen as it provides the largest marginal increase: \({\MI}\left((0,0,1,0,0),\{c_2,c_3\},c_1\right) = (0.3\times10 + 0.7\times0.6\times8 + 0.7\times0.4\times0.7\times7) - (0.6\times8 + 0.4\times0.7\times7)\).
\end{itemize}
Under {\UM} behavior, reported preferences are equal across all feasible decisions since they are exogenous. In contrast, under {\IEUM} and {\CEUM} behaviors, reported preferences depend on first-stage decisions through their effect on perceived admission chances. Notably, {\UM} and {\CEUM} coincide only for $(0,0,1,0,0)$, while {\IEUM} and {\CEUM} coincide only for $(1,0,0,0,0)$, showing that the three behaviors yield distinct reported preferences.

\section{Deferred Acceptance Algorithm}\label{app:da}

We present the Deferred Acceptance algorithm {\DAA}~\citep{gale1962college} in Algorithm~\ref{alg:da}. The algorithm operates for a fixed capacity decision $x \in \mathcal{X}$, behavior $a \in \mathcal{A}$, scenario $w \in \mathcal{W}$, and instance $ \Gamma_{x,w}^a = \langle \mathcal{S}, \mathcal{C}, \succ_{\mathcal{C}}, \succ^{a}_{\mathcal{S},w}, \boldsymbol{\bar{q}} \rangle $, where $ \succ^{a}_{\mathcal{S},w}$ is also fixed according to behavior $a$.
\begin{algorithm}[ht]
    \caption{Deferred Acceptance Algorithm ({\DAA})}\label{alg:da}  
    \small \Input{$x \in \mathcal{X}$, $w \in \mathcal{W}$, $a \in \mathcal{A}$ ${\Gamma_{x,w}^a = \langle \mathcal{S}, \mathcal{C}, \succ_{\mathcal{C}}, \succ^{a}_{\mathcal{S},w}, \boldsymbol{\bar{q}} \rangle}$ ;}\\
    \Output{Student-optimal stable matching;} \\
    \Stepone{Each student applies to their most preferred school. Schools temporarily accept their most preferred students up to their expanded capacity and reject the rest;} \\
    \Steptwo{Each rejected student $s$ proposes to their most preferred school to which they have not applied yet. If the student has proposed to all acceptable schools, they do not apply (are unassigned);}\\
    \Stepthree{If the school's expanded capacity is not yet met, the student’s application is temporarily accepted. If the school prefers the new student $s$ over a temporarily accepted student $s'$, the school temporarily accepts $s$ and rejects $s'$. Conversely, if the school prefers all temporarily accepted students over $s$, then $s$ is rejected.}\\
    \Stepfour{If all students are either assigned to a school or have no remaining schools to apply to, return the current matching. Otherwise, return to Step 2.}\\
\end{algorithm}

\section{Heuristic Algorithms}\label{app:heur-algo}

We present the detailed heuristic algorithms: Assignment-based ({\ASG}) (Algorithm~\ref{alg:asg-heuristic}), Local Search ({\LS}) (Algorithm~\ref{alg:ls-heuristic}), and Simulated Annealing ({\SA}) (Algorithm~\ref{alg:sa-heuristic}).
\begin{algorithm}[ht]
    \caption{{\ASG} Heuristic for Step~1 of the {\SAA} Method under {\UM} Behavior}\label{alg:asg-heuristic}  
    \small \Input{$\mathcal{S},\mathcal{C},\mathcal{B},\mathcal{K},\succ_{\mathcal{C}},\mathcal{W_N}$;}\\
    \Output{$\overline{x}_N^{\UM},\overline{v}_N^{\UM}$;} \\
    \Stepone{Solve the {\ASG} {\SAA} Program~\eqref{prog:asg-saa} with the set $\mathcal{W_N}$ of scenarios to obtain a feasible solution $\overline{x}_N^{\UM} \in \mathcal{X}$;} \\
    \Steptwo{Compute the objective value $\overline{v}_N^{\UM} = \hat{g}^{\UM}_N(\overline{x}_N^{\UM})$ at {\SAA} Program~\eqref{prog:saa} with scenario set $\mathcal{W_N}$;}
\end{algorithm}
\begin{algorithm}[ht]
    \caption{{\LS} Heuristic for Step~1 of the {\SAA} Method}\label{alg:ls-heuristic}  
    \small \Input{$a\in \mathcal{A},\mathcal{S},\mathcal{C},\mathcal{B},\mathcal{K},\succ_{\mathcal{C}},\boldsymbol{q},\boldsymbol{\overline{p}},\mathcal{W_N}$;}\\
    \Output{$\overline{x}_N^a,\overline{v}_N^a$;} \\
    \Stepone{Initialize $\overline{x}_N^a$ as a feasible solution in $\mathcal{X}$ and set $\overline{v}_N^a = \hat{g}_N^a(\overline{x}_N^a)$;}\\
    \Steptwo{Set $\hat{x} \in \arg\min_{x \in \text{NG}_{\LS}(\overline{x}_N^a)}\left\{ \hat{g}_N^a(x) \right\}$;} \\
    \Stepthree{If $\hat{g}_N^a(\hat{x}) < \overline{v}_N^a$, then set $\overline{x}_N^a = \hat{x}$ and $ \overline{v}_N^a = \hat{g}_N^a(\hat{x})$ and return to Step~2, otherwise stop.}
\end{algorithm}
\begin{algorithm}[ht]
\caption{{\SA} Heuristic for Step~1 of the {\SAA} Method}\label{alg:sa-heuristic}
\small
\Input{
$a \in \mathcal{A}, \mathcal{S}, \mathcal{C}, \mathcal{B}, \mathcal{K}, \succ_{\mathcal{C}}, \boldsymbol{q}, \boldsymbol{\overline{p}}, \mathcal{W_N}$, $T_0$, $T_{\min}$, $\zeta$, $I_{\max}$, $I_{T}$;}\\
\Output{$\overline{x}_N^a,\overline{v}_N^a$;}\\
\Stepone{Set $i = 1$, $T =T_0$, and $\overline{x}_N^a$ and $\widetilde{x}_N^a$ as a feasible solution in $\mathcal{X}$, and set $\overline{v}_N^a = \widetilde{v}_N^a = \hat{g}_N^a(x)$;}\\
\Steptwo{Randomly select $\hat{x} \in \text{NG}(\widetilde{x}_N^a)$ and compute $\hat{v} = \hat{g}_N^a(\hat{x})$;}\\
\Stepthree{Sample $u \sim U(0,1)$. If $\hat{v} < \widetilde{v}_N^a$ or $u \leq \exp\left( -(\hat{v} - \widetilde{v}_N^a)/T \right)$, then set $\widetilde{x}_N^a = \hat{x}$ and $\widetilde{v}_N^a = \hat{v}$;}\\
\Stepfour{If $\hat{v} < \overline{v}_N^a$, then set $\overline{x}_N^a = \hat{x}$ and $\overline{v}_N^a = \hat{v}$;}\\
\Stepfive{Set $i = i +i$. If $i$ is a multiple of $I_T$, then set $T = \zeta T$. If $T \geq T_{min}$, return to Step~2; otherwise, stop.}\\
\end{algorithm}

\section{Linear Reformulation of the Mathematical Model Under the {\CEUM} Behavior} \label{app:lin-ceum}

We detail the linearization of variables $\phi_{s,c}^{w,k}$ and $\upsilon_{s,c}^{w,k}$ and Constraints~\eqref{constr:correct-select-schools} when modeling the {\CEUM} behavior in Section~\ref{subsec:strategic-behavior}. We introduce the variables $h_{s,c,j}^{w,k} \geq 0$ for each $c \in \mathcal{C}_{s,w}$, $j \in \mathcal{B}^0$, and $k \in \mathcal{K}$, with ${h_{s}^w = \{h_{s,c,j}^{w,k}\}_{c \in \mathcal{C}_{s,w}, k \in \mathcal{K}, j \in \mathcal{B}^0}}$, representing the nonlinear term ${\left[ \overline{\xi}_{s,c}^{w,k} \cdot \left(x_c^j - x_c^{j+1}\right) \cdot \prod_{c' \succ_{s,w} c} \left(1 - p_{s,c'}^w(x_{c'}) \overline{\xi}_{s,c'}^{w,k}\right)\right]}$. We also define the variables $\overline{h}_{s,c}^{w,k} \geq 0$ for each $c \in \mathcal{C}_{s,w}$ and $k \in \mathcal{K}$ with $\overline{h}_s^w = \{\overline{h}_{s,c}^{w,k}\}_{c \in \mathcal{C}_{s,w}, k \in \mathcal{K}}$ to capture the nonlinear term ${\left[\left(1 - \overline{\xi}_{s,c}^{w,k}\right) \prod_{c' \succ_{s,w} c} \left(1 - p_{s,c'}^w(x_{c'}) \overline{\xi}_{s,c'}^{w,k}\right)\right]}$. The following linear constraints, based on the probability chain technique by~\citet{o2013probability}, model these variables for all $k \in \mathcal{K}$ and $c \in \mathcal{C}_{s,w}$:
\begin{align}\label{constr:recurrence-h}&\overline{h}_{s,c}^{w,k} + \sum_{j \in \mathcal{B}^0} h_{s,c,j}^{w,k} = \overline{h}_{s,c'}^{w,k} + \sum_{j \in \mathcal{B}^0} \left( 1 - \overline{p}_{s,c'}^j \right) h_{s,c',j}^{w,k},   \\ 
    \label{constr:bound-q-xi}& \sum_{j \in \mathcal{B}^0} h_{s,c,j}^{w,k} \leq \overline{\xi}_{s,c}^{w,k}, \\
    \label{constr:bound-barq-xi}& \overline{h}_{s,c}^{w,k} \leq  1 - \overline{\xi}_{s,c}^{w,k}, \\
    \label{constr:bound-barq-x}& h_{s,c,j}^{w,k} \leq x_c^j -x_c^{j+1}, \, \forall j \in \mathcal{B}^0.
\end{align}
In Constraint~\eqref{constr:recurrence-h}, $c' = \delta_{s}^w(c) $ such that $ \delta_{s}^w(c) \neq 0$, where $\delta_{s}^w(c)$ denotes a function that identifies the school $c' \in \mathcal{C}_{s,w}$ immediately preceding school $c$ in the true preference order $\succ_{s,w}$. If $c$ is the most preferred school in $\succ_{s,w}$, then $\delta_{s}^w(c)$ returns zero. Constraints~\eqref{constr:recurrence-h} define the values of the variables $h_{s,c,j}^{w,k}$ and $\overline{h}_{s,c}^{w,k}$ based on student $s$'s belief about not being accepted by any school $c'$ \emph{truly} preferred over school $c$ ($c' \succ_{s,w} c$) that was selected by iteration $k \in \mathcal{K}$ of the {\MIA} to form the set $\mathcal{C}_{s,w}^{\CEUM}$. Constraints~\eqref{constr:bound-q-xi} ensure that the value of $q_{w,s,c}^{j,k}$ is zero if school $c \in \mathcal{C}_{s,w}$ was not selected by iteration $k$ of the {\MIA}. Similarly, Constraints~\eqref{constr:bound-barq-xi} guarantee that $\overline{h}_{s,c}^{w,k}$ equals zero if school $c \in \mathcal{C}_{s,w}$ was already selected by iteration $k$ of the {\MIA}. Constraints~\eqref{constr:bound-barq-x} enforce that $h_{s,c,j}^{w,k}$ equals zero if the number of additional seats allocated to school $c$ is not equal to $j \in \mathcal{B}^0$. We have $\overline{h}_{s,c}^{w,k} = 1 - \overline{\xi}_{s,c}^{w,k}$ and $\sum_{j \in \mathcal{B}} h_{s,c,j}^{w,k} = \overline{\xi}_{s,c}^{w,k}$ when school $c$ is the most preferred, i.e., when $\delta_s^w(c) = 0$. As a result, we can define variables $\phi_{s,c}^{w,k}$ and $ \upsilon_{s,c}^{w,k}$ by the following linear constraints:
\begin{align}\label{constr:def-phi-upsilon}
     & \phi_{s,c}^{w,k} =  \overline{h}_{s,c}^{w,k} + \sum_{j \in \mathcal{B}^0} h_{s,c,j}^{w,k}, \, \forall k \in \mathcal{K}, c \in \mathcal{C}_{s,w} \\ 
     & \upsilon_{s,c}^{w,k} = \sum_{j \in \mathcal{B}^0} \overline{p}_{s,c}^j u_{s,c}^w h_{s,c,j}^{w,k}, \, \forall k \in \mathcal{K}, c \in \mathcal{C}_{s,w}.
\end{align}
Finally, to linearize Constraints~\eqref{constr:correct-select-schools}, we leverage the definition of function $p_{s,c}(x_c)$, which consists solely of binary variables, and derive the following linear constraints:
\begin{align}
    \label{constr:correct-select-schools-lin}&
    \resizebox{0.5\linewidth}{!}{$
    \displaystyle
    L_{s,w}^k - L_{s,w}^{k-1} \geq \overline{p}_{s,c}^j \left[ u_{s,c}^w \left( \phi_{s,c}^{w,k-1} + x_c^j -1 \right)  - \sum_{c' \preceq_{s,w} c}  \upsilon_{s,c'}^{w,k-1} \right]
    $},
\end{align}
for every $k \in \mathcal{K} , c \in \mathcal{C}_{s,w},$ and $j \in \mathcal{B}^0$. Program~\eqref{eq:max-exp-utility} is modeled by the set $\Xi^{s,w}_{{\CEUM}} = \big\{ \xi^w_s \in \{0,1\}^{\mathcal{C}_{s,w}} : \eqref{constr:select-school-it}-\eqref{constr:def-var-L},\eqref{constr:safety-strategy},\eqref{constr:correct-select-schools-lin}, \eqref{constr:recurrence-h}-\eqref{constr:bound-barq-x}, \eqref{constr:def-phi-upsilon}, \, \xi^{\CEUM}_{s,c,w} = \overline{\xi}_{s,c}^{w,K} \big\}$, composed only by linear constraints.

\section{Modified $L$-constraints Formulation for {\CEUM} and {\IEUM} Behaviors} \label{app:mod-lconstr-strategic}

The modified $L$-constraints formulation of Program~\eqref{prog:stmmp-2st} for $w \in \mathcal{W_N}$ and $a \in \{\CEUM,\IEUM\}$ is:
\begin{subequations}\label{Prog:mod-lconstr-second-stage}
    \begin{align}
        \min\mbox{:} &\eqref{constr:objective} \nonumber \\
        \mbox{s.t.:} & \eqref{constr:assign},\eqref{constr:capa},\eqref{constr:respect-report}-\eqref{constr:domain-y}\nonumber\\
        \label{constr:mod-lconstr-stab} &
        \sum_{c' \succeq_{s,w} c } y_{s,c'}^w + \left( 1 - \xi_{s,c,w}^{a} \right) \geq y_{s',c}^w, && \forall (s,c) \in \mathcal{S} \times \mathcal{C}_{s,w}^a, s' \prec_c s,
    \end{align}
\end{subequations}
where $d_{s,c}^w = \min(r_{s,c,w},K)$ and $d_{s,\varnothing}^w = \min(|\mathcal{C}_{s,w}|,K)+1$. The modified $L$-Constraints~\eqref{constr:mod-lconstr-stab} ensure stability in the matching: if student $ s $ includes school $ c $ in their reported preference list (i.e., $ \xi_{s,c,w}^{a} =1 $), then any student $ s' $ who is less preferred by school $ c $ (i.e., $ s' \prec_c s $) can only be assigned to $ c $ if $ s $ is either assigned to $ c $ or to a school $ c' $ that they prefer over $ c $ ($ c' \succ_{s,w} c $). These constraints have a loose continuous relaxation: variables \(\xi_{s,c,w}^{a} = 1\), which indicate students' reported schools, determine whether the constraint is active.

\section{Cutoff Score Formulation for {\UM} Behavior}\label{app:cutoff-nonstrategic}

The cutoff score formulation of Program~\eqref{prog:stmmp-2st} for scenario $w \in \mathcal{W_N}$ and {\UM} behavior is defined as:
\begin{subequations} \label{Prog:cutoff-second-stage-exo}
    \begin{align}
        \min \mbox{: } & \eqref{constr:exo-objective} && \nonumber \\
        \mbox{s.t.: } & \eqref{constr:exo-assign},\eqref{constr:exo-capa},\eqref{constr:exo-domain-y}, \eqref{constr:cutoff-waste2},\eqref{constr:domain-z} \nonumber \\
        \label{constr:exo-cutoff-score1} & z_c^w \leq \big( 1 - y_{s,c}^w \big) \big( n+1 \big) + e_{s,c}, && \forall (s,c) \in \mathcal{S} \times \mathcal{C}_{s,w}^{\UM}\\
        \label{constr:exo-cutoff-score2} &
        e_{s,c} + \epsilon  \leq z_c^w +  \big( n+1 \big)\sum_{c' \succeq_{s,w} c} y_{s,c'}^w, && \forall (s,c) \in \mathcal{S} \times \mathcal{C}_{s,w}^{\UM} \\ 
        \label{constr:exo-cutoff-waste1} & \sum_{j \in \mathcal{B}} x_c^j + B - \hat{q}_c f_c^w \leq \sum_{s \in \mathcal{S}: c \in  \mathcal{C}_{s,w}^{\UM}} y_{s,c}^w, && \forall c \in \mathcal{C}.
    \end{align}
\end{subequations}
Constraints~\eqref{constr:exo-cutoff-score1},~\eqref{constr:exo-cutoff-score2}, and~\eqref{constr:exo-cutoff-waste1} are the direct adaptation of Constraints~\eqref{constr:cutoff-score1},~\eqref{constr:cutoff-score2}, and~\eqref{constr:cutoff-waste1} considering only the precomputed set $\mathcal{C}_{s,w}^{\UM}$ of possible schools, respectively.

\section{Preprocessing}\label{app:preprocess}

We outline the preprocessing for students' reported preferences. Under {\UM} behavior, both $\mathcal{C}_{s,w}^{\UM}$ and $\succ_{s,w}^{\UM}$ are exogenous and thus precomputable. For {\CEUM} and {\IEUM} behaviors, the reported preference set $\mathcal{C}_{s,w}^{a}$ and order $\succ^{a}_{s,w}$ can be precomputed \emph{a priori} when $|\mathcal{C}_{s,w}| \leq K$, in which case they coincide with the true preferences $\mathcal{C}_{s,w}$ and $\succ_{s,w}$ regardless of the first-stage decision $x$. For the {\CEUM} behavior, we also preprocess students' reported preference orders $\succ^{{\CEUM}}_{s,w}$ according to Lemmas~\ref{lemma:preprocess-complex-probability} and~\ref{lemma:preprocess-complex-utility}, which we prove in~\ref{app:proofs}. For the {\IEUM} behavior, our preprocessing is based on Lemma~\ref{lemma:preprocess-simple}, which we also prove in~\ref{app:proofs}.
\begin{lemma}\label{lemma:preprocess-complex-probability}
    Assume ${\CEUM}$ behavior. For any student $s \in \mathcal{S}$ and scenario $w \in \mathcal{W}$, if $\overline{p}_{s,c}^0 = 1$ for some $(s,c) \in \mathcal{S} \times \mathcal{C}_{s,w}$, then under the ${\MIA}$ in Algorithm~\ref{alg:mia}, the following conditions hold:
    (i) If $r_{s,c,w} > K$, then schools $c' \prec_{s,w} c$ are not selected ($c' \notin \mathcal{C}_{s,w}^{\CEUM}$).
    (ii) If $r_{s,c,w} \leq K$, then the top $K$ schools with the best \textbf{true} ranks are selected. Specifically, schools $c'$ where $ r_{s,c',w} \leq K $ are selected ($ c' \in \mathcal{C}_{s,w}^{\CEUM} $). The remaining schools $c''$, where $ r_{s,c'',w} > K $, are not selected ($ c'' \notin \mathcal{C}_{s,w}^{\CEUM} $).
\end{lemma}
\begin{lemma}\label{lemma:preprocess-complex-utility}
    Assume ${\CEUM}$ behavior. For any student $s \in \mathcal{S}$ and scenario $w \in \mathcal{W}$, let $\mathcal{\overline{C}}_{c}^{s,w} = \{c' \succ_{s,w} c : p_{s,c'}^0 u_{s,c'}^w > p_{s,c}^Bu_{s,c}^w\}$ for any acceptable school $c \in \mathcal{C}_{s,w}$. If $|\mathcal{\overline{C}}_{c}^{s,w}| \geq K$, then $c$ is not selected ($c \notin \mathcal{C}_{s,w}^{\CEUM}$).
\end{lemma}
\begin{lemma}\label{lemma:preprocess-simple}
    Assume ${\IEUM}$ behavior. For any student $s \in \mathcal{S}$ and scenario $w \in \mathcal{W}$, let $\mathcal{\hat{C}}_{c}^{-,s,w} = \{ c' \in \mathcal{C}_{s,w}\backslash\{c\}: \overline{p}_{s,c'}^0 u_{s,c'}^w > \overline{p}_{s,c}^Bu_{s,c}^w\}$ and $\mathcal{\hat{C}}_{c}^{+,s,w} = \{ c' \in \mathcal{C}_{s,w}\backslash\{c\}: \overline{p}_{s,c}^0 u_{s,c}^w > \overline{p}_{s,c'}^Bu_{s,c'}^w\}$ where $c \in \mathcal{C}_{s,w}$. If $|\mathcal{\hat{C}}_{c}^{-,s,w}| \geq K$, then $\xi_{s,c,w}^{\IEUM} = 0$. However, if ${|\mathcal{\hat{C}}_{c}^{+,s,w}| \geq |\mathcal{C}_{s,w}| - K}$, then $\xi_{s,c,w}^{\IEUM} = 1$. 
\end{lemma}

\section{Proofs}\label{app:proofs}

In this section, we provide proofs for Lemmas~\ref{lemma:safety-strategy} and~\ref{lemma:preprocess-complex-probability}-\ref{lemma:preprocess-simple}. We also present Lemma~\ref{lemma:not-prefered-chosen-after} and its proof, which form the basis for Lemmas~\ref{lemma:safety-strategy} and~\ref{lemma:preprocess-complex-probability}.
\begin{lemma}\label{lemma:not-prefered-chosen-after}
    Let $x$ be fixed. If $p_{s,c}^w(x_c) = 1$ for some scenario $w \in \mathcal{W}$ and acceptable pair $(s,c) \in \mathcal{S} \times \mathcal{C}_{s,w}$, then the {\MIA} chooses schools $c'' \succeq_{s,w} c$ to compose the endogenous preference $\mathcal{C}_{s,w}^{\CEUM}$ before selecting schools $c' \prec_{s,w} c$.
\end{lemma}
\proof{Proof of Lemma~\ref{lemma:not-prefered-chosen-after}.}
We prove Lemma~\ref{lemma:not-prefered-chosen-after} by recurrence: 

\textbf{Base Case.} At iteration $k=1$ of the {\MIA}, we select $c_1$ that maximizes the marginal utility ${\MI}_s^w(x,\empty,c_1) = p_{s,c_1}^w(x_{c_1}) u_{s,c_1}^w$. Since $p_{s,c}^w(x_c) = 1$ by assumption, and by the given utility condition $u_{s,c}^w > u_{s,c'}^w$ for all $c' \prec_{s,w} c$, it is always better to select school $c$ over any school $c'$. Therefore, at iteration 1, $c_1$ is one of the schools $c'' \succeq_{s,w} c$.

\textbf{Inductive Step.} Assume that at iteration $k \in \mathcal{K} \setminus \{K\}$, all schools $c'''$ chosen before and at iteration $k$ (i.e., the schools in set $C$ as defined in Algorithm~\ref{alg:mia}) respect the preference relation $c''' \succeq_{s,w} c$. At iteration $k+1$, we have two cases:
(i) If school $c$ has already been selected, then any school $c' \prec_{s,w} c$ has zero marginal contribution, meaning that ${\MI}_s^w(x, C \cup \{c'\}) = 0$. On the other hand, for any school $c'' \succeq_{s,w} c$ that has not yet been selected ($c'' \notin C$), we have ${\MI}_s^w(x, C \cup \{c''\}) \geq 0$. Thus, we will select $c_{k+1}$ from the set of schools $c'' \succeq_{s,w} c$ such that $c'' \notin C$. This is valid even if all schools not in $C$ have zero marginal increase, as in this case, we select the one with the best utility.
(ii) If school $c$ has not yet been selected, the contribution of school $c$ is given by ${\MI}_s^w(x, C \cup \{c\}) = p_{s,c}^w(x_c) u_{s,c}^w \prod_{c''' \in C} \left(1 - p_{s,c'''}^w(x_{c'''})\right)$, and the contribution of any school $c' \prec_{s,w} c$ is ${\MI}_s^w(x, C \cup \{c'\}) = p_{s,c'}^w(x_{c'}) u_{s,c'}^w \prod_{c''' \in C} \left(1 - p_{s,c'''}^w(x_{c'''})\right)$. Since $p_{s,c}^w(x_c) u_{s,c}^w > p_{s,c'}^w(x_{c'}) u_{s,c'}^w$ by assumption (due to the higher preference of $c$ and the definition of the utilities), it is always better to select school $c$ before any school $c' \prec_{s,w} c$. Therefore, we will select $c_{k+1}$ from the set of schools $c'' \succeq_{s,w} c$ such that $c'' \notin C$.
Thus, by induction, we have shown that schools $c'' \succeq_{s,w} c$ are always chosen before schools $c' \prec_{s,w} c$ when $p_{s,c}^w(x_c) = 1$.
\endproof

\proof{Proof of Lemma~\ref{lemma:safety-strategy}.}
From Lemma~\ref{lemma:not-prefered-chosen-after}, we know that in the {\MIA}, for student $s \in \mathcal{S}$ in scenario $w \in \mathcal{W}$, schools $c'' \succeq_{s,w} c$ are always selected before schools $c' \prec_{s,w} c$ when $p_{s,c}^w(x_c) = 1$. If $r_{s,c,w} > K$, then we will select $K$ schools from $c'' \succeq_{s,w} c$, and Program~\eqref{eq:max-exp-utility} has not multiple solutions due to the perceived admission of student $s$ to school $c$ in scenario $w$. However, if $r_{s,c,w} \leq K$, the {\MIA} will select all schools $c'' \succeq_{s,w} c$ up to iteration $r_{s,c,w}$. After this iteration, any school $c' \prec_{s,c}$ will have zero marginal increase, and Program~\eqref{eq:max-exp-utility} will have multiple solutions due to the admission chance for school $c$. To comply with Assumption~\ref{assump:safety-strat}, we must select the optimal solution composed of $K$ schools, as specified by the equality constraint in~\eqref{constr:select-school-it}, with schools having the best true utilities. Therefore, Assumption~\ref{assum:strategy} requires the selection of all schools $c' \prec_{s,w} c$ such that $r_{s,c',w} \leq K$, which is guaranteed by Constraints~\eqref{constr:safety-strategy}. In other words, if adding at least $j$ extra seats to school $c$ ensures a perceived admission chance equal to one, then we force the selection of schools $c' \prec_{s,w} c$ with $r_{s,c',w} \leq K$.
\endproof

\proof{Proof of Lemma~\ref{lemma:preprocess-complex-probability}.}
From Assumption~\ref{assum:belief}, if $\overline{p}_{s,c}^0 = 1$, then the probability $p_{s,c}^w(x_c)$ equals one for any feasible solution $x_c$ when $(s,c) \in \mathcal{S} \times  \mathcal{C}_{s,w}$ (indicating that the pair $(s,c)$ is acceptable at the scenario $w$). From Lemma~\ref{lemma:not-prefered-chosen-after}, we know that in the {\MIA}, schools $c'' \succeq_{s,w} c$ are always selected before schools $c' \prec_{s,w} c$. Thus, if $r_{s,c,w} > K$, we have $c' \notin \mathcal{C}_{s,w}^{\CEUM}$, as we are only choosing $K$ schools. If $r_{s,c,w} \leq K$, the {\MIA} will select all schools $c'' \succeq_{s,w} c$ until iteration $r_{s,c,w}$. After this iteration, any school $c' \prec_{s,c}$ will have zero marginal increase. At this point, by Assumption~\ref{assump:safety-strat}, we prioritize selecting schools with the best true utilities. Hence, if $r_{s,c,w} \leq K$, we will select the top $K$ schools with the best true ranks, meaning $c''' \in \mathcal{C}_{s,w}^{\CEUM}$ for all schools $c'''$ such that $r_{s,c''',w} \leq K$, and $c''' \notin \mathcal{C}_{s,w}^{\CEUM}$ for all schools $c''''$ such that $r_{s,c''''} > K$.
\endproof

\proof{Proof of Lemma~\ref{lemma:preprocess-complex-utility}.} For any school $c' \succ_{s,w} c$ with $ p_{s,c'}^0 u_{s,c'}^w > p_{s,c}^B u_{s,c}^w $ such that $c' \notin C$, the marginal increase ${\MI}(x,C,c')$ (see Equation~\eqref{eq:marginal-increase}) satisfies: 
\begin{align}
& {\MI}_{s}^w(x,C,c') \geq \nonumber \\ 
&\resizebox{0.49\linewidth}{!}{$
        \displaystyle  \varrho_{s, c'}^w(x, C) u_{s, c'}^w - p_{s, c'}^w(x_{c'}) \sum_{c'' \in C} \mathbb{I}(u_{s, c}^w, u_{s, c'}^w, u_{s, c''}^w) \varrho_{s, c''}^w(x, C) $} \nonumber \\ &= 
 p_{s,c'}^w(x_{c'})\left[\frac{\varrho_{s, c}^w(x, C)}{p_{s,c}^w(x_c)} u_{s, c'}^w  \sum\limits_{\substack{c'' \in C: \\  u_{s,c''}^w ,< ,u_{s,c}^w }}  \varrho_{s, c''}^w(x, C) u_{s, c''}^w \right] \nonumber \\ &= \overline{{\MI}}_s^w(x,C,c') \nonumber,
\end{align}
where the indicator function \( \mathbb{I}(u_{s, c}^w, u_{s, c'}^w, u_{s, c''}^w) \) selects one of the two possible values based on the relative order of the utilities \( u_{s, c}^w \), \( u_{s, c'}^w \), and \( u_{s, c''}^w \). Specifically, it returns \( u_{s, c'}^w \) when $u_{s, c}^w < u_{s, c''}^w < u_{s, c'}^w$, it returns \( u_{s, c''}^w \) when $u_{s, c''}^w < u_{s, c}^w$. Otherwise, it returns zero. Moreover, we can rearrange the terms in Equation~\eqref{eq:marginal-increase} an write the marginal increase of school $c$ as:
\begin{align}
 \resizebox{0.5\linewidth}{!}{$
        \displaystyle
    {\MI}_{s}^w(x,C,c) =  p_{s,c}^w(x_{c})\left[\frac{\varrho_{s, c}^w(x, C)}{p_{s,c}^w(x_c)} u_{s, c}^w  \sum\limits_{\substack{c'' \in C: \\  u_{s,c''}^w ,< ,u_{s,c}^w }}  \varrho_{s, c''}^w(x, C) u_{s, c''}^w \right] , $} \nonumber
\end{align}
The second term inside the parentheses of $\overline{{\MI}}_s^w(x,C,c')$ and ${\MI}_s^w(x,C,c)$ is identical. Therefore, we can conclude that if $p_{s,c'}^{w}(x_{c'}) > p_{s,c}^{w}(x_c)$, then $\overline{{\MI}}_s^w(x,C,c')- {\MI}_s^w(x,C,c) \geq 0$. This is true because $u_{s,c'}^w > u_{s,c}^w$ by assumption. Similarly, if  $p_{s,c'}^{w}(x_{c'}) < p_{s,c}^{w}(x_c)$, then the inequality still holds. This is because $p_{s,c'}^{w}(x_{c'})u_{s,c'}^w > p_{s,c}^{w}(x_c) u_{s,c}^w$, which is always true as $ p_{s,c'}^0 u_{s,c'}^w > p_{s,c}^B u_{s,c}^w$ by assumption.
\endproof

\proof{Proof of Lemma~\ref{lemma:preprocess-simple}.}
If, for a pair of schools $ c $ and $ c' $, $ \overline{p}^0_{s,c'} u_{s,c'}^w > \overline{p}^B_{s,c} u_{s,c}^w $, this indicates that, regardless of the extra seat allocation, school $ c' $ provides a better individual expected utility than school $ c $. Thus, if at least $ K $ schools $ c' $ satisfy this condition, school $ c $ will never be selected, and we have $c \notin \mathcal{C}_{s,w}^{\IEUM}$.
Moreover, if, for a pair of schools $ c $ and $ c' $, $ \overline{p}^0_{s,c} u_{s,c}^w > \overline{p}^B_{s,c'} u_{s,c'}^w $, then school $ c $ is always preferred over school $ c' $, regardless of the extra seat allocation. Thus, if school $ c $ satisfies this condition for at least $ |\mathcal{C}_{s,w}| - K $ schools $ c' $, school $ c $ will be selected, and we have $c \in \mathcal{C}_{s,w}^{\IEUM}$.
\endproof

\section{Detailed Instance Generation}\label{app:instance-generation}

This section details our instance generation procedure. First, we generate schools' preferences ($\succ_{\mathcal{C}}$) by randomly permuting the set of students $\mathcal{S}$, where the permutation is uniformly distributed. We assume that $\sum_{c \in C} q_c = |\mathcal{S}|$. Thus, we generate the capacities by first allocating one seat to every school and then randomly distributing the remaining $|S| - |C|$ seats among the schools according to the uniform multinomial distribution. As previously explained, we consider students' random utilities to be composed of a deterministic part representing characteristics of students and schools (i.e., euclidian distance in our setting), and a random part following the Gumbel or Type 1 generalized extreme value distribution, i.e., $\boldsymbol{u}_{s,c} = \hat{u}_{s,c} + \varepsilon$ where $\hat{u}_{s,c}$ is deterministic and $\varepsilon \sim Gumbel(0,\beta=4)$. We randomly generate the $\hat{u}_{s,c}$ parameters for $(s,c) \in \mathcal{S} \times \mathcal{C}$. To do this, we randomly generate the positions of schools and students on a 2D grid, with coordinates drawn from a continuous uniform distribution $U(0,10)$. Then, we define $\hat{u}_{s,c}^s = D_{max} - ED_{s,c}$, where $ED_{s,c}$ is the euclidean distance between student $s$ and school $c$, and $D_{max}$ is students' maximum accepted distance, which we assume to be 50\% of the maximum distance within the grid. Hence, $\hat{u}_{s,c}^s$ captures the student's limited information as well as their preference for nearby schools. The maximum accepted distance, $D_{\text{max}}$, penalizes faraway schools, reflecting the student’s limited knowledge of more distant schools, as students commonly have limited information in large market design contexts~\citep{azevedo2019strategy}. It ensures that only schools within a reasonable range are considered. Indeed, distance is a common feature when analysing students utilities~\citep[see, e.g.,][]{agarwal2018demand,agarwal2020revealed}. The definition of $\hat{u}_{s,c}$ through distances induces a correlation between the utilities of students in close proximity, as students located near each other tend to have similar utility values for the same school. Moreover, we chose $\beta = 4$ because it corresponds to approximately 30\% of the range of possible values for $\hat{u}_{s,c}$, thereby introducing a level of uncertainty that is substantial while still allowing the deterministic component to retain significant influence.

We compute students' perceived admission chances using the bootstrapping procedure presented by~\citet{agarwal2018demand} and detailed in Algorithm~\ref{alg:boots}. This algorithm relies on what we call students' deterministic preferences, denoted as $\hat{\succ}_{\mathcal{S}}$, which represents the ranking induced by the deterministic utilities $\hat{u}_{s,c}$. The method consists of bootstrapping students to form a bootstrap set $\mathcal{S}_B^i$, then finding a student-optimal stable matching in the instance $\Gamma = \langle\mathcal{S}_B^i, \mathcal{C}, \hat{\succ}_{\mathcal{S}_B^i}, \succ_{\mathcal{C}}, \boldsymbol{\bar{q}} \rangle$. Notably, $\Gamma$ is scenario-independent, as it is based solely on the deterministic preferences $\hat{\succ}_{\mathcal{S}_B^i}$ of the bootstrap set $\mathcal{S}_B^i$, and the capacities $\bar{\boldsymbol{q}}$ are fixed, meaning they do not depend on first-stage solutions. From the resulting student-optimal stable matching, we compute the cutoff score $\overline{cs}_c^{ij}$ for school $c$ with $j$ additional seats under bootstrap sample $i$. The cutoff score represents the minimum score a student must achieve to be admitted to school $c$. Finally, we use these cutoff scores to estimate students' perceived admission chances $\boldsymbol{\overline{p}}$. In the algorithm, let $BTS=20$ be the number of bootstrapping samples. 

\begin{algorithm}[ht]
\caption{Bootstrap Sampling for Perceived Admission Chances Estimation}\label{alg:boots}
\KwIn{$\mathcal{S}$,$\mathcal{C}$,$\hat{\succ}_{\mathcal{S}}$,$\succ_{\mathcal{C}}$,$\boldsymbol{q}$,$\mathcal{B}$,$\mathcal{B}^0$, $BTS$;}
\KwOut{$\boldsymbol{\overline{p}}$;}
\For{$i \in \{1 \ldots BTS\}$}{
    Sample $n$ students with replacement from $\mathcal{S}$ to create bootstrap set $\mathcal{S}_B^i$\;
    \ForEach{$(c,j) \in \mathcal{C}, \times ,\mathcal{B}^0$}{
        Set $\bar{q}_{c'} \gets q_{c'} \quad \forall c' \in \mathcal{C}\backslash c, \quad \bar{q}_c \gets q_c + j$\;
        Solve the student-optimal stable matching problem with {\DAA} for instance $\Gamma = \langle\mathcal{S}_B^i, \mathcal{C}, \hat{\succ}_{\mathcal{S}_B^i}, \succ_{\mathcal{C}}, \boldsymbol{\bar{q}} \rangle$ to obtain cutoff score $\overline{cs}_c^{ij}$ for school $c$ with $j$ extra seats \;
    }
}
\ForEach{$(s,c,j) \in \mathcal{S} \times \mathcal{C} \times \mathcal{B}^0$}{
    $\overline{p}_{s,c}^j \gets \frac{1}{BTS} \sum_{i=1}^{BTS} \mathbf{1}\{e_{s,c} \geq \overline{cs}_c^{ij}\}$\;
}
\end{algorithm}

\section{Warm-Start Procedure}\label{app:warm-start}

When solving the {\SAA} Program~\eqref{prog:saa} under the {\UM} behavior, we begin by solving Program~\eqref{prog:stmmp-2st} with no additional seats ($x_c^j = 0 \ \forall c \in \mathcal{C},\ j \in \mathcal{B}$) using the {\DAA}. This step yields the matching $\hat{\mu}_w$ for each scenario $w \in \mathcal{W_N}$. Given that school capacities can only increase relative to the no-extra-seats solution, and that student preferences remain fixed in this context, the matching can only improve with higher capacities. Thus, we can set $y_{s,c}^w = 0$ for all schools $c \prec_{s,w} \hat{\mu}_w(s)$, students $s \in \mathcal{S}$, and scenarios $w \in \mathcal{W_N}$. Furthermore, if a school $c$ is unsaturated in all scenarios, that is, $|\hat{\mu}_w(c)| \leq q_c$ for all $w \in \mathcal{W_N}$, it implies that no extra seats are needed for that school and we can set $x_c^j = 0$ for all $j \in \mathcal{B}$. Finally, we proceed with the {\ASG} heuristic and utilize the resulting feasible solution as an initial input for the {\SAA} program. Under the {\CEUM} and {\IEUM} behaviors, we first compute the reported preferences $\succ_{s,w}^{a}$ for the no-additional-seats case using the appropriate algorithm. Next, we apply the {\DAA} to derive a feasible matching. This feasible solution is then used as the initial warm-start for the {\SAA} program. For the exact approach under the {\CEUM} and {\IEUM}, we run the {\LS} heuristic with a 10-minute time limit and use its best feasible solution as an initial solution.

\section{Detailed Results for Computational Performance}\label{app:detailed-results}

This section presents detailed results on the computational performance of the proposed methods under nonstrategic and strategic behaviors.

\subsection{Nonstrategic Behavior}\label{app:perf-nonstrat}

We compare the detailed computational performance of the proposed approaches for solving {\SAA} Program~\eqref{prog:saa} under the {\UM} behavior. We compare the modified $L$-constraints solved with the Gurobi solver, as well as the {\ASG}, {\LR}, {\LS}, and {\SA} heuristics. For the {\LR} heuristic, the subgradient method is used with scalar ${\hat{\tau} = 2.0}$ at the first iteration. This scalar is halved every 15 consecutive iterations without improvement in the lower bound. The algorithm terminates if the current step size $\hat{\lambda}$ is smaller than $0.001/N$, if a maximum of 30 iterations without improvement on the lower bound is achieved, or if a maximum number of 300 iterations is achieved. Moreover, we initialize the Lagrangian multiplier $\hat{\alpha}$ vector at the first iteration with the vector of zeros. For the {\SA} heuristic, the temperature is initialized at $T_0=100$ and decreased by a factor of $\delta=0.95$ every $I_T=25$ iterations. The algorithm terminates when the temperature drops below $T_{min}=0.1$. Finally, we use the same warm-start procedure for all approaches (see~\ref{app:warm-start}).

Table~\ref{tab:perf-um} presents the detailed results for large instances. All runs had a time limit of two hours. The table reports the computational time in seconds and the optimality gap [\({\text{gap}_{opt} = 100\% \cdot (ub - lb) / ub}\)] for each approach and instance setting (budget $B$ and size limit $K$), where \(ub\) and \(lb\) represent the best upper and lower bounds obtained by the approach on the optimal value of the {\SAA} Program~\eqref{prog:saa}, respectively. We also report the upper bound gap [\({\text{gap}_{ub^*} = 100\% \cdot (ub - ub^*) / ub}\)], where \(ub\) is the upper bound obtained by the considered approach, and \(ub^*\) is the best upper bound found across all approaches within the time limit. Neither the {\LS} nor the {\SA} heuristics provide lower bounds, and therefore, optimality gaps cannot be computed in these cases. 
\begin{table}[ht]
\caption{Computational performance for large instances of approaches for {\SAA} Program~\eqref{prog:saa} under {\UM} behavior.} \label{tab:perf-um}
\resizebox{\textwidth}{!}{%
\begin{tabular}{cc|ccc|ccc|ccc|cc|cc}
\hline 
\multirow{2}{*}{$B$} & \multirow{2}{*}{$K$} & \multicolumn{3}{c|}{Modified $L$-constraints}             & \multicolumn{3}{c|}{{\ASG} Heuristic} & \multicolumn{3}{c|}{{\LR} Heuristic}     & \multicolumn{2}{c|}{{\LS} Heuristic} & \multicolumn{2}{c}{{\SA} Heuristic} \\ 
                     &                      & $gap_{opt}$ (\%) & Time (Seconds)     & $gap_{ub^*}$ (\%) & $gap_{opt}$ (\%)  & Time (Seconds) & $gap_{ub^*}$ (\%) & $gap_{opt}$ (\%) & Time (Seconds)     & $gap_{ub^*}$ (\%) & Time (Seconds)             & $gap_{ub^*}$ (\%)        & Time (Seconds)            & $gap_{ub^*}$ (\%)          \\ \hline   
\multirow{3}{*}{1}   & 2                    & 0.006$\pm$0.001    & 10.967$\pm$1.583     & 0.002$\pm$0.001     & 4.386$\pm$0.353     & 1.9$\pm$0.345    & 0.013$\pm$0.005     & 1.173$\pm$0.134    & 302.4$\pm$39.42      & 0.002$\pm$0.002     & 35.267$\pm$7.721             & 0.0$\pm$0.0                & 1485.967$\pm$211.222        & 0.0$\pm$0.0                \\
                     & 3                    & 0.009$\pm$0.001    & 27.067$\pm$5.654     & 0.0$\pm$0.0         & 11.226$\pm$0.387    & 2.6$\pm$0.423    & 0.037$\pm$0.009     & 4.278$\pm$0.347    & 528.867$\pm$72.724   & 0.021$\pm$0.008     & 51.067$\pm$11.735            & 0.0$\pm$0.0                & 2184.267$\pm$328.612        & 0.003$\pm$0.004            \\
                     & 4                    & 0.009$\pm$0.001    & 369.6$\pm$150.955    & 0.0$\pm$0.0         & 17.181$\pm$0.521    & 3.667$\pm$0.63   & 0.07$\pm$0.02       & 8.521$\pm$0.465    & 663.3$\pm$90.621     & 0.041$\pm$0.012     & 64.267$\pm$15.337            & 0.0$\pm$0.0                & 2837.967$\pm$496.344        & 0.004$\pm$0.007              \\ \hline   
\multirow{3}{*}{5}   & 2                    & 0.143$\pm$0.032    & 6298.833$\pm$740.375 & 0.003$\pm$0.002     & 5.127$\pm$0.241     & 1.933$\pm$0.276  & 0.045$\pm$0.011     & 1.582$\pm$0.121    & 727.567$\pm$174.602  & 0.017$\pm$0.005     & 348.033$\pm$91.151           & 0.005$\pm$0.004            & 1442.5$\pm$200.225          & 0.032$\pm$0.012            \\
                     & 3                    & 1.321$\pm$0.077    & 7207.0$\pm$1.122     & 0.097$\pm$0.019     & 11.684$\pm$0.319    & 3.0$\pm$0.404    & 0.12$\pm$0.021      & 5.047$\pm$0.266    & 911.7$\pm$151.647    & 0.051$\pm$0.014     & 493.533$\pm$112.387          & 0.005$\pm$0.005            & 2093.0$\pm$315.857          & 0.035$\pm$0.013            \\
                     & 4                    & 3.084$\pm$0.159    & 7210.1$\pm$1.801     & 0.202$\pm$0.039     & 17.06$\pm$0.6       & 3.767$\pm$0.625  & 0.194$\pm$0.037     & 9.309$\pm$0.302    & 1056.0$\pm$167.953   & 0.12$\pm$0.028      & 820.433$\pm$235.901          & 0.01$\pm$0.008             & 2691.033$\pm$480.344        & 0.041$\pm$0.016            \\ \hline 
\multirow{3}{*}{15}  & 2                    & 0.891$\pm$0.058    & 7207.133$\pm$1.537   & 0.042$\pm$0.01      & 5.746$\pm$0.15      & 2.367$\pm$0.422  & 0.074$\pm$0.012     & 2.148$\pm$0.085    & 865.4$\pm$157.137    & 0.036$\pm$0.01      & 1696.533$\pm$488.695         & 0.011$\pm$0.008            & 1465.667$\pm$197.783        & 0.099$\pm$0.022            \\
                     & 3                    & 3.031$\pm$0.136    & 7209.1$\pm$1.342     & 0.152$\pm$0.026     & 11.533$\pm$0.477    & 4.133$\pm$0.897  & 0.155$\pm$0.027     & 5.735$\pm$0.134    & 1421.6$\pm$270.113   & 0.066$\pm$0.017     & 2544.533$\pm$650.84          & 0.024$\pm$0.017            & 1995.967$\pm$309.646        & 0.096$\pm$0.032            \\
                     & 4                    & 5.182$\pm$0.352    & 7211.767$\pm$2.059   & 0.207$\pm$0.042     & 15.587$\pm$1.024    & 4.8$\pm$0.781    & 0.231$\pm$0.046     & 9.265$\pm$0.412    & 1679.933$\pm$302.926 & 0.097$\pm$0.023     & 3429.8$\pm$869.12            & 0.041$\pm$0.025            & 2406.9$\pm$416.328          & 0.067$\pm$0.035            \\ \hline 
\multirow{3}{*}{30}  & 2                    & 1.228$\pm$0.083    & 7207.267$\pm$1.271   & 0.05$\pm$0.014      & 5.615$\pm$0.272     & 2.433$\pm$0.401  & 0.088$\pm$0.02      & 2.442$\pm$0.071    & 975.567$\pm$187.272  & 0.027$\pm$0.012     & 3799.767$\pm$866.16          & 0.101$\pm$0.095            & 1445.367$\pm$220.658        & 0.14$\pm$0.028             \\
                     & 3                    & 3.389$\pm$0.291    & 7210.767$\pm$1.928   & 0.107$\pm$0.026     & 10.074$\pm$0.853    & 4.733$\pm$1.393  & 0.104$\pm$0.025     & 5.546$\pm$0.331    & 1688.8$\pm$355.357   & 0.021$\pm$0.012     & 4861.9$\pm$947.643           & 0.753$\pm$0.39             & 1874.9$\pm$290.44           & 0.132$\pm$0.048            \\
                     & 4                    & 4.915$\pm$0.592    & 7213.033$\pm$2.154   & 0.138$\pm$0.033     & 12.494$\pm$1.371    & 5.633$\pm$1.714  & 0.142$\pm$0.027     & 7.907$\pm$0.7      & 2096.167$\pm$476.071 & 0.019$\pm$0.012     & 5165.633$\pm$863.277         & 1.549$\pm$0.732            & 2123.6$\pm$367.858          & 0.088$\pm$0.037              \\ \hline  
\multirow{3}{*}{45}  & 2                    & 1.204$\pm$0.132    & 7207.533$\pm$1.628   & 0.036$\pm$0.014     & 4.974$\pm$0.429     & 2.267$\pm$0.459  & 0.075$\pm$0.013     & 2.358$\pm$0.142    & 1059.0$\pm$216.939   & 0.008$\pm$0.005     & 5076.8$\pm$916.008           & 0.735$\pm$0.414            & 1342.3$\pm$183.886          & 0.176$\pm$0.04             \\
                     & 3                    & 2.912$\pm$0.412    & 7209.9$\pm$1.645     & 0.095$\pm$0.024     & 8.168$\pm$1.045     & 3.633$\pm$0.669  & 0.11$\pm$0.023      & 4.746$\pm$0.483    & 1708.633$\pm$419.832 & 0.006$\pm$0.006     & 5513.6$\pm$813.567           & 2.86$\pm$1.08              & 1694.9$\pm$272.711          & 0.206$\pm$0.056            \\
                     & 4                    & 3.872$\pm$0.658    & 7212.767$\pm$2.179   & 0.123$\pm$0.033     & 9.462$\pm$1.387     & 4.6$\pm$1.063    & 0.125$\pm$0.028     & 6.253$\pm$0.784    & 2009.433$\pm$495.263 & 0.011$\pm$0.01      & 6030.067$\pm$745.995         & 4.821$\pm$1.85             & 1922.967$\pm$330.74         & 0.152$\pm$0.061              \\ \hline  
\multirow{3}{*}{60}  & 2                    & 1.046$\pm$0.18     & 7207.133$\pm$1.48    & 0.04$\pm$0.013      & 4.191$\pm$0.54      & 2.133$\pm$0.389  & 0.092$\pm$0.018     & 2.103$\pm$0.208    & 1021.867$\pm$230.506 & 0.007$\pm$0.004     & 5606.267$\pm$851.909         & 2.077$\pm$0.835            & 1285.6$\pm$189.804          & 0.214$\pm$0.059            \\
                     & 3                    & 2.279$\pm$0.436    & 7209.6$\pm$1.625     & 0.08$\pm$0.021      & 6.319$\pm$1.052     & 3.533$\pm$0.994  & 0.115$\pm$0.022     & 3.841$\pm$0.534    & 1619.233$\pm$415.588 & 0.008$\pm$0.006     & 5985.667$\pm$734.266         & 5.17$\pm$1.89              & 1570.667$\pm$240.972        & 0.192$\pm$0.052            \\
                     & 4                    & 2.825$\pm$0.6      & 7212.233$\pm$2.253   & 0.095$\pm$0.023     & 6.978$\pm$1.256     & 4.167$\pm$0.827  & 0.112$\pm$0.024     & 4.761$\pm$0.749    & 1817.267$\pm$465.427 & 0.003$\pm$0.003     & 6376.8$\pm$673.306           & 8.39$\pm$2.733             & 1832.467$\pm$325.762        & 0.208$\pm$0.049                \\ \hline 
\multicolumn{2}{c|}{Mean}                   & 2.075$\pm$0.154    & 5980.1$\pm$225.431   & 0.082$\pm$0.007     & 9.323$\pm$0.395     & 3.406$\pm$0.201  & 0.106$\pm$0.007     & 4.834$\pm$0.235    & 1230.707$\pm$79.399  & 0.031$\pm$0.004     & 3216.665$\pm$247.704         & 1.475$\pm$0.293            & 1872.002$\pm$79.213         & 0.105$\pm$0.01               \\
\multicolumn{2}{c|}{Min.}                   & 0.000            & 5.000              & 0.000             & 1.572             & 0.000          & 0.000             & 0.589            & 136.000            & 0.000             & 8.000                      & 0.000                    & 500.000                   & 0.000                    \\
\multicolumn{2}{c|}{Max.}                   & 7.627            & 7223.000           & 0.495             & 20.721            & 27.000         & 0.572             & 11.306           & 4798.000           & 0.328             & 7263.000                   & 18.411                   & 4627.000                  & 0.783                       \\ \hline
\end{tabular}%
}
\caption*{\centering The values are averages over the considered instances, with a 95\% confidence interval.}
\end{table}

\subsection{Strategic Behaviors}\label{app:perf-strat}

We compare the computational performance of the proposed methods for solving the {\SAA} Program~\eqref{prog:saa} under the {\CEUM} and {\IEUM} behaviors. The {\SA} heuristic is run with the same parameters as in the previous section for both small and large instances. Table~\ref{tab:perf-ceum-ieum-small} reports results for the small instances, including computational time, optimality gap, and upper bound gap (as defined previously), for all methods under both behaviors. All runs were subject to an eight-hour time limit. As noted, the {\LS} and {\SA} heuristics do not yield optimality gaps. Table~\ref{tab:perf-ceum-ieum-large} presents the corresponding results for the large instances, reporting computational time and upper bound gap for the heuristics, as the exact approach failed to produce good results---based on the investigated gap measures---within the time limit.
\begin{table}[ht]
\centering
\caption{Computational performance for small instances of approaches for {\SAA} Program~\eqref{prog:saa} under {\CEUM} and {\IEUM} behaviors.} \label{tab:perf-ceum-ieum-small}
\resizebox{\textwidth}{!}{%
\begin{tabular}{cc|ccccccc|ccccccc}
\hline 
\multicolumn{1}{l}{\multirow{3}{*}{$B$}} & \multicolumn{1}{l|}{\multirow{3}{*}{$K$}} & \multicolumn{7}{c|}{{\CEUM} Behavior}                                                                                                                                             & \multicolumn{7}{c}{{\IEUM} Behavior}                                                                                                                                              
    \\  \cline{3-16} 
\multicolumn{1}{l}{}                     & \multicolumn{1}{l|}{}                     & \multicolumn{3}{c|}{Cutoff Score}                                                & \multicolumn{2}{c|}{{\LS} Heuristic}   & \multicolumn{2}{c|}{{\SA} Heuristic} & \multicolumn{3}{c|}{Cutoff Score}                                               & \multicolumn{2}{c|}{{\LS} Heuristic}   & \multicolumn{2}{c}{{\SA} Heuristic}      \\ 
\multicolumn{1}{l}{}                     & \multicolumn{1}{l|}{}                     & $gap_{opt}$ (\%) & Time (Seconds)       & \multicolumn{1}{c|}{$gap_{ub^*}$ (\%)} & Time (Seconds) & \multicolumn{1}{c|}{$gap_{ub^*}$ (\%)} & Time (Seconds)           & $gap_{ub^*}$ (\%)          & $gap_{opt}$ (\%) & Time (Seconds)      & \multicolumn{1}{c|}{$gap_{ub^*}$ (\%)} & Time (Seconds) & \multicolumn{1}{c|}{$gap_{ub^*}$ (\%)} & Time (Seconds)          & $gap_{ub^*}$ (\%)             \\ \hline
\multirow{3}{*}{1}                       & 2                                         & 0.0$\pm$0.0        & 29.6$\pm$14.626        & \multicolumn{1}{c|}{0.0$\pm$0.0}         & 0.2$\pm$0.302    & \multicolumn{1}{c|}{0.0$\pm$0.0}         & 43.9$\pm$14.109            & 0.0$\pm$0.0                  & 0.0$\pm$0.0        & 2.5$\pm$0.843         & \multicolumn{1}{c|}{0.0$\pm$0.0}         & 0.0$\pm$0.0      & \multicolumn{1}{c|}{0.0$\pm$0.0}         & 28.4$\pm$9.249            & 0.0$\pm$0.0                   \\
                                         & 3                                         & 0.0$\pm$0.0        & 116.8$\pm$59.284       & \multicolumn{1}{c|}{0.0$\pm$0.0}         & 0.3$\pm$0.346    & \multicolumn{1}{c|}{0.0$\pm$0.0}         & 60.7$\pm$19.669            & 0.0$\pm$0.0                  & 0.0$\pm$0.0        & 5.2$\pm$2.857         & \multicolumn{1}{c|}{0.0$\pm$0.0}         & 0.1$\pm$0.226    & \multicolumn{1}{c|}{0.0$\pm$0.0}         & 40.5$\pm$13.848           & 0.0$\pm$0.0                  \\
                                         & 4                                         & 0.0$\pm$0.0        & 136.7$\pm$68.952       & \multicolumn{1}{c|}{0.0$\pm$0.0}         & 0.4$\pm$0.369    & \multicolumn{1}{c|}{0.0$\pm$0.0}         & 70.2$\pm$23.307            & 0.0$\pm$0.0                  & 0.001$\pm$0.002    & 6.8$\pm$4.072         & \multicolumn{1}{c|}{0.0$\pm$0.0}         & 0.2$\pm$0.302    & \multicolumn{1}{c|}{0.0$\pm$0.0}         & 53.0$\pm$19.186           & 0.0$\pm$0.0                    \\ \hline  
\multirow{3}{*}{3}                       & 2                                         & 0.001$\pm$0.002    & 2898.1$\pm$1047.096    & \multicolumn{1}{c|}{0.0$\pm$0.0}         & 0.6$\pm$0.369    & \multicolumn{1}{c|}{0.089$\pm$0.107}     & 43.6$\pm$14.104            & 0.0$\pm$0.0                  & 0.0$\pm$0.0        & 609.3$\pm$424.694     & \multicolumn{1}{c|}{0.0$\pm$0.0}         & 0.5$\pm$0.377    & \multicolumn{1}{c|}{0.179$\pm$0.173}     & 28.5$\pm$9.298            & 0.0$\pm$0.0                  \\
                                         & 3                                         & 1.752$\pm$1.689    & 15798.5$\pm$8576.876   & \multicolumn{1}{c|}{0.04$\pm$0.077}      & 0.6$\pm$0.5      & \multicolumn{1}{c|}{0.105$\pm$0.112}     & 58.7$\pm$18.721            & 0.0$\pm$0.0                  & 0.0$\pm$0.0        & 1899.3$\pm$1567.417   & \multicolumn{1}{c|}{0.0$\pm$0.0}         & 0.5$\pm$0.377    & \multicolumn{1}{c|}{0.049$\pm$0.076}     & 41.0$\pm$14.351           & 0.0$\pm$0.0                  \\
                                         & 4                                         & 1.078$\pm$1.357    & 14259.8$\pm$9104.322   & \multicolumn{1}{c|}{0.0$\pm$0.0}         & 0.7$\pm$0.346    & \multicolumn{1}{c|}{0.016$\pm$0.022}     & 66.1$\pm$21.125            & 0.0$\pm$0.0                  & 0.0$\pm$0.0        & 2725.1$\pm$2613.811   & \multicolumn{1}{c|}{0.0$\pm$0.0}         & 0.9$\pm$0.406    & \multicolumn{1}{c|}{0.0$\pm$0.0}         & 52.4$\pm$18.685           & 0.0$\pm$0.0                    \\ \hline  
\multirow{3}{*}{3}                       & 2                                         & 1.139$\pm$2.07     & 16473.5$\pm$7810.223   & \multicolumn{1}{c|}{0.061$\pm$0.138}     & 1.1$\pm$0.528    & \multicolumn{1}{c|}{0.186$\pm$0.159}     & 42.3$\pm$13.586            & 0.0$\pm$0.0                  & 0.001$\pm$0.001    & 7081.2$\pm$4403.039   & \multicolumn{1}{c|}{0.0$\pm$0.0}         & 0.6$\pm$0.369    & \multicolumn{1}{c|}{0.17$\pm$0.207}      & 28.9$\pm$9.805            & 0.0$\pm$0.0                  \\
                                         & 3                                         & 4.505$\pm$2.937    & 26596.7$\pm$4162.015   & \multicolumn{1}{c|}{0.287$\pm$0.584}     & 1.4$\pm$0.5      & \multicolumn{1}{c|}{0.316$\pm$0.579}     & 56.9$\pm$18.221            & 0.0$\pm$0.0                  & 1.233$\pm$1.494    & 17281.9$\pm$7592.597  & \multicolumn{1}{c|}{0.0$\pm$0.0}         & 1.0$\pm$0.337    & \multicolumn{1}{c|}{0.165$\pm$0.159}     & 40.5$\pm$13.97            & 0.0$\pm$0.0                  \\
                                         & 4                                         & 2.659$\pm$2.03     & 19308.7$\pm$7453.271   & \multicolumn{1}{c|}{0.082$\pm$0.116}     & 1.4$\pm$0.369    & \multicolumn{1}{c|}{0.082$\pm$0.116}     & 64.4$\pm$21.432            & 0.0$\pm$0.0                  & 2.605$\pm$2.583    & 17418.8$\pm$7672.289  & \multicolumn{1}{c|}{0.075$\pm$0.101}     & 1.1$\pm$0.406    & \multicolumn{1}{c|}{0.246$\pm$0.168}     & 52.5$\pm$19.002           & 0.0$\pm$0.0                    \\ \hline  
\multirow{3}{*}{10}                      & 2                                         & 5.507$\pm$2.283    & 27713.3$\pm$1820.562   & \multicolumn{1}{c|}{0.4$\pm$0.355}       & 2.1$\pm$0.787    & \multicolumn{1}{c|}{0.407$\pm$0.35}      & 40.3$\pm$12.761            & 0.0$\pm$0.0                  & 3.455$\pm$1.395    & 28387.9$\pm$936.005   & \multicolumn{1}{c|}{0.151$\pm$0.117}     & 1.6$\pm$0.5      & \multicolumn{1}{c|}{0.338$\pm$0.217}     & 28.8$\pm$9.968            & 0.015$\pm$0.034              \\
                                         & 3                                         & 5.496$\pm$1.879    & 28802.9$\pm$0.98       & \multicolumn{1}{c|}{0.605$\pm$0.398}     & 2.9$\pm$0.98     & \multicolumn{1}{c|}{0.605$\pm$0.398}     & 54.0$\pm$17.441            & 0.0$\pm$0.0                  & 5.19$\pm$1.647     & 28802.2$\pm$0.739     & \multicolumn{1}{c|}{0.305$\pm$0.218}     & 2.0$\pm$0.674    & \multicolumn{1}{c|}{0.342$\pm$0.208}     & 39.7$\pm$13.59            & 0.017$\pm$0.038              \\
                                         & 4                                         & 3.02$\pm$1.446     & 28803.1$\pm$0.787      & \multicolumn{1}{c|}{0.204$\pm$0.228}     & 3.0$\pm$1.012    & \multicolumn{1}{c|}{0.217$\pm$0.224}     & 60.5$\pm$19.533            & 0.0$\pm$0.0                  & 5.257$\pm$2.423    & 28803.1$\pm$1.037     & \multicolumn{1}{c|}{0.431$\pm$0.367}     & 2.4$\pm$0.84     & \multicolumn{1}{c|}{0.45$\pm$0.357}      & 49.7$\pm$16.79            & 0.0$\pm$0.0                     \\ \hline  
\multirow{3}{*}{15}                      & 2                                         & 4.535$\pm$1.643    & 28803.5$\pm$1.08       & \multicolumn{1}{c|}{0.612$\pm$0.537}     & 3.1$\pm$1.037    & \multicolumn{1}{c|}{0.616$\pm$0.535}     & 39.0$\pm$12.275            & 0.0$\pm$0.0                  & 3.428$\pm$1.111    & 28803.0$\pm$0.754     & \multicolumn{1}{c|}{0.557$\pm$0.455}     & 2.3$\pm$0.679    & \multicolumn{1}{c|}{0.557$\pm$0.455}     & 27.4$\pm$8.904            & 0.012$\pm$0.026              \\
                                         & 3                                         & 3.304$\pm$1.377    & 28804.4$\pm$1.478      & \multicolumn{1}{c|}{0.147$\pm$0.165}     & 4.2$\pm$1.575    & \multicolumn{1}{c|}{0.147$\pm$0.165}     & 50.3$\pm$14.719            & 0.0$\pm$0.0                  & 3.958$\pm$1.22     & 28803.2$\pm$1.0       & \multicolumn{1}{c|}{0.47$\pm$0.478}      & 3.0$\pm$1.118    & \multicolumn{1}{c|}{0.47$\pm$0.478}      & 38.5$\pm$13.452           & 0.0$\pm$0.0                  \\
                                         & 4                                         & 1.592$\pm$0.909    & 27435.5$\pm$3098.779   & \multicolumn{1}{c|}{0.046$\pm$0.061}     & 4.6$\pm$1.624    & \multicolumn{1}{c|}{0.046$\pm$0.061}     & 56.8$\pm$17.313            & 0.0$\pm$0.0                  & 3.881$\pm$1.492    & 28803.9$\pm$1.448     & \multicolumn{1}{c|}{0.373$\pm$0.387}     & 3.7$\pm$1.308    & \multicolumn{1}{c|}{0.373$\pm$0.387}     & 49.3$\pm$17.112           & 0.0$\pm$0.0                       \\ \hline 
\multicolumn{2}{c|}{Mean}                                                            & 2.306$\pm$0.465    & 17732.073$\pm$2077.769 & \multicolumn{1}{c|}{0.166$\pm$0.066}     & 1.773$\pm$0.289  & \multicolumn{1}{c|}{0.189$\pm$0.065}     & 53.847$\pm$4.081           & 0.0$\pm$0.0                  & 1.934$\pm$0.426    & 14628.893$\pm$2157.63 & \multicolumn{1}{c|}{0.157$\pm$0.059}     & 1.327$\pm$0.222  & \multicolumn{1}{c|}{0.223$\pm$0.061}     & 39.94$\pm$3.425           & 0.003$\pm$0.003              \\
\multicolumn{2}{c|}{Min.}                                                            & 0.000            & 8.000                & \multicolumn{1}{c|}{0.000}             & 0.000          & \multicolumn{1}{c|}{0.000}             & 21.000                   & 0.000                      & 0.000            & 1.000               & \multicolumn{1}{c|}{0.000}             & 0.000          & \multicolumn{1}{c|}{0.000}             & 15.000                  & 0.000                      \\
\multicolumn{2}{c|}{Max.}                                                            & 10.093           & 28809.000            & \multicolumn{1}{c|}{2.605}             & 8.000          & \multicolumn{1}{c|}{2.605}             & 109.000                  & 0.000                      & 11.349           & 28806.000           & \multicolumn{1}{c|}{2.305}             & 6.000          & \multicolumn{1}{c|}{2.305}             & 83.000                  & 0.168                        \\ \hline
\end{tabular}%
}
\caption*{\centering The values are averages over the considered instances, with a 95\% confidence interval.}
\end{table}
\begin{table}[]
\caption{Computational performance for large instances of approaches for {\SAA} Program~\eqref{prog:saa} under {\CEUM} and {\IEUM} behaviors.} \label{tab:perf-ceum-ieum-large}
\centering
\begin{adjustbox}{width=0.8\textwidth,center}%
\centering
\begin{tabular}{cc|cccc|cccc}
\hline     
\multirow{3}{*}{$B$} & \multirow{3}{*}{$K$} & \multicolumn{4}{c|}{{\CEUM} Behavior}                                                                & \multicolumn{4}{c}{{\IEUM} Behavior}                                                                  \\ \cline{3-10}  
                     &                      & \multicolumn{2}{c|}{{\LS} Heuristic}         & \multicolumn{2}{c|}{{\SA} Heuristic} & \multicolumn{2}{c|}{{\LS} Heuristic}         & \multicolumn{2}{c}{{\SA} Heuristic} \\
                     &                      & Time (Seconds)       & \multicolumn{1}{c|}{$gap_{ub^*}$ (\%)} & Time (Seconds)             & $gap_{ub^*}$ (\%)        & Time (Seconds)       & \multicolumn{1}{c|}{$gap_{ub^*}$ (\%)} & Time (Seconds)             & $gap_{ub^*}$ (\%)         \\ \hline  
\multirow{3}{*}{1}   & 2                    & 70.7$\pm$17.56         & \multicolumn{1}{c|}{0.0$\pm$0.0}         & 2907.6$\pm$463.243           & 0.003$\pm$0.006            & 37.133$\pm$9.194       & \multicolumn{1}{c|}{0.0$\pm$0.0}         & 1610.333$\pm$273.005         & 0.001$\pm$0.001           \\
                     & 3                    & 116.667$\pm$28.842     & \multicolumn{1}{c|}{0.0$\pm$0.0}         & 4603.3$\pm$716.794           & 0.019$\pm$0.026            & 59.233$\pm$15.932      & \multicolumn{1}{c|}{0.0$\pm$0.0}         & 2386.8$\pm$416.27            & 0.002$\pm$0.002           \\
                     & 4                    & 156.8$\pm$41.345       & \multicolumn{1}{c|}{0.0$\pm$0.0}         & 5882.633$\pm$938.821         & 0.034$\pm$0.049            & 80.267$\pm$19.935      & \multicolumn{1}{c|}{0.0$\pm$0.0}         & 3559.4$\pm$636.435           & 0.0$\pm$0.0                 \\ \hline  
\multirow{3}{*}{5}   & 2                    & 638.967$\pm$174.994    & \multicolumn{1}{c|}{0.054$\pm$0.076}     & 3060.767$\pm$447.84          & 0.166$\pm$0.084            & 350.733$\pm$96.593     & \multicolumn{1}{c|}{0.002$\pm$0.004}     & 1786.7$\pm$293.726           & 0.061$\pm$0.022           \\
                     & 3                    & 1266.567$\pm$403.176   & \multicolumn{1}{c|}{0.018$\pm$0.015}     & 4668.033$\pm$700.284         & 0.239$\pm$0.229            & 626.7$\pm$197.677      & \multicolumn{1}{c|}{0.004$\pm$0.007}     & 2701.567$\pm$441.912         & 0.072$\pm$0.039           \\
                     & 4                    & 1810.433$\pm$537.023   & \multicolumn{1}{c|}{0.018$\pm$0.011}     & 5757.233$\pm$931.982         & 0.094$\pm$0.048            & 893.6$\pm$243.759      & \multicolumn{1}{c|}{0.012$\pm$0.01}      & 3720.1$\pm$650.273           & 0.071$\pm$0.032             \\ \hline  
\multirow{3}{*}{15}  & 2                    & 4227.133$\pm$1169.752  & \multicolumn{1}{c|}{0.093$\pm$0.078}     & 3339.6$\pm$519.21            & 0.232$\pm$0.126            & 2192.3$\pm$679.222     & \multicolumn{1}{c|}{0.017$\pm$0.022}     & 2158.933$\pm$332.251         & 0.2$\pm$0.069             \\
                     & 3                    & 7209.7$\pm$2034.277    & \multicolumn{1}{c|}{0.11$\pm$0.079}      & 4793.3$\pm$780.669           & 0.146$\pm$0.094            & 3724.333$\pm$1137.161  & \multicolumn{1}{c|}{0.028$\pm$0.038}     & 3389.067$\pm$536.096         & 0.128$\pm$0.054           \\
                     & 4                    & 9354.433$\pm$2876.633  & \multicolumn{1}{c|}{0.071$\pm$0.036}     & 5805.5$\pm$895.761           & 0.123$\pm$0.073            & 5675.9$\pm$1834.722    & \multicolumn{1}{c|}{0.039$\pm$0.029}     & 4365.6$\pm$657.509           & 0.111$\pm$0.04              \\ \hline  
\multirow{3}{*}{30}  & 2                    & 13263.6$\pm$3667.107   & \multicolumn{1}{c|}{0.215$\pm$0.193}     & 3694.2$\pm$599.153           & 0.352$\pm$0.22             & 7147.367$\pm$2701.971  & \multicolumn{1}{c|}{0.028$\pm$0.034}     & 2711.567$\pm$445.771         & 0.441$\pm$0.133           \\
                     & 3                    & 16843.8$\pm$3969.567   & \multicolumn{1}{c|}{0.625$\pm$0.437}     & 4966.767$\pm$820.169         & 0.115$\pm$0.109            & 11422.267$\pm$3202.756 & \multicolumn{1}{c|}{0.075$\pm$0.068}     & 4203.333$\pm$671.925         & 0.433$\pm$0.16            \\
                     & 4                    & 18917.033$\pm$4066.063 & \multicolumn{1}{c|}{1.415$\pm$0.885}     & 6169.933$\pm$1084.511        & 0.034$\pm$0.026            & 14792.6$\pm$3770.887   & \multicolumn{1}{c|}{0.231$\pm$0.25}      & 5518.5$\pm$886.001           & 0.331$\pm$0.148             \\ \hline  
\multirow{3}{*}{45}  & 2                    & 17802.2$\pm$4108.084   & \multicolumn{1}{c|}{0.617$\pm$0.366}     & 3834.2$\pm$663.004           & 0.232$\pm$0.144            & 11943.4$\pm$3641.285   & \multicolumn{1}{c|}{0.069$\pm$0.078}     & 3108.567$\pm$539.173         & 0.601$\pm$0.218           \\
                     & 3                    & 20179.633$\pm$3894.567 & \multicolumn{1}{c|}{3.086$\pm$1.524}     & 5132.733$\pm$912.672         & 0.04$\pm$0.045             & 16767.267$\pm$3935.0   & \multicolumn{1}{c|}{0.558$\pm$0.385}     & 4674.767$\pm$759.907         & 0.5$\pm$0.254             \\
                     & 4                    & 21179.8$\pm$3848.994   & \multicolumn{1}{c|}{4.68$\pm$2.102}      & 6438.467$\pm$1207.701        & 0.012$\pm$0.013            & 19496.2$\pm$4087.287   & \multicolumn{1}{c|}{1.657$\pm$0.937}     & 5928.433$\pm$995.973         & 0.246$\pm$0.152             \\ \hline  
\multirow{3}{*}{60}  & 2                    & 19738.733$\pm$3961.868 & \multicolumn{1}{c|}{1.92$\pm$0.767}      & 3952.967$\pm$716.434         & 0.08$\pm$0.08              & 15660.6$\pm$4000.346   & \multicolumn{1}{c|}{0.389$\pm$0.302}     & 3360.8$\pm$596.7             & 0.72$\pm$0.279            \\
                     & 3                    & 21826.433$\pm$3679.658 & \multicolumn{1}{c|}{5.992$\pm$2.642}     & 5179.667$\pm$964.602         & 0.028$\pm$0.025            & 19462.8$\pm$3897.446   & \multicolumn{1}{c|}{1.821$\pm$0.952}     & 4966.7$\pm$883.452           & 0.5$\pm$0.422             \\
                     & 4                    & 22403.333$\pm$3556.696 & \multicolumn{1}{c|}{7.682$\pm$2.939}     & 6403.4$\pm$1280.114          & 0.016$\pm$0.014            & 21370.133$\pm$3773.619 & \multicolumn{1}{c|}{3.958$\pm$1.712}     & 6218.467$\pm$1155.359        & 0.379$\pm$0.285             \\ \hline 
\multicolumn{2}{c|}{Mean}                   & 10944.776$\pm$976.174  & \multicolumn{1}{c|}{1.478$\pm$0.322}     & 4810.572$\pm$210.512         & 0.109$\pm$0.024            & 8427.935$\pm$879.608   & \multicolumn{1}{c|}{0.494$\pm$0.145}     & 3687.202$\pm$186.754         & 0.266$\pm$0.043  \\
\multicolumn{2}{c|}{Min.}                   & 16.000               & \multicolumn{1}{c|}{0.000}             & 1271.000                   & 0.000                    & 9.000                & \multicolumn{1}{c|}{0.000}             & 788.000                    & 0.000                   \\
\multicolumn{2}{c|}{Max.}                   & 28987.000            & \multicolumn{1}{c|}{20.253}            & 14383.000                  & 3.287                    & 28958.000            & \multicolumn{1}{c|}{13.385}            & 12515.000                  & 5.462                      \\ \hline
\end{tabular}%
\end{adjustbox}
\caption*{\centering The values are averages over the considered instances, with a 95\% confidence interval.}
\end{table}

\section{Detailed Results for Value of the Stochastic Solution}\label{app:vss}

This section presents the detailed results for the estimators of the {\VSS} defined in Section~\ref{subsec:vss}. To compute these estimators, the {\EV} problem under behavior $a \in \mathcal{A}$ is formulated as $\min\{ Q^a(x, \overline{u}) : x \in \mathcal{X}\}$, where $ \overline{u} = \{\overline{u}\}_{s \in \mathcal{S}, c \in \mathcal{C}} $ represents the vector of average utilities. Therefore, the procedures introduced in Section~\ref{sec:methodology} for the {\SAA} program can be applied to solve the {\EV} problem, but with the distinction that only the average-case scenario is considered, rather than the set $\mathcal{W_N}$ of sampled scenarios. As highlighted in Section~\ref{subsec:vss}, our {\EV} problem is an approximation rather than the standard formulation in stochastic programming. The typical {\EV} problem would consider the expected values of $\boldsymbol{\succ}^{a}_{\mathcal{S}}$. In contrast, we apply the transformation function that determines $\boldsymbol{\succ}^{a}_{\mathcal{S}}$ directly to the average $\overline{u}$ of $\boldsymbol{u}$ to obtain $\overline{\succ}^a_{\mathcal{S}}$. As a result, $\overline{\succ}^a_{\mathcal{S}}$ is not the average of $\boldsymbol{\succ}^{a}_{\mathcal{S}}$. Nonetheless, this is necessary as the endogenous probability distribution of the reported preferences contains numerous scenarios and lacks a straightforward closed form.

Table~\ref{tab:vss} reports the average values of the estimators $\overline{{\VSS}}^a$, $\overline{{\VSSENTER}}^a$, and $\overline{{\VSSIMPROV}}^a$ for each student behavior $a \in \mathcal{A}$, categorized by budget $B$ and maximum rank limit $K$ of the instances. In addition, Figure~\ref{fig:diff-rank-small} displays the values of the estimator $\overline{{\texttt{DIFF-RANK}}}^k$ across different values of $K$ and $B$, averaged over all instances with the same parameters, for budgets $B \in {1, 5, 15, 30}$. The values of the estimator $\overline{{\texttt{DIFF-RANK}}}^k$ for these budgets follow the same pattern observed for budgets $B = 45$ and $B = 60$.
\begin{table}[]
\caption{Average $\overline{{\VSS}}$, $\overline{{\VSSENTER}}$, and $\overline{{\VSSIMPROV}}$ per list size limit $K$ and budget $B$ for large instances.} \label{tab:vss}
\begin{adjustbox}{width=0.8\textwidth,center}%
\begin{tabular}{cc|ccc|ccc|ccc}
\hline
\multirow{2}{*}{$B$} & \multirow{2}{*}{$K$} & \multicolumn{3}{c|}{{\UM} Behavior} & \multicolumn{3}{c|}{{\CEUM} Behavior} & \multicolumn{3}{c}{{\IEUM} Behavior}   \\ \cline{3-11}  
                     &                      & $\overline{{\VSS}}^{\UM}$        & $\overline{{\VSSENTER}}^{\UM}$  & $\overline{{\VSSIMPROV}}^{\UM}$  & $\overline{{\VSS}}^{\CEUM}$        & $\overline{{\VSSENTER}}^{\CEUM}$    & $\overline{{\VSSIMPROV}}^{\CEUM}$  & $\overline{{\VSS}}^{\IEUM}$       & $\overline{{\VSSENTER}}^{\IEUM}$   & $\overline{{\VSSIMPROV}}^{\IEUM}$    \\ \hline  
\multirow{3}{*}{1}   & 2                    & 0.044$\pm$0.024   & 17.244$\pm$9.092  & 19.911$\pm$8.739   & 0.265$\pm$0.092   & 53.472$\pm$10.554   & 43.171$\pm$11.467  & 0.086$\pm$0.03    & 14.914$\pm$10.186  & 13.279$\pm$10.909  \\
                     & 3                    & 0.041$\pm$0.023   & 11.177$\pm$6.49   & 11.609$\pm$6.486   & 0.398$\pm$0.206   & 33.053$\pm$12.396   & 28.329$\pm$10.51   & 0.149$\pm$0.053   & 14.69$\pm$5.522    & 12.81$\pm$6.146    \\
                     & 4                    & 0.036$\pm$0.025   & 5.69$\pm$4.509    & 7.337$\pm$4.263    & 0.266$\pm$0.182   & 15.953$\pm$9.372    & 12.646$\pm$6.136   & 0.191$\pm$0.07    & 13.83$\pm$7.127    & 15.589$\pm$6.281      \\ \hline  
\multirow{3}{*}{5}   & 2                    & 0.323$\pm$0.089   & 26.019$\pm$6.814  & 27.895$\pm$6.734   & 0.808$\pm$0.235   & 45.858$\pm$6.465    & 34.753$\pm$6.52    & 0.349$\pm$0.078   & 7.122$\pm$6.862    & 6.429$\pm$7.145    \\
                     & 3                    & 0.377$\pm$0.133   & 15.603$\pm$5.2    & 17.402$\pm$5.334   & 0.988$\pm$0.258   & 27.351$\pm$6.417    & 21.256$\pm$5.018   & 0.521$\pm$0.101   & 11.069$\pm$4.154   & 10.631$\pm$4.235   \\
                     & 4                    & 0.301$\pm$0.122   & 7.351$\pm$2.435   & 9.249$\pm$2.876    & 0.95$\pm$0.342    & 14.249$\pm$4.206    & 12.601$\pm$3.61    & 0.571$\pm$0.14    & 8.96$\pm$2.936     & 7.87$\pm$2.725       \\ \hline  
\multirow{3}{*}{15}  & 2                    & 1.534$\pm$0.325   & 39.376$\pm$6.615  & 42.006$\pm$6.689   & 1.927$\pm$0.385   & 41.094$\pm$5.157    & 32.747$\pm$5.043   & 0.96$\pm$0.179    & 5.213$\pm$5.332    & 6.128$\pm$5.443    \\
                     & 3                    & 1.904$\pm$0.439   & 26.249$\pm$5.962  & 30.141$\pm$6.197   & 2.638$\pm$0.454   & 18.816$\pm$3.746    & 21.054$\pm$3.161   & 1.472$\pm$0.296   & 10.401$\pm$3.471   & 10.31$\pm$3.534    \\
                     & 4                    & 2.118$\pm$0.623   & 17.702$\pm$5.6    & 22.588$\pm$6.249   & 2.485$\pm$0.548   & 5.753$\pm$1.417     & 12.7$\pm$2.305     & 1.877$\pm$0.378   & 11.154$\pm$2.383   & 11.899$\pm$2.644      \\ \hline  
\multirow{3}{*}{30}  & 2                    & 3.708$\pm$0.538   & 47.048$\pm$6.05   & 51.579$\pm$5.877   & 3.962$\pm$0.842   & 37.775$\pm$4.653    & 34.998$\pm$4.732   & 1.869$\pm$0.314   & 8.002$\pm$4.646    & 8.576$\pm$4.725    \\
                     & 3                    & 5.198$\pm$0.919   & 34.079$\pm$5.738  & 41.793$\pm$6.231   & 4.853$\pm$0.808   & 10.496$\pm$2.964    & 20.401$\pm$2.785   & 2.693$\pm$0.525   & 10.647$\pm$2.492   & 12.341$\pm$2.786   \\
                     & 4                    & 5.91$\pm$1.12     & 20.788$\pm$4.372  & 32.225$\pm$5.502   & 4.374$\pm$0.758   & 2.24$\pm$0.742      & 13.42$\pm$1.928    & 3.506$\pm$0.584   & 9.661$\pm$1.897    & 12.795$\pm$2.464     \\ \hline  
\multirow{3}{*}{45}  & 2                    & 5.566$\pm$0.659   & 47.009$\pm$5.31   & 53.621$\pm$5.066   & 5.519$\pm$1.061   & 31.296$\pm$4.742    & 34.479$\pm$4.148   & 2.782$\pm$0.428   & 9.35$\pm$3.643     & 10.707$\pm$3.923   \\ 
                     & 3                    & 7.732$\pm$1.035   & 31.85$\pm$4.587   & 43.666$\pm$4.658   & 6.807$\pm$0.967   & 6.512$\pm$2.427     & 21.316$\pm$2.62    & 4.109$\pm$0.775   & 10.036$\pm$2.285   & 13.188$\pm$2.789   \\
                     & 4                    & 9.336$\pm$1.179   & 20.581$\pm$3.777  & 37.513$\pm$4.491   & 6.15$\pm$0.901    & 1.088$\pm$0.404     & 14.748$\pm$2.061   & 5.131$\pm$0.869   & 9.08$\pm$1.997     & 14.614$\pm$2.277      \\ \hline  
\multirow{3}{*}{60}  & 2                    & 7.024$\pm$0.708   & 44.569$\pm$5.405  & 53.194$\pm$4.954   & 6.878$\pm$1.094   & 25.314$\pm$4.942    & 33.895$\pm$4.02    & 3.842$\pm$0.671   & 10.167$\pm$3.492   & 12.533$\pm$4.09    \\
                     & 3                    & 9.824$\pm$1.093   & 29.741$\pm$5.202  & 44.934$\pm$4.921   & 7.804$\pm$1.117   & 3.877$\pm$1.752     & 20.124$\pm$2.839   & 5.523$\pm$1.077   & 10.526$\pm$2.488   & 14.945$\pm$3.146   \\
                     & 4                    & 11.149$\pm$1.107  & 16.337$\pm$3.601  & 36.828$\pm$3.827   & 6.921$\pm$0.892   & 0.584$\pm$0.251     & 14.379$\pm$1.984   & 6.49$\pm$1.01     & 7.569$\pm$1.775    & 14.329$\pm$2.345      \\ \hline
\end{tabular}%
\end{adjustbox}
\caption*{\centering The values are averages over the considered instances, with a 95\% confidence interval.}
\end{table}
\begin{figure}[ht]
    \caption{Average $\overline{{\texttt{DIFF-RANK}}}^k$ per list size limit $K$ and budget $B$ ($B\in\{1,5,15,30,45\}$) for large instances.}\label{fig:diff-rank-small}
    \centering
    \begin{subfigure}{0.9\textwidth}
        \centering
        \includegraphics[width=0.7\textwidth]{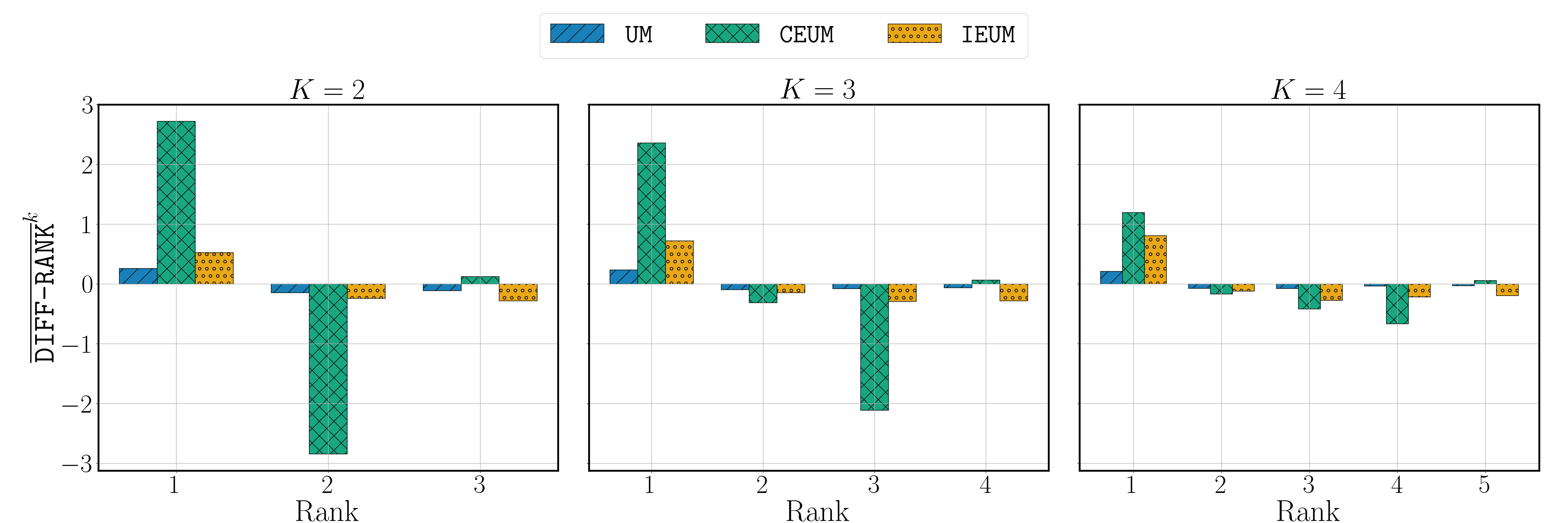}
        \caption{\footnotesize $B=1$.}
    \end{subfigure}
    \hfill 
    \begin{subfigure}{0.9\textwidth}
        \centering
        \includegraphics[width=0.7\textwidth]{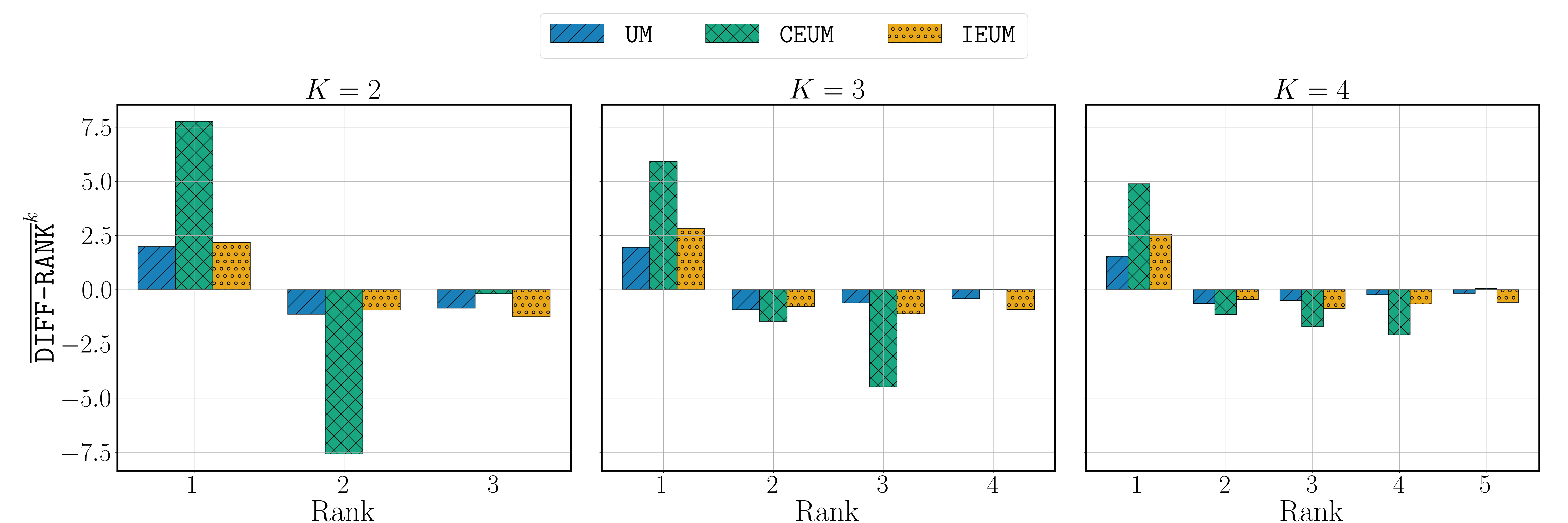}
        \caption{\footnotesize $B=5$.}
    \end{subfigure}
    \hfill
    \begin{subfigure}{0.9\textwidth}
        \centering
        \includegraphics[width=0.7\textwidth]{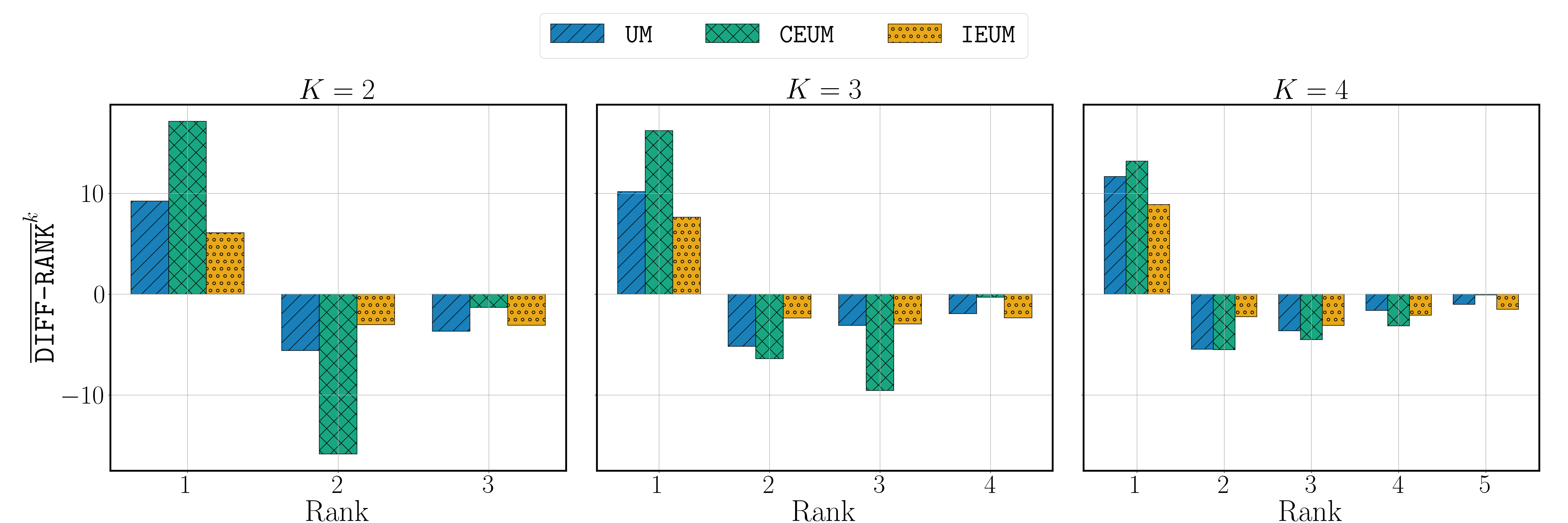}
        \caption{\footnotesize $B=15$.}
    \end{subfigure}
    \hfill
    \begin{subfigure}{0.9\textwidth}
        \centering
        \includegraphics[width=0.7\textwidth]{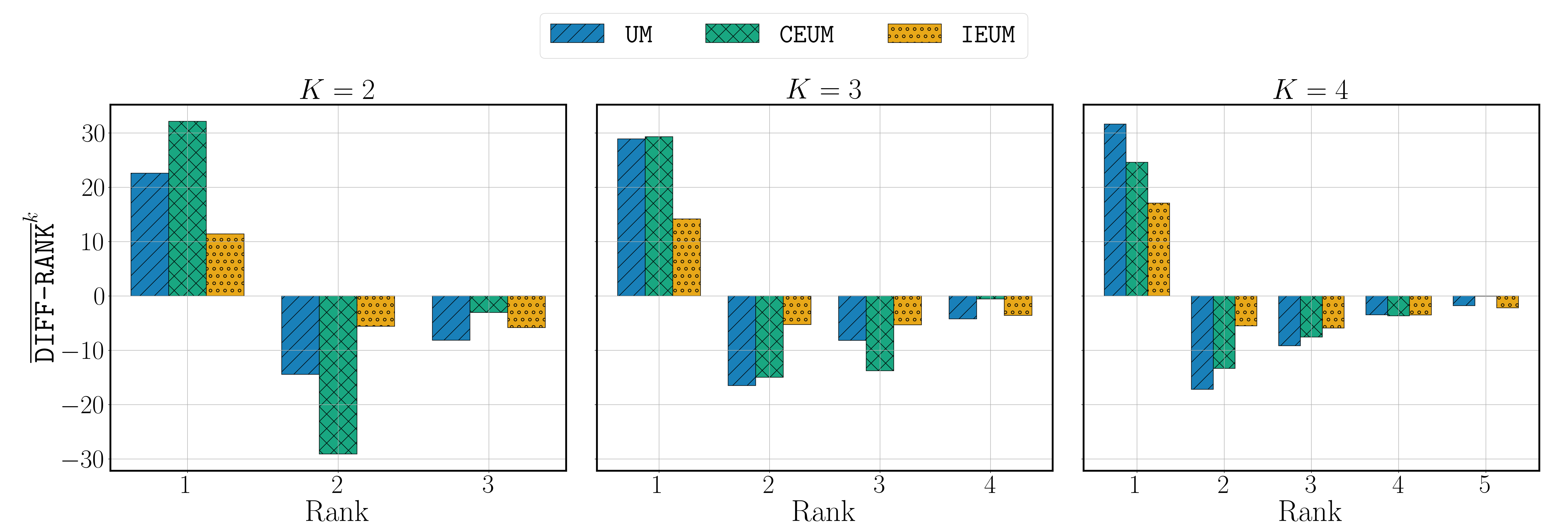}
        \caption{\footnotesize $B=30$.}
    \end{subfigure}
    \begin{subfigure}{0.9\textwidth}
        \centering
        \includegraphics[width=0.7\textwidth]{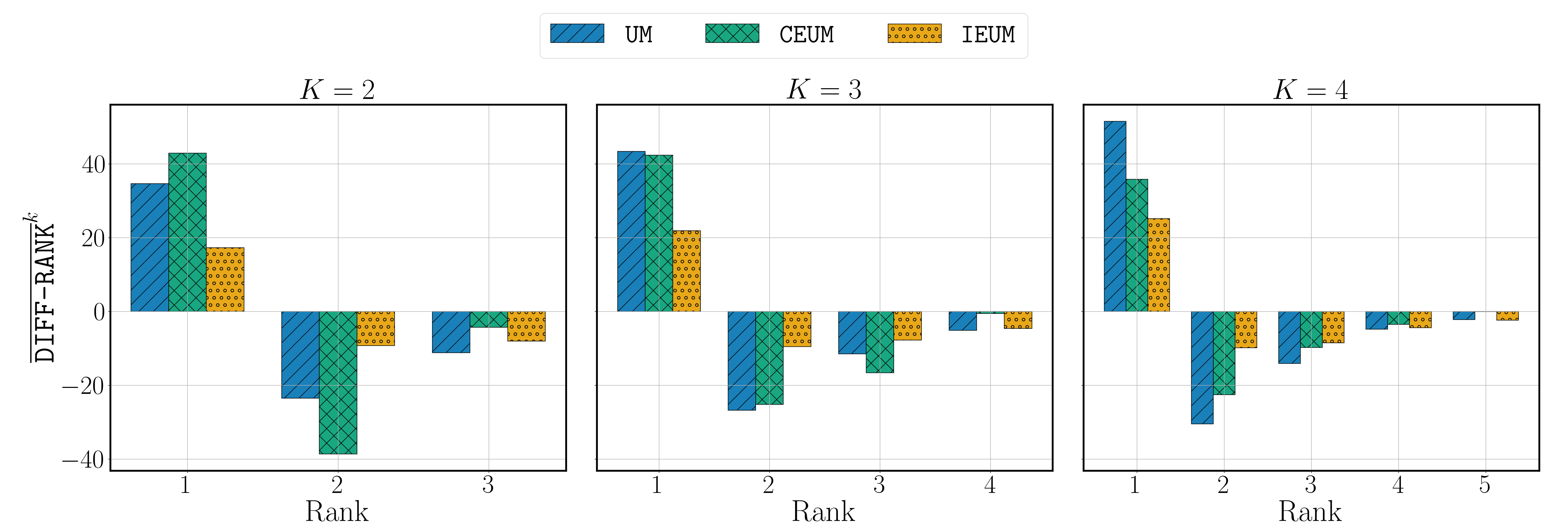}
        \caption{\footnotesize $B=45$.}
    \end{subfigure}
\end{figure}

\section{Detailed Results for Comparison of Students’ Behavior}\label{app:comparison-strategic}

This section presents detailed results comparing the capacity expansion decisions obtained under different student behavior assumptions. Table~\ref{tab:strat} reports both stochastic and deterministic behavior gaps ($a'$--$a$) for two key metrics: students' matching preferences ($\overline{\texttt{gap}}_{N'}^{a'-a}$ and $\overline{\texttt{gap}}_{ev,N'}^{a'-a}$) and the number of students who benefit from additional capacity ($\overline{\texttt{gap-b}}_{N'}^{a'-a}$ and $\overline{\texttt{gap-b}}_{ev,N'}^{a'-a}$). In particular, we focus on the behavioral gaps between ({\UM}--{\CEUM}), ({\UM}--{\IEUM}), and ({\IEUM}--{\CEUM}).
\begin{table}
\centering
\caption{Stochastic and deterministic behavior gaps per list size limit $K$ and budget $B$ for large instances.}\label{tab:strat}
\resizebox{\textwidth}{!}{%
\begin{tabular}{cc|cccccc|cccccc}
\hline
\multirow{3}{*}{$B$} & \multirow{3}{*}{$K$} & \multicolumn{6}{c|}{Stochastic}                                                                                                                                         & \multicolumn{6}{c}{Deterministic}                                                                                                                                         \\ \cline{3-14}  
                     &                      & \multicolumn{2}{c|}{{\UM}--{\CEUM}}                              & \multicolumn{2}{c|}{{\UM}-{\IEUM}}                                & \multicolumn{2}{c|}{{\IEUM}--{\CEUM}}               & \multicolumn{2}{c|}{{\UM}--{\CEUM}}                               & \multicolumn{2}{c|}{{\UM}--{\IEUM}}                               & \multicolumn{2}{c}{{\IEUM}--{\CEUM}}               \\
                     &                      & \multicolumn{1}{c}{$\overline{\texttt{gap}}_{N'}^{{\UM}-{\CEUM}}$} & \multicolumn{1}{c|}{$\overline{\texttt{gap-b}}_{N'}^{{\UM}-{\CEUM}}$}               & \multicolumn{1}{c}{$\overline{\texttt{gap}}_{N'}^{{\UM}-{\IEUM}}$} & \multicolumn{1}{c|}{$\overline{\texttt{gap-b}}_{N'}^{{\UM}-{\IEUM}}$}                & \multicolumn{1}{c}{$\overline{\texttt{gap}}_{N'}^{{\IEUM}-{\CEUM}}$} & \multicolumn{1}{c|}{$\overline{\texttt{gap-b}}_{N'}^{{\IEUM}-{\CEUM}}$} & \multicolumn{1}{c}{$\overline{\texttt{gap}}_{ev,N'}^{{\UM}-{\CEUM}}$} & \multicolumn{1}{c|}{$\overline{\texttt{gap-b}}_{ev,N'}^{{\UM}-{\CEUM}}$}                & \multicolumn{1}{c}{$\overline{\texttt{gap}}_{ev,N'}^{{\UM}-{\IEUM}}$} & \multicolumn{1}{c|}{$\overline{\texttt{gap-b}}_{ev,N'}^{{\UM}-{\IEUM}}$}               & \multicolumn{1}{c}{$\overline{\texttt{gap}}_{ev,N'}^{{\IEUM}-{\CEUM}}$} & \multicolumn{1}{c}{$\overline{\texttt{gap-B}}_{ev,N'}^{{\IEUM}-{\CEUM}}$}    \\ \hline  
\multirow{3}{*}{1}   & 2                    & 0.289$\pm$0.087        & \multicolumn{1}{l|}{50.995$\pm$9.251} & 0.091$\pm$0.031        & \multicolumn{1}{l|}{24.315$\pm$11.666} & 0.468$\pm$1.79         & 85.508$\pm$3.283        & 0.287$\pm$0.088        & \multicolumn{1}{l|}{53.89$\pm$7.222}   & 0.089$\pm$0.031        & \multicolumn{1}{l|}{22.69$\pm$9.501}  & 0.464$\pm$1.794        & 84.266$\pm$3.69        \\
                     & 3                    & 0.348$\pm$0.205        & \multicolumn{1}{l|}{28.784$\pm$9.595} & 0.118$\pm$0.052        & \multicolumn{1}{l|}{21.679$\pm$9.027}  & 2.419$\pm$2.416        & 91.012$\pm$2.16         & 0.395$\pm$0.204        & \multicolumn{1}{l|}{29.097$\pm$10.303} & 0.125$\pm$0.045        & \multicolumn{1}{l|}{18.872$\pm$9.354} & 2.304$\pm$2.418        & 83.776$\pm$3.433       \\
                     & 4                    & 0.257$\pm$0.186        & \multicolumn{1}{l|}{14.168$\pm$6.977} & 0.156$\pm$0.068        & \multicolumn{1}{l|}{14.285$\pm$7.78}   & 5.831$\pm$3.022        & 86.841$\pm$4.13         & 0.317$\pm$0.2          & \multicolumn{1}{l|}{18.304$\pm$7.002}  & 0.171$\pm$0.066        & \multicolumn{1}{l|}{19.829$\pm$8.258} & 5.705$\pm$3.018        & 82.052$\pm$3.894          \\ \hline  
\multirow{3}{*}{5}   & 2                    & 0.773$\pm$0.216        & \multicolumn{1}{l|}{37.673$\pm$6.446} & 0.337$\pm$0.07         & \multicolumn{1}{l|}{17.432$\pm$5.959}  & 2.709$\pm$1.391        & 69.244$\pm$4.313        & 0.855$\pm$0.226        & \multicolumn{1}{l|}{44.119$\pm$5.595}  & 0.289$\pm$0.049        & \multicolumn{1}{l|}{16.194$\pm$5.93}  & 2.805$\pm$1.416        & 68.907$\pm$4.534       \\
                     & 3                    & 0.729$\pm$0.246        & \multicolumn{1}{l|}{16.632$\pm$4.919} & 0.447$\pm$0.119        & \multicolumn{1}{l|}{15.085$\pm$4.768}  & 7.019$\pm$2.136        & 75.237$\pm$3.475        & 1.087$\pm$0.272        & \multicolumn{1}{l|}{22.93$\pm$4.954}   & 0.38$\pm$0.093         & \multicolumn{1}{l|}{13.577$\pm$4.444} & 6.954$\pm$2.141        & 73.466$\pm$3.925       \\
                     & 4                    & 0.508$\pm$0.319        & \multicolumn{1}{l|}{7.199$\pm$3.5}    & 0.546$\pm$0.144        & \multicolumn{1}{l|}{12.859$\pm$3.309}  & 12.498$\pm$2.858       & 74.3$\pm$3.957          & 0.986$\pm$0.408        & \multicolumn{1}{l|}{12.998$\pm$4.384}  & 0.55$\pm$0.143         & \multicolumn{1}{l|}{14.261$\pm$3.788} & 12.299$\pm$2.793       & 71.633$\pm$4.632          \\ \hline  
\multirow{3}{*}{15}  & 2                    & 1.684$\pm$0.375        & \multicolumn{1}{l|}{31.746$\pm$4.905} & 0.927$\pm$0.241        & \multicolumn{1}{l|}{16.605$\pm$4.967}  & 5.568$\pm$1.25         & 50.26$\pm$5.194         & 2.048$\pm$0.501        & \multicolumn{1}{l|}{39.684$\pm$4.962}  & 0.707$\pm$0.129        & \multicolumn{1}{l|}{16.572$\pm$4.609} & 6.016$\pm$1.315        & 50.94$\pm$5.79         \\
                     & 3                    & 1.461$\pm$0.381        & \multicolumn{1}{l|}{12.715$\pm$3.379} & 1.37$\pm$0.355         & \multicolumn{1}{l|}{15.143$\pm$3.202}  & 13.023$\pm$2.289       & 55.355$\pm$6.028        & 2.802$\pm$0.629        & \multicolumn{1}{l|}{21.849$\pm$4.188}  & 1.084$\pm$0.224        & \multicolumn{1}{l|}{15.19$\pm$4.073}  & 13.31$\pm$2.35         & 54.795$\pm$6.38        \\
                     & 4                    & 0.564$\pm$0.355        & \multicolumn{1}{l|}{3.214$\pm$1.737}  & 1.774$\pm$0.452        & \multicolumn{1}{l|}{14.04$\pm$2.66}    & 18.254$\pm$3.426       & 47.815$\pm$7.565        & 2.586$\pm$0.599        & \multicolumn{1}{l|}{13.735$\pm$3.441}  & 1.495$\pm$0.289        & \multicolumn{1}{l|}{13.829$\pm$2.867} & 18.445$\pm$3.418       & 47.557$\pm$7.824         \\ \hline  
\multirow{3}{*}{30}  & 2                    & 3.221$\pm$0.74         & \multicolumn{1}{l|}{29.385$\pm$4.184} & 1.882$\pm$0.485        & \multicolumn{1}{l|}{16.724$\pm$4.956}  & 6.433$\pm$1.476        & 38.615$\pm$4.991        & 4.109$\pm$0.921        & \multicolumn{1}{l|}{40.16$\pm$4.464}   & 1.385$\pm$0.279        & \multicolumn{1}{l|}{20.216$\pm$3.627} & 7.097$\pm$1.653        & 38.67$\pm$5.872        \\
                     & 3                    & 2.49$\pm$0.762         & \multicolumn{1}{l|}{10.494$\pm$2.776} & 2.715$\pm$0.767        & \multicolumn{1}{l|}{15.254$\pm$3.147}  & 12.727$\pm$2.61        & 35.077$\pm$6.191        & 6.09$\pm$0.913         & \multicolumn{1}{l|}{25.908$\pm$3.46}   & 2.026$\pm$0.41         & \multicolumn{1}{l|}{16.964$\pm$3.148} & 13.164$\pm$2.753       & 34.937$\pm$6.741       \\
                     & 4                    & 0.955$\pm$0.556        & \multicolumn{1}{l|}{2.664$\pm$1.355}  & 3.494$\pm$0.911        & \multicolumn{1}{l|}{14.19$\pm$2.846}   & 14.788$\pm$3.256       & 26.217$\pm$6.143        & 6.462$\pm$0.955        & \multicolumn{1}{l|}{20.229$\pm$3.558}  & 2.743$\pm$0.627        & \multicolumn{1}{l|}{15.792$\pm$2.64}  & 13.433$\pm$3.551       & 22.012$\pm$6.636          \\ \hline  
\multirow{3}{*}{45}  & 2                    & 4.296$\pm$0.895        & \multicolumn{1}{l|}{26.669$\pm$4.081} & 2.867$\pm$0.743        & \multicolumn{1}{l|}{17.693$\pm$4.422}  & 6.248$\pm$1.479        & 32.633$\pm$4.089        & 5.759$\pm$1.197        & \multicolumn{1}{l|}{38.286$\pm$4.335}  & 2.131$\pm$0.492        & \multicolumn{1}{l|}{22.598$\pm$3.868} & 6.678$\pm$1.653        & 30.679$\pm$4.811       \\
                     & 3                    & 3.093$\pm$0.918        & \multicolumn{1}{l|}{9.122$\pm$2.375}  & 4.134$\pm$1.079        & \multicolumn{1}{l|}{16.019$\pm$3.255}  & 11.309$\pm$2.012       & 26.941$\pm$3.861        & 8.612$\pm$1.006        & \multicolumn{1}{l|}{26.665$\pm$2.981}  & 3.09$\pm$0.682         & \multicolumn{1}{l|}{18.587$\pm$3.035} & 10.022$\pm$2.384       & 22.455$\pm$4.902       \\
                     & 4                    & 1.116$\pm$0.604        & \multicolumn{1}{l|}{2.332$\pm$1.008}  & 5.19$\pm$1.206         & \multicolumn{1}{l|}{15.025$\pm$2.692}  & 11.55$\pm$1.654        & 19.144$\pm$3.287        & 9.092$\pm$1.026        & \multicolumn{1}{l|}{22.42$\pm$3.578}   & 4.046$\pm$0.767        & \multicolumn{1}{l|}{16.354$\pm$2.172} & 8.736$\pm$2.387        & 13.093$\pm$3.668          \\ \hline  
\multirow{3}{*}{60}  & 2                    & 5.477$\pm$0.971        & \multicolumn{1}{l|}{25.634$\pm$3.814} & 3.888$\pm$0.981        & \multicolumn{1}{l|}{18.417$\pm$4.216}  & 6.387$\pm$1.23         & 29.012$\pm$3.228        & 7.434$\pm$1.183        & \multicolumn{1}{l|}{37.853$\pm$4.157}  & 2.784$\pm$0.663        & \multicolumn{1}{l|}{23.156$\pm$3.821} & 6.898$\pm$1.539        & 27.689$\pm$4.129       \\
                     & 3                    & 3.43$\pm$1.058         & \multicolumn{1}{l|}{7.998$\pm$2.163}  & 5.461$\pm$1.309        & \multicolumn{1}{l|}{16.984$\pm$3.118}  & 10.045$\pm$1.187       & 21.886$\pm$2.664        & 10.253$\pm$1.264       & \multicolumn{1}{l|}{26.065$\pm$3.154}  & 4.281$\pm$0.983        & \multicolumn{1}{l|}{20.198$\pm$3.011} & 7.668$\pm$1.529        & 15.627$\pm$3.013       \\
                     & 4                    & 1.106$\pm$0.558        & \multicolumn{1}{l|}{1.919$\pm$0.808}  & 6.572$\pm$1.389        & \multicolumn{1}{l|}{15.413$\pm$2.651}  & 10.425$\pm$1.367       & 17.837$\pm$3.949        & 10.534$\pm$1.113       & \multicolumn{1}{l|}{21.709$\pm$2.929}  & 5.297$\pm$1.161        & \multicolumn{1}{l|}{17.59$\pm$2.697}  & 6.845$\pm$1.253        & 10.432$\pm$2.55          \\ \hline
\end{tabular}%
}
\caption*{\centering The values are averages over the considered instances, with a 95\% confidence interval.}
\end{table}

\end{document}